\documentclass[12pt]{article}
\pdfoutput=1
\usepackage[utf8x]{inputenc}
\usepackage[colorlinks,linkcolor=blue,citecolor=blue,bookmarks,bookmarksnumbered]{hyperref}
\usepackage[scaled=0.85]{helvet}
\usepackage{amsmath,amssymb,accents,mathrsfs,XoohmE}
\usepackage{graphicx,color}
\usepackage{booktabs}
\usepackage{multirow}
\usepackage{placeins}
\usepackage{amsmath}

\usepackage{enumitem}
\usepackage[aligntableaux=center,boxsize=1.2em]{ytableau}

\usepackage{XoohmE}

\definecolor{Green}  {rgb}{0.10,0.70,0.10} 
\definecolor{Orange} {rgb}{1.00,0.50,0.15} 
\definecolor{Red}    {rgb}{0.90,0.00,0.12} 
\definecolor{Purple} {rgb}{0.50,0.25,0.55} 
\definecolor{Turque} {rgb}{0.00,0.65,0.85} 
\definecolor{Blue}   {rgb}{0.00,0.00,1.00} 
\definecolor{Magenta}{rgb}{1.00,0.00,1.00} 
\definecolor{Gold}   {rgb}{1.00,0.75,0.25} 
\definecolor{Seaweed}{rgb}{0.01,0.24,0.09} 
\definecolor{Brown}  {rgb}{0.43,0.26,0.32} 
\definecolor{grey1}  {rgb}{0.20,0.20,0.20} 
\definecolor{grey2}  {rgb}{0.40,0.40,0.40} 
\definecolor{grey3}  {rgb}{0.60,0.60,0.60} 
\definecolor{grey4}  {rgb}{0.80,0.80,0.80} 
\definecolor{grey5}  {rgb}{0.90,0.90,0.90} 
\def\C#1#2{{\ifcase#1\or
             \color{Green}\or \color{Orange}\or \color{Red}\or
              \color{Purple}\or \color{Turque}\or \color{Blue}\or
               \color{Magenta}\or \color{Gold}\or \color{Seaweed}\or
                \color{Brown}\or\color{grey1}\or\color{grey2}\or
                 \color{grey3}\else\color{grey4}\fi#2}}

\definecolor{Slate} {rgb}{0.00,0.45,0.55}

\usepackage[enableskew,vcentermath]{youngtab}
\let\TC=\textcolor
\definecolor{Hey}{rgb}{.9,.05,.4}
\definecolor{orange}{rgb}{1,.5,0}
\definecolor{plum}{rgb}{.4,0,.6}
\definecolor{R}{rgb}{1,0,0}
\definecolor{G}{rgb}{0.1,0.7,0}
\definecolor{B}{rgb}{0,0,1}

\long\def\CMTred#1{\leavevmode\TC{red}{\sf#1}}

\long\def\CMTR#1{\leavevmode\TC{R}{\sf#1}}
\long\def\CMTG#1{\leavevmode\TC{G}{\sf#1}}
\long\def\CMTB#1{\leavevmode\TC{B}{\sf#1}}

\usepackage{lipsum}
\usepackage{listings}
\definecolor{MyDarkGreen}{rgb}{0.0,0.4,0.0} 
\def\fracm#1#2{\hbox{\large{${\frac{{#1}}{{#2}}}$}}}

\def\be{\begin{equation}}
\def\ee{\end{equation}}
\newcommand{\bea}{\begin{eqnarray}}
\newcommand{\eea}{\end{eqnarray}}
\newcommand{\ena}{\end{eqnarray}}

\def\pp{{\mathchoice
          {
              \kern 1pt%
              \raise 1pt
              \vbox{\hrule width5pt height0.4pt depth0pt
                    \kern -2pt
                    \hbox{\kern 2.3pt
                          \vrule width0.4pt height6pt depth0pt
                          }
                    \kern -2pt
                    \hrule width5pt height0.4pt depth0pt}%
                    \kern 1pt
           }
            {
              \kern 1pt%
              \raise 1pt
              \vbox{\hrule width4.3pt height0.4pt depth0pt
                    \kern -1.8pt
                    \hbox{\kern 1.95pt
                          \vrule width0.4pt height5.4pt depth0pt
                          }
                    \kern -1.8pt
                    \hrule width4.3pt height0.4pt depth0pt}%
                    \kern 1pt
            }
            {
              \kern 0.5pt%
              \raise 1pt
              \vbox{\hrule width4.0pt height0.3pt depth0pt
                    \kern -1.9pt  
                    \hbox{\kern 1.85pt
                          \vrule width0.3pt height5.7pt depth0pt
                          }
                    \kern -1.9pt
                    \hrule width4.0pt height0.3pt depth0pt}%
                    \kern 0.5pt
            }
            {
              \kern 0.5pt%
              \raise 1pt
              \vbox{\hrule width3.6pt height0.3pt depth0pt
                    \kern -1.5pt
                    \hbox{\kern 1.65pt
                          \vrule width0.3pt height4.5pt depth0pt
                          }
                    \kern -1.5pt
                    \hrule width3.6pt height0.3pt depth0pt}%
                    \kern 0.5pt
            }
        }}

\def\mm{{\mathchoice
                       {
                             \kern 1pt
               \raise 1pt    \vbox{\hrule width5pt height0.4pt depth0pt
                                  \kern 2pt
                                  \hrule width5pt height0.4pt depth0pt}
                             \kern 1pt}
                       {
                            \kern 1pt
               \raise 1pt \vbox{\hrule width4.3pt height0.4pt depth0pt
                                  \kern 1.8pt
                                  \hrule width4.3pt height0.4pt depth0pt}
                             \kern 1pt}
                       {
                            \kern 0.5pt
               \raise 1pt
                            \vbox{\hrule width4.0pt height0.3pt depth0pt
                                  \kern 1.9pt
                                  \hrule width4.0pt height0.3pt depth0pt}
                            \kern 1pt}
                       {
                           \kern 0.5pt
             \raise 1pt  \vbox{\hrule width3.6pt height0.3pt depth0pt
                                  \kern 1.5pt
                                  \hrule width3.6pt height0.3pt depth0pt}
                           \kern 0.5pt}
                       }}

\def\ad{{\kern0.5pt
                   \alpha \kern-5.05pt \raise5.8pt\hbox{$\textstyle.$}\kern
0.5pt}}

\def\bd{{\kern0.5pt
                   \beta \kern-5.05pt \raise5.8pt\hbox{$\textstyle.$}\kern
0.5pt}}

\def\qd{{\kern0.5pt
                   q \kern-5.05pt \raise5.8pt\hbox{$\textstyle.$}\kern
0.5pt}}
\def\Dot#1{{\kern0.5pt
     {#1} \kern-5.05pt \raise5.8pt\hbox{$\textstyle.$}\kern
0.5pt}}

\catcode`@=11
\def\un#1{\relax\ifmmode\@@underline#1\else
        $\@@underline{\hbox{#1}}$\relax\fi}
\catcode`@=12

\def\a{\alpha}
\def\b{\beta}
\def\c{\chi}
\def\d{\delta}
\def\e{\epsilon}

\def\g{\gamma}

\def\i{\iota}
\def\j{\psi}
\def\k{\kappa}
\def\l{\lambda}

\def\o{\omega}

\def\r{\rho}
\def\s{\sigma}
\def\t{\tau}

\def\L{\Lambda}
\def\O{\Omega}
\def\P{\Pi}

\def\S{\Sigma}

\def\ca{{\cal A}}

\def\cf{{\cal F}}

\def\dslash{\not{\hbox{\kern-2pt $\partial$}}}
\def\Dslash{\not{\hbox{\kern-4pt $D$}}}
\def\pslash{\not{\hbox{\kern-2.3pt $p$}}}
 \newtoks\slashfraction
 \slashfraction={.13}
 \def\slash#1{\setbox0\hbox{$ #1 $}
 \setbox0\hbox to \the\slashfraction\wd0{\hss \box0}/\box0 }
 
\def\kcr{{\hbox{\ro \char'170}}}                
\def\ktl{{\hbox{\ro \char'170}}}        
\def\ktr{{\hbox{\ro \char'170}}}        
\def\kbl{{\hbox{\ro \char'170}}}        
\def\kbr{{\hbox{\ro \char'170}}}        

\def\plpl{\raise-2pt\hbox{$\raise3pt\hbox{$_+$}\hskip-6.67pt\raise0.0pt
\hbox{$^+$}\hskip 0.01pt$}}
\def\mimi{\raise-2pt\hbox{$\raise3pt\hbox{$_-$}\hskip-6.67pt\raise0.0pt
\hbox{$^-$}\hskip 0.01pt$}}

\def\bo{{\raise.15ex\hbox{\large$\Box$}}}               
\def\pa{\partial}                                       
\def\TH{{\raise.2ex\hbox{$\displaystyle \bigodot$}\mskip-4.7mu \llap H \;}}
\def\face{{\raise.2ex\hbox{$\displaystyle \bigodot$}\mskip-2.2mu \llap {$\ddot
        \smile$}}}                                      

\def\dt#1{\on{\hbox{\bf .}}{#1}}                
\def\Dot#1{\dt{#1}}

\def\Hat#1{\widehat{#1}}                        
\def\Bar#1{\overline{#1}}                       
\def\leftrightarrowfill{$\mathsurround=0pt \mathord\leftarrow \mkern-6mu
        \cleaders\hbox{$\mkern-2mu \mathord- \mkern-2mu$}\hfill
        \mkern-6mu \mathord\rightarrow$}
\def\dvec#1{\vbox{\ialign{##\crcr
        \leftrightarrowfill\crcr\noalign{\kern-1pt\nointerlineskip}
        $\hfil\displaystyle{#1}\hfil$\crcr}}}           
\def\dt#1{{\buildrel {\hbox{\LARGE .}} \over {#1}}}     

\def\fracm#1#2{\hbox{\large{${\frac{{#1}}{{#2}}}$}}}
\def\sfrac#1#2{{\vphantom1\smash{\lower.5ex\hbox{\small$#1$}}\over
        \vphantom1\smash{\raise.4ex\hbox{\small$#2$}}}} 
\def\bfrac#1#2{{\vphantom1\smash{\lower.5ex\hbox{$#1$}}\over
        \vphantom1\smash{\raise.3ex\hbox{$#2$}}}}       
\def\afrac#1#2{{\vphantom1\smash{\lower.5ex\hbox{$#1$}}\over#2}}    

\def\pa{\partial}      
\let\bm\relax
\newcommand{\bm}[1]{{\boldsymbol{#1}}}

\def\ad{{\Dot{\alpha}}}
\def\bd{{\Dot{\beta}}}

 \font\rOpe=cmsy10                        
 \def\ktl{{\hbox{\rOpe\char'170}}}        
 \def\kbl{{\hbox{\rOpe\char'170}}}        
 \def\kcr{{\reflectbox{\rOpe\char'170}}}        
 \def\ktr{{\reflectbox{\rOpe\char'170}}}        
 \def\kbr{{\reflectbox{\rOpe\char'170}}}        
 \def\Border{\vbox{\hsize0pt
        \setlength{\unitlength}{1mm}
        \newcount\xco
        \newcount\yco
        \xco=-21
        \yco=12
        \begin{picture}(0,0)(-7.5,0)
        \put(\xco,\yco){$\ktl$}
        \advance\yco by-1
        {\loop
        \put(\xco,\yco){$\kcr$}
        \advance\yco by-2
        \ifnum\yco>-240
        \repeat
        \put(\xco,\yco){$\kbl$}}
        \xco=170
        \yco=12
        \put(\xco,\yco){$\ktr$}
        \advance\yco by-1
        {\loop
        \put(\xco,\yco){$\kcr$}
        \advance\yco by-2
        \ifnum\yco>-240
        \repeat
        \put(\xco,\yco){$\kbr$}}
        \put(-19.5,13){\scalebox{.6065}{%
         University of Maryland Center for String and Particle  Theory \&\ Physics Department%
        |University of Maryland Center for String and Particle  Theory \&\ Physics Department}}
        \put(-19.5,-241.5){\scalebox{.5835}{%
         ****University of Maryland * Center for String and
         Particle  Theory* Physics Department****University of Maryland *Center
        for String and Particle  Theory* Physics Department}}
        \end{picture}
        \par\vskip-8mm}}
\definecolor{UMred}{rgb}{.9,.05,.2}
\definecolor{HUblue}{rgb}{.0,.3,.7}

\definecolor{Red}    {rgb}{0.90,0.00,0.12} 
\definecolor{Blue}   {rgb}{0.00,0.00,1.00} 
\definecolor{Green}  {rgb}{0.10,0.70,0.10} 
\definecolor{Turque} {rgb}{0.00,0.65,0.85} 
\definecolor{Orange} {rgb}{1.00,0.50,0.15} 
\definecolor{Magenta}{rgb}{1.00,0.00,1.00} 
\definecolor{Gold}   {rgb}{1.00,0.75,0.25} 
\definecolor{Seaweed}{rgb}{0.01,0.24,0.09} 
\definecolor{Purple} {rgb}{0.50,0.25,0.55} 
\definecolor{Brown}  {rgb}{0.43,0.26,0.32} 
\definecolor{grey1}  {rgb}{0.20,0.20,0.20} 
\definecolor{grey2}  {rgb}{0.40,0.40,0.40} 
\definecolor{grey3}  {rgb}{0.60,0.60,0.60} 
\definecolor{grey4}  {rgb}{0.80,0.80,0.80} 
\definecolor{grey5}  {rgb}{0.90,0.90,0.90} 
\def\C#1#2{{\ifcase#1\or
             \color{Red}\or \color{Green}\or \color{Blue}\or\
              \color{Turque}\or \color{Orange}\or \color{Magenta}\or
               \color{Gold}\or \color{Seaweed}\or \color{Purple}\or
                \color{Brown}\or\color{grey1}\or\color{grey2}\or
                 \color{grey3}\else\color{grey4}\fi#2}}

\definecolor{Slate} {rgb}{0.00,0.45,0.55}

\newdimen\parshift\parshift=\parindent
\catcode`@=11
 \long\def\@footnotetext#1{\insert\footins{\reset@font\footnotesize
           \interlinepenalty\interfootnotelinepenalty\splittopskip%
            \footnotesep\splitmaxdepth\dp\strutbox\floatingpenalty\@MM%
             \hsize\columnwidth\addtolength{\hsize}{-2\parindent}
              \@parboxrestore\protected@edef\@currentlabel%
              {\csname p@footnote\endcsname\@thefnmark}%
                \color@begingroup%
                 \@makefntext{\rule\z@\footnotesep\ignorespaces#1%
                  \@finalstrut\strutbox}%
                \color@endgroup}}
 \long\def\@makefntext#1{\hglue\parshift%
           \vbox{\noindent\baselineskip=11pt plus.5pt minus.5pt\hb@xt@0em{\hss\@makefnmark\kern1pt}#1}}
\catcode`@=12

\newskip\humongous \humongous=0pt plus 1000pt minus 1000pt
\def\caja{\mathsurround=0pt}
\def\eqalign#1{\,\vcenter{\openup2\jot \caja
        \ialign{\strut \hfil$\displaystyle{##}$&$
        \displaystyle{{}##}$\hfil\crcr#1\crcr}}\,}
\newif\ifdtup

\makeatletter
\def\section{\@startsection{section}{1}{\z@}
        {3ex plus-1ex minus-.2ex}{1pt plus1pt}{\large\sf\bfseries\boldmath}}
\def\subsection{\@startsection{subsection}{2}{\z@}
         {1.5ex plus-1ex minus-.2ex}{0.01pt plus1pt}{\sf\slshape}}
\def\subsubsection{\@startsection{subsubsection}{3}{\z@}
          {1.5ex plus-1ex minus-.2ex}{0.01pt plus0.2pt}{\sf\boldmath}}
\def\paragraph{\@startsection{paragraph}{4}{\z@}
           {.75ex \@plus.5ex \@minus.2ex}{-2mm}{\sf\bfseries\boldmath}}
\makeatother

 \allowdisplaybreaks
 \seceq

\newcommand{\aone}{{\un a}_1}
\newcommand{\bone}{{\un b}_1}
\newcommand{\cone}{{\un c}_1}
\newcommand{\done}{{\un d}_1}
\newcommand{\eone}{{\un e}_1}

\newcommand{\atwo}{{\un a}_2}
\newcommand{\btwo}{{\un b}_2}
\newcommand{\ctwo}{{\un c}_2}

\newcommand{\athree}{{\un a}_3}

\newcommand{\afour}{{\un a}_4}

\definecolor{skyblue}{rgb}{0.12, 0.46, 1.00}
\definecolor{brightpink}{rgb}{1.0, 0.0, 0.5}
\definecolor{darkgreen}{rgb}{0.10, 0.75, 0.24}

\begin{document}

\thispagestyle{empty}
\noindent{\small
\hfill{$~~$}  \\ 
{}
}
\begin{center}
{\large \bf
Weyl Covariance, and Proposals for   \vskip0.02in
Superconformal Prepotentials in 10D Superspaces
}   \\   [8mm]
{\large {
S.\ James Gates, Jr.\footnote{sylvester$_-$gates@brown.edu}${}^{,a, b}$,
Yangrui Hu\footnote{yangrui$_-$hu@brown.edu}${}^{,a,b}$, and
S.-N. Hazel Mak\footnote{sze$_-$ning$_-$mak@brown.edu}${}^{,a,b}$
}}
\\*[6mm]
\emph{
\centering
$^{a}$Brown Theoretical Physics Center,
\\[1pt]
Box S, 340 Brook Street, Barus Hall,
Providence, RI 02912, USA
\\[10pt]
$^{b}$Department of Physics, Brown University,
\\[1pt]
Box 1843, 182 Hope Street, Barus \& Holley,
Providence, RI 02912, USA
}
 \\*[60mm]
{ ABSTRACT}\\[5mm]
\parbox{142mm}{\parindent=2pc\indent\baselineskip=14pt plus1pt
Proposals are made to describe the Weyl scaling transformation laws 
of supercovariant derivatives $\nabla{}_{\un A}$, the torsion supertensors 
$T{}_{{\un A} \, {\un B}}{}^{{\un C}}$, and curvature supertensors 
$R{}_{{\un A} \, {\un B}}{}_{\, \un c} {}^{\un d}$ in 10D superspaces.
Starting from the proposal that an unconstrained supergravity prepotential
for the 11D, $\cal N$ = 1 theory is described by a scalar superfield,
considerations for supergravity prepotentials in the 10D theories are
enumerated.  We derive infinitesimal 10D superspace Weyl transformation
laws and discover ten possible 10D, $\cal N$ = 1 superfield supergravity
prepotentials. The first identification of all off-shell ten dimensional
supergeometrical Weyl field strength tensors, constructed from
respective torsions, is presented.
} \end{center}
\vfill
\vfill
\clearpage

\newpage
{\hypersetup{linkcolor=black}
\tableofcontents
}

\newpage
\section{Introduction}
\label{sec:NTRO}

The first explicit discussions in the literature on the topic of Weyl symmetry in superspace
were initiated among the works in \cite{HW1,HW2} by Howe et al..  The subject of the interplay between 
conformal symmetry and the constraints of superspace descriptions of Poincar\' e supergravity, 
has long been of fascination to one of the authors \cite{SC1,SC2,SC3,SC4,SC5}.  Of course, 
numbers of other authors have also pursued this subject.  A special area of these considerations 
involves the context of 11D, $\cal N$ = 1 superspace \cite{crD11a,crD11b} related to M-Theory 
\cite{MTh}.  After its beginning, a literature (e.g. \cite{Howe:1997,CGNN,Howe:2003sa,Howe:2003cy,10DScLR,10DConFRM,PVW1,PVW2,PVW3}) has been built up 
including discussions in superspace and also at the level of component fields.  These are but 
a small selection and the interested readers should look at the references in these works for a 
more complete listing of such works.

Near the end of a 1996 investigation \cite{2MT0}, the following paragraph can be found.
\begin{quotation}\itshape
For although we believe our observation is important, we know of at least two arguments 
that suggest that there must exist at least one other tensor superfield that will
be required to have a completely off-shell formalism.  This is to be expected even from
the structure of the non-minimal 4D, ${\cal N} = 1$ supergravity.  There it is known that there
are three algebraically independent tensors $W{}_{{\a \, {\b} \, {\g}}}$, $G{}_{\un a}$
and $T{}_{\a}$,  so apparently it remains
to find the eleven dimensional analog of the $G{}_{\un a}$.  In future works, these aspects of the
eleven dimensional theory will require further study.
\end{quotation}
From our present perspective, the ``strong form'' of the constraints given in a 2000 work by
Cederwall, Gran, Nielsen, and Nilsson \cite{CGNN}, appears to provide the solution
of our 1996 dilemma.  These authors introduced a dimension zero tensor $X_{[5]}{}^{\un{b}}$
that, from our present understanding, is the missing analog to the 4D, $\cal N $ = 1
$G{}_{\un a}$-tensor.

In the work of \cite{M2}, an analysis based on prepotential superfields was undertaken regarding
the scale compensating superfield $\Psi$ and conformal semi-prepotential ${\Hat {\rm H}}{}_{\a}
{}^{\un a}$.  It was shown these play an interesting role with respect to the emergence of Weyl 
scaling covariance in 11D, $\cal N$ = 1 superspace.  Three {\it {key}} points were noted in this 
context:
\begin{quote}
\begin{enumerate}[label={\rm (\alph*.)}] \itshape
\item When a sufficiency of conventional constraints are imposed upon the 11D, ${\cal N} = 1$ 
\un{Poincar\' e} superspace supergravity covariant derivative operators 
$(\nabla_{\a}, \nabla_{\un a})$ so that
$\Psi$ and ${\Hat {\rm H}}{}_{\a}{}^{\un a} $ are the \un{only} 
independent superfields within them, a spinorial connection 
field ${\cal J}{}_{\a}{}^{(+)}$ (for Weyl scaling in superspace) emerges 
among the dimension one-half 
\un{Poincar\'e} supergravity supertensor components.
\item The sufficiency of conventional constraints that leads to the existence of
${\cal J}{}_{\a}{}^{(+)}$ is 
also sufficient to lead to the emergence of a vectorial connection field 
${\cal J}{}_{\un a}{}^{(+)}$ for Weyl scaling in superspace. 
\item The existence of ${\cal J}_{\a}{}^{(+)}$ and ${\cal J}_{\un a}{}^{(+)}$
together with the existence of the superspace supergravity covariant derivative operators
$(\nabla{}_{\a}, \nabla{}_{\un a})$ imply the existence of modified supergravity covariant derivative operators
$(\Hat{\nabla}_{\a}, \Hat{\nabla}_{\un a})$ that transform covariantly with respect 
to Weyl scaling.
\end{enumerate}
\end{quote}
In fact, precisely these three points are at the foundation of a paper written in 1991 within the 
context of 10D, $\cal N$ = 1 superspace \cite{SC5}.

As the focus of this 
analysis is the modified supergravity covariant derivative operators (and their related field 
strengths, Bianchi identities, etc.), this approach is mute on implications for the superfields 
needed to construct the superframe superfields.  This is true for most of the citations among 
\cite{HW1} - \cite{PVW3} as they are not focused upon the superfield variables that are ``inside'' 
of (${ {\nabla}}{}_{\a}$, ${ {\nabla}}{}_{\un a}$), i.\ e. the prepotentials.

We will make an observation about the relation of the 11D, $\cal N$ = 1 Nordstr\"om theory 
and the infinitesimal super Weyl transformation laws of the complete non-linear Poincar\'e
superspace supergravity derivatives (${ {\nabla}}{}_{\a}$, ${ {\nabla}}{}_{\un a}$) as motivating a 
pathway for the derivation of similar results in all ten dimensional superspaces.  
Inspired by the success of relating the emergence of Weyl symmetry in 11D, $\cal N$ = 1
superspace, together with our recent studies of supergravity in eleven and ten dimensions
\cite{GHM1,GHM2,GHM3,GHM4}, we are motivated to extend the discussion and results of 
\cite{M2} into the domain of {\it {all}} ten dimensional supergravity theories in superspace.  This 
is one purpose of this paper.  However, we also wish to push beyond this boundary.

The structure of 4D, N = 2 supergravity at the linearized level is very informative for any reader 
who is concerned about how the gauge degrees of the supergravity 3-form can arise from the 
use of prepotentials.  We will in the following use the work shown in \cite{N2} to discuss the
mechanism seen for how this is to occur.

The relevance of this past work provided the first superspace description involving 
prepotentials where the supergravity supermultiplet contains (in the Poincar\' e limit) 
a gauge propagating physical degrees of freedom.  In this case this field is a 1-form.  
It is analogous to the 3-form in the 11D case.

The work of \cite{N2} showed that 4D, N = 2 superfield supergravity required
three different prepotentials:  
\newline \indent
(a.) a conformal Weyl prepotential,
\newline \indent
(b.) a conformal compensating prepotential, and
\newline \indent
(c.) an SU(2) compensating prepotential.
\newline \noindent
The last of these is required as there is an SU(2) subgroup of the superconformal
group appropriate to this theory.   As in 11D current indications are that there
is no such similar subgroup and thus it seems a reasonable speculation a
corresponding compensator is absent.  On this basis, we have made the
assumption that the eventual formulation of the 11D theory in a standard
Salam-Strathdee superspace will not contain the third type of prepotential.

The work of \cite{N2} also reveals another fact.  It was shown possible in that
work to study the geometry of the superspace in a limit where only the Weyl
prepotential is present.  This is the limit under study in our efforts in both ten
and eleven dimensional theories.

Many years ago, a then surprising discovery about the formulation of supergravity
in superspace was revealed.  The dynamics of the on-shell systems can be derived 
simply for using the superspace supergravity Bianchi and imposing a set of 
constraints on the torsion and curvature super tensors.  This was first shown in 
the works of \cite{GWZ1,GWZ2}.  Our efforts are directed along this historical arc 
and do not introduce any assumption about about augmenting the structure of 
Salam-Strathdee superspace such as assumptions in other formulations
\cite{CdWLL}
(e.\ g.\
 ``pure spinors, etc.'').  If a complete analysis without such added structures should 
fail, that would constitute a proof of the necessity of such structures.  But to our
knowledge no such complete analysis exists in the literature.

In an effort to be completely clear about this point, we wish to expand on
the meaning we assign to the term ``Salam-Strathdee superspace'' to be
contrasted with the term ``Wess-Zumino superspace.''  We also want to make
precise what we mean by additional assumptions.  A superspace supergravity
covariant derivative operator $\nabla{}_{\un {A}}$ = 
$(\nabla{}_{\a}, \nabla{}_{\un a})$ will satisfy an equation of the form
\be 
{\big [} \nabla{}_{\a} \, , \, \nabla{}_{\b} {\big ]} ~=~ T{}_{\a \, \b}{}^{\g} \,
\nabla{}_{\g} ~+~ T{}_{\a \, \b}{}^{\un c} \, \nabla{}_{\un c}  ~+~
\fracm 12 \, R{}_{\a \, \b}{}_{\un c}{}^{\un d} \, {\cal M}{}_{\un d}{}^{\un c}  ~~~.
\ee
When we use the term ``Salam-Strathdee'' superspace theory of supergravity, we consider the
additional assumption that
\be
({\g}_{\un d}) {}^{\a \, \b} \, 
 T{}_{\a \, \b}{}^{\un c} ~\propto ~ \d{}_{\un d}{}^{\un c}  ~~~,
\label{S-Ssp8c}
\ee
along other ``conventional constraints'' that algebraically relate one superfield to another.
A general discussion of ``conventional constraints'' for supergravity can be seen in the works 
of \cite{SC2,SC4} and for the particular case of 11D in \cite{2MT}.

The constraint in (\ref{S-Ssp8c}) is a weaker constraint than
\be
T{}_{\a \, \b}{}^{\un c} ~\propto ~ ({\g}^{\un c}) {}_{\a \, \b} ~~~.
\label{W-Zsp8c}
\ee
In this latter case we refer to this as a ``Wess-Zumino superspace theory of supergravity.''  As discussed
in the work of \cite{CdWLL}, in flat superspace both of these definitions coincide
and by the introduction of a pure spinor superfield $\l{}^{\a}$ to multiply the
spinorial supercovariant derivative $\nabla{}_{\a}$ one can form the product
$\l{}^{\a} \nabla{}_{\a}$.  This last operator is a BRST charge.  However, with the
assumption of (\ref{S-Ssp8c}), this product is not a BRST charge.  Therefore, in this regard
(\ref{W-Zsp8c}) makes additional assumptions over and above those in (\ref{S-Ssp8c}).
 
If a complete analysis without such added structures implied by  (\ref{W-Zsp8c}) should fail,
that would constitute a proof of the necessity of such structures.  But to our
knowledge no such complete analysis exists in the literature.

In our recent works of \cite{GHM2,GHM3,GHM4}, we developed algorithmically 
based techniques that permit the investigation of the component field contents of 
possible supergravity (SG) prepotentials.  This is a capacity that never before 
existed to our knowledge in the context of high dimensional superfield SG theories. 
Based on an assumption about the 11D, $\cal N$ = 1 SG prepotential superfield, this
allows us to identify candidate 10D, $\cal N$ = 1 superfields that are most likely
the 10D, $\cal N$ = 1 SG prepotential superfield and four additional 10D, $\cal 
N$ = 1 superfields that are the most likely candidates to describe the matter gravitino
multiplet prepotential required to formulate Type-II theories.  Identifying such candidate 
prepotentials is the major other purpose of this work.

In every superspace formulation where the prepotential formulation is completely 
understood, the supergravity prepotential transforms as the first superspace spinorial 
derivative of a parameter superfield\footnote{However, such parameter superfields 
may themselves be expressed as spinor derivative operators of more fundamental 
parameter superfields.} that is consistent with the Lorentz structure of the supergravity 
prepotential and the presence of the superpace spinorial derivative.  Thus, in a case 
where the supergravity prepotential transforms as a scalar under the associated Lorentz
group, its gauge transformation parameter superfield is a spinor.  Thus, we conjecture 
this sort of rule holds for the cases under discussion.  This will be work for the future  
enabled by the ability to ``scan'' superfields as described in chapter five.

\newpage
\section{Review of 11D, $\cal N$ = 1 Supergravity Derivatives \& Scale Transformations}
\label{sec:primer}

The spinorial 11D frame operator may be parametrized in the following 
equations\footnote{Roughly speaking, this is the superspace supergravity equivalent to the
ADM \cite{ADM,ADMv,ADMv2} formulation but for the spinorial superframe.},
\be \eqalign{ 
{\rm E}{}_\a &=~ \Psi^{1/2} \, \Big[ \, \exp \left( \fracm 12 {\cal A}^{{\un a} \, {\un b}} 
\g_{{\un a} \, {\un b}} \right) \Big]{}_\a{}^{\b} \,
\Big[ \, {\cal N}{}_{\b}{}^{\g}  ~ \Big] ~ 
\Big[ \, {\rm D}{}_{\g} ~+~ {\Hat {\rm H}}{}_{\g}{}^{\un b} \, \pa_{\un b} \, \Big]  ~~, 
\cr  
{~~~~~~~~} {\cal N}{}_{\a}{}^{\b} & \equiv ~ \Big[ \, {\bm {\rm I}} \,+\,  {\cal A}^{\un a} 
\g{}_{\un a} \,+\,  \fracm 1{3!} {\cal A}^{ [ {\un 3} ] }  \g_{[ {\un 3} ]} \,
+\,\fracm 1{4!} {\cal A}^{[ {\un 4} ]} \g_{[ 4 ]} \,+\,  \fracm 1{5!} {\cal A}^{[ 5 ]}
\g_{[{\un 5}]} \, \Big]{}_{\a}{}^{\b}  
~~~,{~~~~~~} 
 } \label{eq:01}  \ee 
where $ \Psi$ is the ``scale compensator,'' ${\cal A}^{{\un a} \, {\un b}} $ is the 
``Lorentz compensator,'' ${\cal N}{}_{\a}{}^{\b}$ is a ``coset factor,'' and $ {\Hat {\rm 
H}}{}_\b{}^{\un b}$ is the ``conformal graviton semi-prepotential.''  The parametrization 
of ${\cal N}{}_{\a}{}^{\b}$ shown in (\ref{eq:01}) is strongly dependent on the dimensionality 
of the superspace.  For example, in simple supergravity theory in two dimensions\footnote{To 
our knowledge, the case of D = 2 is the only one in existence where the dependence \cite{GN2d} 
of ${\cal N}{}_{\a}{}^{\b}$ on ${\rm H}{}_{\a}{}^{\un a}$ has been explicitly presented.  However,
other complete solutions for constaints implicitly contain such information.}, 
only the first two terms are present \cite{GN2d}.

A slightly different, but equivalent parametrization was introduced in a previous work \cite{2MT}.  
Without loss of generality, a slightly different semi-prepotential can be written in the form
\be{
{\Hat {\rm H}}{}_\b{}^{\un b} ~=~ {\rm H}{}_{\b}{}^{\un b} ~-~ \fracm 1{\rm D}\, (\, \g
{}^{\un b} \g{}_{\un d} \, ){}_{\b}{}^{\d} \, {\rm H}{}_{\d}{}^{\un d} ~~~,
}\label{eq:02} \ee
where D = 11 for this eleven dimensional case.   This also implies there is a local
symmetry of the form
\be
{\rm H}{}_{\b}{}^{\un b} ~\to~ {\rm H}{}_{\b}{}^{\un b} ~+~ ( \g{}^{\un b} ){}_{\b}{}^{\a}  
\L{}_{\a} ~~~, 
\label{eq:02a} \ee
that applies to (\ref{eq:02}).

All previous experience in superspace supergravity implies the vectorial 11D frame operator 
${\rm E}{}_{\un a}$, the spinorial spin connection ${\o}{}_{\a \, {\un c} {\un d}}$, and the vectorial 
spin connection ${\o}{}_{{\un a} \, {\un c} {\un d}}$ must be determined in terms of the content 
of the spinorial 11D frame field.  This is done by the imposition of ``conventional constraints'' 
on the torsion and curvature supertensors defined in (\ref{eq:TR}).  However, as first noted in 
the work \cite{GN2d}, ``conventional constraints'' must also be imposed to determine ${\cal N}
{}_{\a}{}^{\b}$ {\it \un {solely}} in terms of ${\rm H}{}_{\b}{}^{\un b}$, but totally {\it \un 
{independent}} of $\Psi$. This has powerful implications for constraints in superfield 
supergravity.

A Poincar\' e supergeometry of the 11D theory demands the introduction of superspace 
supergravity covariant derivatives $\nabla{}_{\a}$ and $\nabla{}_{\un a}$ defined by
\be {\eqalign{
\nabla{}_{\a} ~=&~ {\rm E}{}_{\a} ~+~ \fracm 12 \, {\o}{}_{{\a} \, {\un d}}{}^{\un e}{\cal M}
{}_{\un e}{}^{\un d}
~~~, \cr
\nabla{}_{\un a} ~=&~ {\rm E}{}_{\un a} ~+~ \fracm 12 \, {\o}{}_{{\un a} \, {\un d}}{}^{\un e}{\cal M}
{}_{\un e}{}^{\un d} ~~~.
}} \label{eq:NBLA}
\ee
Upon calculating the graded commutator of these leads to superspace torsions
and curvatures according to,
\be \label{eq:TR}
\eqalign{
[ \,\nabla_{\alpha}~,~\nabla_{\beta} \, \} ~=&~ T_{\alpha\beta}{}^{\un{c}} \, \nabla_{\un{c}} ~+~
T_{\alpha\beta}{}^{\gamma} \, \nabla_{\gamma} ~+~ \fracm{1}{2}R_{\alpha\beta \,\un{
d}}{}^{\un{e}}\, \mathcal{M}_{\un{e}}{}^{\un{d}}  ~~~, \cr
[ \, \nabla_{\alpha}~,~\nabla_{\un{b}} \, \} ~=&~ T_{\alpha\un{b}}{}^{\un{c}}\, \nabla_{\un{c}}\, ~+~
T_{\alpha\un{b}}{}^{\gamma}\, \nabla_{\gamma} ~+~ \fracm{1}{2}R_{\alpha\un{b}\, \un{
d}}{}^{\un{e}} \, \mathcal{M}_{\un{e}}{}^{\un{d}}  ~~~~,  \cr
[ \, \nabla_{\un{a}}~,~ \nabla_{\un{b}} \, \} ~=&~ T_{\un{a}\un{b}}{}^{\un{c}} \, \nabla_{\un{c}}\,\, ~+~
T_{\un{a}\un{b}}{}^{\gamma} \, \nabla_{\gamma} ~+~ \fracm{1}{2}R_{\un{a}\un{b} \, \un{
d}}{}^{\un{e}}\, \mathcal{M}_{\un{e}}{}^{\un{d}}
~~~~ . }
\ee

Based on an analysis of constraints for the 11D vielbein, the work in \cite{M2} concluded the set
\be {
\eqalign{
 i \,  \fracm 1{32} \,  (\g_{\un a})^{\a \b} \, T_{\a \b}{}^{\un b} &=~   \d_{\un a} \, {}^{\un b} 
~~\,~~~~,~~~~~~~~~~~~~~~~~~~~~~ (\g_{\un a})^{\a \b} \, T_{\a \b}{}^{\g} ~=~ 0 ~~~~~~~,
~~~~~
 \cr 
{~~~~~~~~~~}
T_{\a \, [ {\un d} {\un e}]} ~-~ \fracm 2{55} \, (\g_{{\un d} {\un e}})_{\a}{}^{\g}  
\, T_{\g {\un b}}{}^{\un b} ~&=~0 ~~~~\,~~~\,~, {~~}~~~~~~~~~~~~~~~~~~~
(\g_{\un a})^{\a \b} \, R_{\a \b}{}^{{\un d}{\un e}} ~=~ 0 ~~~~~~~,~~~ ~~
\cr
(\g_{{\un a}  {\un b}  {\un c}  {\un d} {\un e}})^{\a \b} \, T_{\a \b}{ }^{\un e} &=~ 0 ~~~\,~~~\,~~, 
~~~~~~~~~~~~~~~~
(\g_{[  {\un a}  {\un b}  {\un c}  {\un d}  {\un e}  |   })^{\a \b} \, 
T_{\a \b}{}_{  | {\un f} ]  } ~=~ 0  ~~~~~~~,
\cr 
(\g_{{\un a}{\un b}})^{\a \b} \,T_{\a \b}{}^{\un b} ~&=~ 0 ~\,~
~~\,~~~~,{~~}~~~~~~~~~~~~~\,~ ~~~(\g_{[{\un a} {\un b}|})^{\a \b} \, T_{\a \b}{}_{ | {\un c}]} ~=~ 0 
~~~~~~~,
} 
\label{eq:constrts} }
\ee
can be enforced without implying any dynamical consequences on the components fields that remain 
with the vielbein.  The first two constraints in (\ref{eq:constrts}) ensure that the graviton and 
gravitino, respectively, that appear at first order in the $\theta$-expansion of in $\nabla{}_{\a}$ 
are identified with the self-same fields that appear at zeroth order in $\nabla{}_{\un a}$.  The 
next two constraints have the same effects by removing the Lorentz connection of the theory (as an 
independent variable), and defining it in terms of the anholonomy associated with the component frame field.
These second two constraints of (\ref{eq:constrts}) also ensure that the spin-connection that appears 
at first order in the $\theta$-expansion of in $\nabla{}_{\a}$ is identified with the self-same field 
appearing at zeroth order in $\nabla{}_{\un a}$.  The last four constraints in (\ref{eq:constrts}) 
remove all the superfields that appear in ${\cal N}{}_{\a}{}^{\b}$ as independent quantities and makes 
them non-trivially dependent only on ${\rm H}{}_{\a}{}^{\un a}$.  It is directly possible to prove 
at the linearized order.  However, beyond this order the process is expected to be horribly 
non-linear as in the 2D, $\cal N$ = 1 supergravity theory.

Recent calculations \cite{GHM3} inspired by adinkras have yielded IT-based techniques providing 
unprecedented access to the component field composition of superfields in 11D.  Based on
this, we have proposed the set of constraints above may be strengthened 
by the replacement as indicated below,
\be {
\eqalign{
 i \,  \fracm 1{32} \,  (\g_{\un a})^{\a \b} \, T_{\a \b}{}^{\un b} &=~   \d_{\un a} \, {}^{\un b} 
~~\,~~~~,{~~}~~~~~~~~~~~~~~~~~~~~ (\g_{\un a})^{\a \b} \, T_{\a \b}{}^{\g} ~=~ 0 ~~~~~~~,
~~~~~
 \cr 
{~~~~~~~~~~}
T_{\a \, [ {\un d} {\un e}]} ~-~ \fracm 2{55} \, (\g_{{\un d} {\un e}})_{\a}{}^{\g}  
\, T_{\g {\un b}}{}^{\un b} ~&=~0 ~~~~\,~~~\,~, {~~}~~~~~~~~~~~~~~~~~~~
(\g_{\un a})^{\a \b} \, R_{\a \b}{}^{{\un d}{\un e}} ~=~ 0 ~~~~~~~,~~~ ~~
\cr
(\g_{{\un a}  {\un b}  {\un c}  {\un d} {\un e}})^{\a \b} \, T_{\a \b}{ }^{\un e} &=~ 0 ~~~\,~~~\,~~, 
~~~~~~~~~~~~~~~~
(\g_{[  {\un a}  {\un b}  {\un c}  {\un d}  {\un e}  |   })^{\a \b} \, 
T_{\a \b}{}_{  | {\un f} ]  } ~=~ 0  ~~~~~~~,
\cr 
 ~&~
~~~~~ (\g_{{\un a}{\un b}})^{\a \b} \,T_{\a \b}{}^{\un c} ~=~ 0 ~~~~,
} 
\label{eq:constrts2} }
\ee
as this set, ``CGNN constraints'' \cite{CGNN}, is consistent with the choice of 
a scalar superfield $\cal V$ as the supergravity prepotential.  This implies the 
superfunction ${\rm H}{}_{\b}{}^{\un c}$ = ${\rm H}{}_{\b}{}^{\un c}({\cal V})$ must 
admit a functional dependence that satisfies this equation.  This functional dependence 
must involve fifteen powers of ${\rm D}{}_{\a}$ acting on ${\cal V}$.  

We highlight the following two points which support the complete vanishing of the 
final constraint shown in (2.7) in the limit where only the proposed scalar supergravity 
prepotential is retained:
\newline \indent
(a.)
a complete parametrization of all possible degrees of freedom of the 11D
\newline \indent $~~~~$
spinorial supervielbein was given by (2.1) in our paper.  The constraints
\newline \indent $~~~~$
shown in (2.6) determine all of the $\cal A$-superfields within ${\cal N}{}_{\a}
{}^{\b}$ in terms of the 
\newline \indent $~~~~$
$\Hat H$ superfield based on linearized
calculations reported in the work of \cite{2MT}.
  
(b.)
the $\Hat H$-superfield is a ``semi-prepotential'' along the lines of such
superfields \newline \indent $~~~~$
discussed in the work of \cite{N2}.  In turn, however, our work
of \cite{GHM2} using \newline \indent $~~~~$
our ``scanning" techniques showed that the representation required for
\newline \indent $~~~~$
the final 
line of (2.7) to be non-zero  is absent from the proposed scalar 
\newline \indent $~~~~$
supergavity 
prepotential.  Thus, we conclude that torsion must vanish.
\newline \noindent
We invite any skeptic unfamiliar with the workings of supergravity prepotential
formulations to undertake the proof this final constraint to show the accuracy 
of our statements.

A solution to (\ref{eq:constrts2}) is provided by making the ``shift'' $\Psi$ $\to$
$1 \,+\,\Psi$ (Euqations (12) and (18) in \cite{M2}) so that
\be \eqalign{
\nabla_{\a} ~=&~  {\rm D}{}_{\a} ~+~ \fracm 12 \, \Psi \, {\rm D}{}_{\a}
~+~ \fracm 1{10} \, ({\rm D}{}_{\b} \Psi) (\g^{{\un d} {\un e}})_{\a}{}^{\b} 
{\cal M}_{ {\un d} {\un e}}  ~~~, \cr
\nabla_{\un a} ~=&~  \pa_{\un a} ~+~ \Psi \, \pa_{\un a} ~+~ i\, \fracm 14 \,
(\g{}_{\un a})^{\a \, \b} \, ({\rm D}{}_{\a} \Psi ) \,  {\rm D}{}_{\b} ~+~ \fracm 15 \, 
(\pa{}_{\un c} \Psi) \, {\cal M}{}_{\un a}{}^{\un c}  \cr
& ~+~ i \fracm 1{160} \, (\g_{\un a}^{~{\un d} {\un e}} )^{\a \b} \, ({\rm D}{}_{\a}
{\rm D}{}_{\b} \Psi)  \, {\cal M}_{{\un d} {\un e}}  ~~~,
}  \label{eq:Clss-1}  \ee
where $\Psi$ in (\ref{eq:Clss-1}) is an infinitesimal superfield and we set the
superfield $ {\rm H}{}_{\b}{}^{\un c}$ = 0.  It should also be noted  we can
regard the supervector fields here as describing a Nordstr\" om supergravity theory
as in our previous work \cite{GHM1}, although we used a different set of constraints in \cite{GHM1}.

In turn, (\ref{eq:Clss-1}) suggests the definition of a set of Weyl scaling transformations of 
the full superspace covariant derivatives given by
\be \eqalign{
\d{}_{S} \nabla_{\a} &=~\fracm 12 \, L \, {\nabla}{}_{\a}
~+~ \fracm 1{10} \, ({\nabla}{}_{\b} L) (\g^{{\un d} {\un e}})_{\a}{}^{\b} 
{\cal M}_{ {\un d} {\un e}}  ~~~, \cr
\d{}_{S} \nabla_{\un a} &=~  L \, {\nabla}{}_{\un a} ~+~ i\, \fracm 14 \,
(\g{}_{\un a})^{\a \, \b} \, ({\nabla}{}_{\a} L ) \,  {\nabla}{}_{\b} ~+~ \fracm 15 \, 
(\nabla{}_{\un c} L) \, {\cal M}{}_{\un a}{}^{\un c}  \cr
&{~~~}~+~i \fracm 1{160} \, (\g_{\un a}^{~{\un d} {\un e}} )^{\a \b} \, ({\nabla}{}_{\a}
{\nabla}{}_{\b} L)  \, {\cal M}_{{\un d} {\un e}}  ~~~.
}  \label{eq:ScaleLW}  \ee
In the results shown in (\ref{eq:ScaleLW}), the superspace covariant derivatives
$\nabla_{\a}$ and $\nabla_{\un a}$ are ``full'' superspace covariant derivatives 
where the only remaining independent superfield variables are $\Psi$ and 
${\rm H}{}_{\b}{}^{\un c}({\cal V})$.

It should be noted there is a lesson to learn from (\ref{eq:Clss-1}) and
(\ref{eq:ScaleLW}).  A set of full Weyl scale transformation laws on the 
superspace supergravity supercovariant derivative operators can be obtained 
by starting from the formulation of a Nordstr\" om superspace supergravity 
theory, acting with a superspace scaling operation $\d{}_{S}$ on the 
Nordstr\" om scalar field in the theory, calling $\d{}_{S} \Psi$ = $L$,
and finally replaces all ``bare'' derivative operators by full superspace 
supergravity supercovariant derivative operators.

In turn this implies we are able to analyze the Weyl scaling properties of all the 
superspace torsion and curvature supertensors with engineering dimensions of less than 
three-halves to find,
\be \eqalign{
\d_S T_{\a \b}{}^{\un c} \, &=~ 0 ~~~, 
{~~~~~~~~~~~~~~~~~~~~~~~~~~~~~~~~~~~~~~~~~~~~~~~~~}
{~~~~~~~~~~~~~~}
}  \label{eq:WeylT1}  \ee

\be \eqalign{
\d_S T_{\a \b}{}^{\g} \, &=~ \fracm 12 L \, T_{\a \b}{}^{\g} ~-~ i 
\fracm 14 \, T_{\a \b}{}^{\un c} \, (\g_{\un c})^{\d \g} \,(\nabla_{\d} L) 
~+~\fracm {\,1\,}{2} (\nabla_{( \a} L)\, \d_{\b)}{}^{\g} \cr 
&{~~~}~+~ \fracm {\,1\,}{20}  (\nabla_{\d} L) \, (\g^{[2]})_{(\a}{
}^{\d} \, (\g_{[2]})_{\b)}{}^{\g} ~~~,{~~~~~~~~} 
{~~~~~~~~} {~~~~~~~~} {~~~~~}    }   \label{eq:WeylT2}   \ee
\be \eqalign{
\d_S T_{\a {\un b}}{}^{\un c} \, &=~  \fracm 12 L \, T_{\a {\un b}}{}^{\un c} ~-~ i \fracm 14 
\, (\g_{\un b} )^{\g \d} \, (\nabla_{\g} L) \, T_{\a \d}{}^{\un c} ~+~ (\nabla_{
\a} L) \, \d_{\un b} {}^{\un c}   
{~~~~~~~~~~}
 \cr 
&{~~~}~+~ \fracm 15 \, (\nabla_{\g} L) \, (\g_{\un b} {}^{\un c})_{\a}{}^{\g} 
~~~,  
}    \label{eq:WeylT3}    
\ee

\be \eqalign{
{~~~~~} \d_S T_{\a {\un b}}{}^{\g} \, &=~ L \, T_{\a {\un b}}{}^{\g} ~-~ i \fracm 
14 \, (\g_{\un b})^{\d \e} \, (\nabla_{\d} L) \, T_{\a \e}{}^{\g} ~-~
i \fracm 14 \, T_{\a {\un b}}{}^{\un c} \, (\g_{\un c})^{\d \g} \, (\nabla_{\d} L) \cr 
&{~~~} ~-~ \fracm 12 \, (\nabla_{\un b} L) \, \d_{\a}{}^{\g} ~-~ \fracm 
{\,1\,}{10} \, (\nabla_{\un d} L) \, (\g_{\un b} {}^{\un d})_{\a}{}^{\g}  \cr 
&{~~~} ~+~ i \fracm 14 \, (\g_{\un b})^{\d \g} \, (\nabla_{\a} \nabla_{\d}
L) ~-~ i \fracm 1{320} \, (\g_{\un b} {}^{{\un d} {\un e}})^{\d \e} \, (\nabla_{\d} \nabla_{\e}
L) \, (\g_{{\un d} {\un e}})_{\a}{}^{\g}  ~~~, } 
 \label{eq:WeylT4}  
\ee
\be \eqalign{
\d_S T_{{\un a} {\un b}}{}^{\un c} \, &=~ L \, T_{{\un a} {\un b}}{}^{\un c} ~+~ i \fracm 14 
(\g_{[{\un a}|})^{\d \e} \, (\nabla_{\d} L) \, T_{\e |{\un b} ]} {}^{\un c}
~+~ \fracm 65 \, (\nabla_{[{\un a}|} L )\,
\d_{|{\un b}  ]}{}^{\un c}   {~~~~~~~}
 \cr
&{~~~}~+~ i \fracm 1{40} \, (\g_{{\un a} {\un b}}{}^{\un c})^{\d \e} (\nabla_{\d} 
\nabla_{\e} L) ~~~,
}\label{eq:WeylT5} \ee
\be \eqalign{
\d_S R_{\a \b \,}{}^{{\un d} {\un e}} \, &=~ L \, R_{\a \b \, }{}^{{\un d} {\un e}} ~+~
\fracm 15 \, T_{\a \b}{}^{[{\un d}} \, (\nabla^{{\un e}]} L) ~+~
\fracm 15 \, T_{\a \b}{}^{\d} \, (\nabla_{\g} L) \, (\g^{{\un d} {\un e}})_{\d}{}^{\g} \cr
&{~~~}~-~ \fracm 15 (\nabla_{(\a |} \nabla_{\g} L) \, (\g^{{\un d} {\un e}})_{|\b)}{}^{\g} 
~+~ i \fracm {\,1\,}{80} \, T_{\a \b}{}^{\un c} (\g_{\un c} {}^{{\un d} {\un e}} )^{\d \e} \, (\nabla_{\d} \nabla_{\e} L)
 ~~~.
}\label{eq:WeylR} \ee

The constraints on the last three lines of  (\ref{eq:constrts2}) have a solution given by 
\cite{CGNN}\footnote{The first indication of the need for modification to the superspace 
torsion proportional to $\g$/$\s$ matrices was noted in 1983 \cite{G&G} in the context of
4D, $\cal N \, > $ 4 conformal supergravity in superspace.}
\be
T_{\a \b}{}^{\un c} ~=~ i \, (\g^{\un c}){}_{\a \b} ~+~  i \, \fracm 1{\, 32 \, \cdot \, 5! \,} \, (\g^{{\un b}{}_1
\, \cdots \, {\un b}{}_5}){}_{\a \b} \, X{}_{{\un b}{}_1\, \cdots \, {\un b}{}_5}{}^{\un c}
~~~~,~~~~ X{}_{{\un a} {\un b} {\un c} {\un d} {\un e} }{}^{\un e} ~=~
X{}_{[ {\un a} {\un b} {\un c} {\un d} {\un e} }{}_{{\un f}]}  ~=~ 0 ~~~,
\label{eq:DIS} \ee
where $X{}_{{\un a}{}_1\, \cdots \, {\un a}{}_5}{}^{\un b} ~ \equiv~ i 
\, (\g_{{\un a}{}_1 \,  \cdots \,  {\un a}{}_5})^{\a \b} \, T_{\a \b}{}^{\un b}~$ 
alternatively\footnote{Notice that the normalization of the $X_{[5]}{}^{\un{b}}$ 
differs by a factor of 32 with that in the work of \cite{M2}, 
i.e. the factor $\frac{1}{32}$ is put in Equation (\ref{eq:DIS}) 
instead of the definition of $X_{[5]}{}^{\un{b}}$.}.
The expression of $T_{\a \b}{}^{\un c}$ may be substituted in (\ref{eq:WeylT2}) and (\ref{eq:WeylT3}). Contraction
of the $\b$ and $\g$ indices in the former and the $\un b$ and $\un c$ indices in 
the latter yields respectively
\be \eqalign{
{~~~~~} \d_S T_{\a \b}{}^{\b} \, &=~ \fracm 12 L \, T_{\a \b}{}^{\b} ~-~ i \, 
\fracm 14 \, \left[ \, i \, (\g^{\un c}){}_{\a \b} ~+~  i \, \fracm 1{32 \,\cdot\, 5!} \, 
(\g^{{\un b}{}_1 \, \cdots \, {\un b}{}_5}){}_{\a \b} \, X{}_{{\un b}{}_1\, \cdots 
\, {\un b}{}_5}{}^{\un c}  \, \right] \, (\g_{\un c})^{\d \b} \,(\nabla_{\d} L)   \cr 
&{~~~} ~+~\fracm {\,1\,}{2} (\nabla_{( \a} L)\, \d_{\b)}{}^{\b} ~+~ \fracm {\,1\,}{20}  
(\nabla_{\d} L) \, (\g^{[2]})_{(\a}{}^{\d} \, (\g_{[2]})_{\b)}{}^{\b} ~~~{~~~~~~~~} 
{~~~~~~~~} {~~~~~~~~} {~~~~~}  \cr
&=~ \fracm 12 L \, T_{\a \b}{}^{\b} ~+~  \, \fracm 14 \, (\g^{\un c}){}_{\a \b}  \,
(\g_{\un c})^{\d \b} \,(\nabla_{\d} L) ~+~\fracm {\,1\,}{2} (\nabla_{( \a} L)\, 
\d_{\b)}{}^{\b} 
\cr 
&{~~~} 
~+~ \fracm {\,1\,}{20} (\nabla_{\d} L) \, (\g^{[2]})_{(\a}{}^{\d} \, (\g_{[2]})_{\b)}
{}^{\b}  ~~~ \cr
&=~ \fracm 12 L \, T_{\a \b}{}^{\b} ~+~ \fracm {33}4 \,(\nabla_{\a} L) ~~~, }   
\label{eq:WeylTx2}   \ee
\be \eqalign{
{~~~~~~} \d_S T_{\a {\un b}}{}^{\un b} \, &=~  \fracm 12 L \, T_{\a {\un b}}{}^{\un b} ~+~  \fracm 14 \, (\g_{\un b} )^{\g \d} \, (\nabla_{\g} L) \, \, \left[ \,  (\g^{\un 
b}){}_{\a \d} ~+~ \fracm 1{32 \,\cdot\, 5!} \, (\g^{{\un b}{}_1 \, \cdots \, {\un b}
{}_5}){}_{\a \d} \, X{}_{{\un b}{}_1\, \cdots \, {\un b}{}_5}{}^{\un b}  \, \right]  
 ~+~ (\nabla_{\a} L) \, \d_{\un b} {}^{\un b}   {~~~~~~~~~~} \cr
 &=~  \fracm 12 L \, T_{\a {\un b}}{}^{\un b} ~+~  \fracm 14 \, (\g_{\un b} )^{\g \d} \, 
(\nabla_{\g} L) \, (\g^{\un b}){}_{\a \d}   ~+~ (\nabla_{\a} L) \, \d_{\un b} {}^{\un b}   
\cr 
&=~ \fracm 12 L \, T_{\a \un b}{}^{\un b} ~+~ \fracm {33}4 \,(\nabla_{\a} L) ~~~.
}    \label{eq:WeylTx3}    \ee
The equation in the final portion of the results shown in (\ref{eq:DIS}) is responsible
for the disappearance of the $X$-tensors from the second lines in (\ref{eq:WeylTx2})
and (\ref{eq:WeylTx3}).

If we make the definitions
\be
{\cal J}{}_{\a}{}^{(1)} ~=~ \fracm 4{33} \, T_{\a \b}{}^{\b}
~~~,~~~ 
{\cal J}{}_{\a}{}^{(2)} ~=~ \fracm 4{33} \, T_{\a {\un b}}{}^{\un b} ~~~,
\label{eq:CNXs1} 
\ee
then we see
\be  {
 \d_S \, {\cal J}{}_{\a}{}^{(1)} ~=~    \fracm 12 L \,  {\cal J}{}_{\a}{}^{(1)} ~+~
 (\nabla_{\a} L)
 ~~~,~~~
  \d_S \, {\cal J}{}_{\a}{}^{(2)} ~=~   \fracm 12 L \,   {\cal J}{}_{\a}{}^{(2)} ~+~
  (\nabla_{\a} L) ~~~.
} \label{eq:CNXs2}\ee
Upon changing basis for these superfields by defining 
\be {
{\cal J}{}_{\a}{}^{(\pm)}  ~=~  \fracm 12 \,
 \left[  \,    {\cal J}{}_{\a}{}^{(1)} ~\pm ~ {\cal J}{}_{\a}{}^{(2)} 
  \, \right] ~~~,
} \label{eq:CNXs3}\ee
we observe that
\be {
\d_S \, {\cal J}{}_{\a}{}^{(+)} ~=~    \fracm 12 L \,   {\cal J}{}_{\a}{}^{(+)} ~+~
 (\nabla_{\a} L)
 ~~~,~~~
\d_S \, {\cal J}{}_{\a}{}^{(-)} ~=~    \fracm 12 L \,  {\cal J}{}_{\a}{}^{(-)}
~~~.
} \label{eq:CNXs4}\ee
While ${\cal J}{}_{\a}{}^{(-)}$ transforms like a scale covariant tensor of weight $\fracm 12$,
the quantity ${\cal J}{}_{\a}{}^{(+)}$ transforms with a scale weight of $\fracm 12$ while
being a spinorial gauge connection under a scaling transformation!  The work of \cite{2MT0}
indicated that such a dimension one-half supertorsion tensor was required for an off-shell 
description of 11D, $\cal N$ = 1 Poincar\' e supergravity.

According to the analysis in \cite{M2}, there is also necessarily an engineering dimension
one superfield ${\cal W}_{{\un a} {\un b} {\un c} {\un d}}$ with Weyl weight one 
(i.e. $\d_S \, {\cal W}_{{\un a} {\un b} {\un c} {\un d}} \,=\, L \, 
{\cal W}_{{\un a} {\un b} {\un c} {\un d}}$) defined by\footnote{Notice that the Weyl tensor ${\cal W}_{\un{a}\un{b}\un{c}\un{d}}$ here differs by an overall factor of $i$ with that in the work of \cite{M2}. It is because in our convention here, all $\g$-matrices are real, while in \cite{M2}, $(\g^{[1]})_{\a\b}$ is real but $(\g^{[4]})_{\a\b}$ is imaginary. Thus ${\cal W}_{\un{a}\un{b}\un{c}\un{d}}$ is real in both cases.}
\be \eqalign{ {~~}
{~~~~~}{\cal W}_{{\un a} {\un b} {\un c} {\un d}} 
 &\equiv~ \fracm 1{32}\, \Big[\, ( \g^{\un e} \g_{{\un a} {\un b} {\un c} {\un d}}){}_{\g}{}^{\a} \, T_{\a \, {\un e}}{}^{\g} 
~+~ i\, \fracm {11}4 \,  (\g_{{\un a} {\un b} {\un c} {\un d}}){}^{\a \b}\,  \big( \, \nabla_{\a} 
{\cal J}{}_{\b}{}^{(+)} \,-\, \fracm {23}{220} \,  
{\cal J}{}_{\a}{}^{(+)} \, {\cal J}{}_{\b}{}^{(+)}  \,\big) ~ \Big] ~~~,
} \label{eq:WyLSF} \ee
and which contains all the on-shell degrees of freedom.  In particular, the lowest
order term in this superfield (i.e. setting $\theta$ = 0) is the supercovariantized
field strength of the component level 3-form gauge field. Moreover, the 11D 
supercovarinatized Weyl-gravitino field strength is contained at the first order 
in the $\theta$-expansion, and the supercovariantized Weyl tensor of the 11D bosonic 
spacetime is contained at the second order in the $\theta$-expansion of ${\cal 
W}_{{\un a} {\un b} {\un c} {\un d}}$.

Thus, we conclude the most likely path forward for an off-shell, {\em{in the sense
of the pioneering work in}}  \cite{GWZ1}, 
11D, $\cal N$ = 1
superspace supergravity theory 
must be the construction, as we suspected in 1996, on
the basis of three superfields now known to be ${\cal W}_{{\un a} {\un b} {\un c} {\un d}}$,
$X{}_{{\un a}_ 1 \cdots {\un a}_ 5}{}^{\un b}$, and ${\cal J}{}_{\a}$.  The ultimate
deepest reason to expect this decomposition is the structure of gravity itself.
When one considers the Riemann Tensor of ordinary gravity (which must be embedded within
any supergravity theory), it can be decomposed into the: \newline \indent $~~~~$
(a.) Weyl, \newline \indent $~~~~$
(b.) Ricci, and \newline \indent $~~~~$ 
(c.) Scalar Curvature \newline
portions.  In our opinion, any superspace construction that ignores this tripartite 
division is unlikely to be adequate.

\newpage
\section{Supergravity Derivatives \& Scale Transformations in 10D}
\label{sec:analytical0}

The content of the previous chapter about the superspace supergravity supercovariant 
derivatives $\nabla{}_{\un A} = (\nabla{}_{\a},\,\nabla{}_{\un a}$), the torsion supertensors 
$T{}_{{\un A} \, {\un B}}{}^{{\un C}}$, and curvature supertensors $R{}_{{\un A} \, {\un B}} 
{}_{\, \un c} {}^{\un d}$ can be directly used to derive similar results for the superspaces 
associated with Type-IIA and and Type-I superstring theories.  However, there is much more 
efficient and rapid way in this can be done.  This alternative route is also directly 
applicable to the Type-IIB case.  So rather than going the route of the dimensional reduction 
of these geometrical quantities, we choose to take a less obvious pathway.

The bottom line message from the last chapter is that the scalar compensator, which
appears in the parametrization of the framefield (\ref{eq:01}), is isomorphic to the
concept of a scalar theory (i.e. Nordstr\" om gravitation).  This means explorations
\cite{GHM1} of Nordstr\" om supergravity are equivalent to the exploration of the limit 
of a full superframe where only the compensator is retained.  This has the implication
that investigations of Nordstr\" om supergravity theories fix the dependences of the 
superspace supergravity supercovariant derivatives $\nabla{}_{\un A}$, the torsion supertensors 
$T{}_{{\un A} \, {\un B}}{}^{{\un C}}$ and curvature supertensors $R{}_{{\un A} \, {\un 
B}} {}_{\, \un c} {}^{\un d}$ on the scale compensating superfield.

The constraints used in the following results can be derived from equations found
in the work of \cite{GHM1}.  Take the following indicated equations from that work
which are linear in the first derivative of the scale compensator, use appropriate 
algebraic operations to express the first derivatives in terms of torsion, and substitute 
those expressions back into the equations.  For the ${\cal N}\,=\,\, $IIA supergeometry, 
the appropriate equations from which to start the derivation are given in (6.10) - (6.19), 
(6.22), and (6.25).  For the ${\cal N}\,=\, $IIB supergeometry, the appropriate equations 
from which to start the derivation are given in (7.11) - (7.20), (7.23), and (7.26).  Finally, for the 
${\cal N}\,=\, $I supergeometry, the appropriate equations from which to start the derivation 
are given in (5.8), (5.9), (5.10), and (5.12).  These are found to be equivalent to results
given previously in \cite{SC5}.

\subsection{10D, ${\cal N}~=~$IIA Supergravity Derivatives \& Scale Transformations}
\label{sec:analyticalA}

The covariant derivatives (Equations (8.51) - (8.53) in \cite{GHM1}) linear in the real conformal compensator $\Psi$ 
are given by
\begin{align}
    \nabla_{\alpha} ~=&~ {\rm D}_{\alpha} ~+~ \fracm{1}{2}\Psi {\rm D}_{\alpha}~+~\, 
    \fracm 1{10}  \,(\sigma^{\un{a}\un{b}})_{\alpha}{}^{\beta}({\rm D}_{\beta}\Psi){\cal M}_{\un{a}\un{b}}  ~~~, 
    \label{eq:N2A1}  \\
    \nabla_{\Dot{\alpha}} ~=&~ {\rm D}_{\Dot{\alpha}} ~+~ \fracm{1}{2}\Psi {\rm D}_{\Dot{\alpha}} ~+~\, \fracm 1{10}  \,(\sigma ^{\un{a}\un{b}})_{\Dot{\alpha}}{}^{\Dot{\beta}}
    ({\rm D}_{\Dot{\beta}}\Psi){\cal M}_{\un{a}\un{b}} ~~~, 
    \label{eq:N2A2} \\
    \nabla_{\un{a}} ~=&~ \pa_{\un{a}}~+~\Psi\pa_{\un{a}} ~-~ i  \fracm 1{5}(\sigma_{\un{a}})^{\delta \gamma}({\rm D}_{\delta}\Psi){\rm D}_{\gamma} ~-~ i \fracm 1{5}(\sigma_{\un{a}})^{\Dot{\delta}
    \Dot{ \gamma}}({\rm D}_{\Dot{\delta}}\Psi){\rm D}_{\Dot{\gamma}}~-~ (\pa_{\un{c}}\Psi){\cal M}_{\un{a}}{}^{\un{c}} ~~~,
    \label{eq:N2A3}
\end{align}
where the ``bare'' algebra of the supersymmetry covariant derivatives and the
spacetime partial derivative takes the forms 
\be
\{ \, {\rm D}_{\alpha}~,~{\rm D}_{\beta} \, \} ~=~ i \, (\sigma^{\un{a}})_{\alpha\beta}
\, \pa_{\un{a}}
~~~,~~~
\{ \, {\rm D}_{\Dot{\alpha}}~,~{\rm D}_{\Dot{\beta}} \, \} ~=~ i \, (\sigma^{\un{a}}
)_{\Dot{\alpha}\Dot{\beta}} \, \pa_{\un{a}} ~~~,~~~
\{ \, {\rm D}_{\alpha}~,~{\rm D}_{\Dot{\beta}} \, \} ~=~ 0 ~~~,
\label{eq:N2A4}
\ee
which indicate the non-vanishing values of the torsion supertensors must be given
by
\begin{align}
T_{\alpha\beta}^{\ \ \un{c}} ~=~ i(\sigma^{\un{c}})_{\alpha\beta} ~~~,~~~
T_{\Dot{\alpha}\Dot{\beta}}{}^{ \un{c}} ~=~ i(\sigma^{\un{c}})_{\Dot{\alpha}\Dot{\beta}}
~~~,~~~
T_{{\a} {\Dot \b}}{}^{\un{c}} ~=~ 0 ~~~,
\label{eq:N2A5}
\end{align}
and noting $[\pa{}_{\un a} \, ,\, \pa{}_{\un b}]$ = 0.

Mimicking results shown in going from (\ref{eq:Clss-1}) to (\ref{eq:ScaleLW}), we conclude 
the super scale transformation laws for the full superspace supercovariant supergravity derivative 
operators must take the forms which follow from the results in (\ref{eq:N2A1}), (\ref{eq:N2A2}), 
and (\ref{eq:N2A3}).  Thus, we write
\begin{align}
    \d{}_{S} \nabla_{\alpha} ~=&~ \fracm{1}{2}\,L \,\nabla{}_{\alpha}~+~\, 
    \fracm 1{10}  \,(\sigma^{\un{a}\un{b}})_{\alpha}{}^{ \beta}(\nabla{}_{\beta}L){\cal M}_{\un{a}\un{b}}  ~~~, 
    \label{eq:N2A1FL}   \\
    \d{}_{S} \nabla_{\Dot{\alpha}} ~=&~ \fracm{1}{2} \,L\, \nabla{}_{\Dot{\alpha}} 
    ~+~ \, \fracm 1{10}  \,(\sigma ^{\un{a}\un{b}})_{\Dot{\alpha}}{}^{\Dot{\beta}}(\nabla{}_{\Dot{\beta}}L){\cal M}_{\un{a}\un{b}} ~~~, 
    \label{eq:N2A2FL} \\
    \d{}_{S} \nabla_{\un{a}} ~=&~ L\, \nabla{}_{\un{a}} ~-~ i  \fracm 1{5}(\sigma_{\un{a}})^{\delta\gamma}
    (\nabla{}_{\delta}L)\nabla{}_{\gamma} ~-~ i \fracm 1{5}(\sigma_{\un{a}})^{\Dot{\delta}\Dot{ \gamma}}
    (\nabla{}_{\Dot{\delta}}L)\nabla{}_{\Dot{\gamma}}~-~ (\nabla{}_{\un{c}}L){\cal M}_{\un{a}}{}^{\un{c}} ~~~,
    \label{eq:N2A3FL}
\end{align}
as the super Weyl transformation laws of superspace supergravity derivatives for
10D, $\cal N$ = IIA theories.  

Upon using the definitions of torsion and curvature tensors 
\be \label{eq:TR1IIA}
\eqalign{
[ \,\nabla_{\a}~,~\nabla_{\b} \, \} ~=&~ T_{\a\b}{}^{\un{c}} \, \nabla_{\un{c}} 
~+~ T_{\a\b}{}^{\g} \, \nabla_{\g}
~+~ T_{\a\b}{}^{\Dot \g} \, \nabla_{\Dot \g} ~+~ \fracm{1}{2}R_{\a\b \,\un{
d}}{}^{\un{e}}\, \mathcal{M}_{\un{e}}{}^{\un{d}}  ~~~, \cr
[ \,\nabla_{\a}~,~\nabla_{{\Dot \b}} \, \} ~=&~ T_{\a{\Dot \b}}{}^{\un{c}} \, \nabla_{\un{c}} 
~+~ T_{\a{\Dot \b}}{}^{\g} \, \nabla_{\g} ~+~ T_{\a{\Dot \b}}{}^{\Dot \g} \, \nabla_{\Dot \g}
~+~ \fracm{1}{2}R_{\a{\Dot \b} \,\un{
d}}{}^{\un{e}}\, \mathcal{M}_{\un{e}}{}^{\un{d}}  ~~~, \cr
[ \,\nabla_{{\Dot \a}}~,~\nabla_{{\Dot \b}} \, \} ~=&~ T_{{\Dot \a}{\Dot \b}}{}^{\un{c}} \, \nabla_{\un{c}} 
~+~ T_{{\Dot \a}{\Dot \b}}{}^{\g} \, \nabla_{\g} ~+~ T_{{\Dot \a}{\Dot \b}}{}^{\Dot \g} \, \nabla_{\Dot \g}
~+~ \fracm{1}{2}R_{{\Dot \a}{\Dot \b} \,\un{
d}}{}^{\un{e}}\, \mathcal{M}_{\un{e}}{}^{\un{d}}  ~~~, \cr
[ \, \nabla_{\a}~,~\nabla_{\un{b}} \, \} ~=&~ T_{\a\un{b}}{}^{\un{c}}\, \nabla_{\un{c}}\, 
~+~ T_{\a\un{b}}{}^{\g}\, \nabla_{\g} ~+~ T_{\a{\un b}}{}^{\Dot \g} \, \nabla_{\Dot \g}
 ~+~ \fracm{1}{2}R_{\a\un{b}\, \un{
d}}{}^{\un{e}} \, \mathcal{M}_{\un{e}}{}^{\un{d}}  ~~~~,  \cr
[ \, \nabla_{{\Dot \a}}~,~\nabla_{\un{b}} \, \} ~=&~ T_{{\Dot \a}\un{b}}{}^{\un{c}}\, \nabla_{\un{c}}\, 
~+~ T_{{\Dot \a}\un{b}}{}^{\g}\, \nabla_{\g} ~+~ T_{{\Dot \a}{\un b}}{}^{\Dot \g} \, \nabla_{\Dot \g}
~+~ \fracm{1}{2}R_{{\Dot \a}\un{b}\, \un{
d}}{}^{\un{e}} \, \mathcal{M}_{\un{e}}{}^{\un{d}}    ~~~~,  \cr
[ \, \nabla_{\un{a}}~,~ \nabla_{\un{b}} \, \} ~=&~ T_{\un{a}\un{b}}{}^{\un{c}} \, \nabla_{\un{c}}\,\, 
~+~ T_{\un{a}\un{b}}{}^{\g} \, \nabla_{\g} ~+~ T_{{\un a}{\un b}}{}^{\Dot \g} \, \nabla_{\Dot \g}
~+~ \fracm{1}{2}R_{\un{a}\un{b} \, \un{
d}}{}^{\un{e}}\, \mathcal{M}_{\un{e}}{}^{\un{d}} ~~~~ , }
\ee
appropriate to the case of the 10, $\cal N$ = IIA theory, we find the Weyl scaling properties of 
all the superspace torsion and curvature supertensors with weights of less than three-halves 
as below. Weyl scaling properties of all other superspace torsion and curvature supertensors 
with weights of equal and larger than three-halves will be presented in Appendix \ref{appen:Weyl-IIA}.
Grouping according to scale. we find,
\begin{align}
\d_S T_{\a \b}{}^{\un c} \, =&~ 0 ~~~,{~~~~~~~~~~~~~~~~~~~~~~~~~~~~~~~~~~~~~~~~~~}
{~~~~~~~~~~~~~~~~~~~~~~~~~~~~~~~~~~~~~~~~~~~~~~~~~~~} \\
\d_S T_{\a \Dot{\b}}{}^{\un c} \, =&~ 0 ~~~, \\
\d_S T_{\Dot{\a} \Dot{\b}}{}^{\un c} \, =&~ 0 ~~~, 
\end{align}

\begin{align}
\d_S T_{\a \b}{}^{\g} \, =&~  \fracm12 LT_{\a \b}{}^{\g} + i\fracm15 T_{\a \b}{}^{\un c} (\s_{\un c})^{\g\d}(\nabla_{\d}L) +~ \fracm12(\nabla_{(\a}L)\d_{\b)}{}^{\g}
~ +\fracm{1}{20}(\s^{[2]})_{(\a}{}^{\d}(\s_{[2]})_{\b)}{}^{\g}(\nabla_{\d}L) ~~~, \\
\d_S T_{\a \b}{}^{\Dot{\g}} \, =&~ \fracm12 LT_{\a \b}{}^{\Dot{\g}} + i\fracm15 T_{\a \b}{}^{\un c} (\s_{\un{c}})^{\Dot{\d}\Dot{\g}}(\nabla_{\Dot{\d}}L) ~~~, \\  
\d_S T_{\Dot{\a} \Dot{\b}}{}^{\Dot{\g}} \, =&~ \fracm12 LT_{\Dot{\a} \Dot{\b}}{}^{\Dot{\g}} + i\fracm15 T_{\Dot{\a} \Dot{\b}}{}^{\un c} (\s_{\un c})^{\Dot{\g} \Dot{\d}}(\nabla_{\Dot{\d}}L) + \fracm12(\nabla_{(\Dot{\a}}L)\d_{\Dot{\b})}{}^{\Dot{\g}} 
~ +\fracm{1}{20}(\s^{[2]})_{(\Dot{\a}}{}^{\Dot{\d}}(\s_{[2]})_{\Dot{\b})}{}^{\Dot{\g}}(\nabla_{\Dot{\d}}L) ~~~, \\
\d_S T_{\Dot{\a} \Dot{\b}}{}^{\g} \, =&~  \fracm12 LT_{\Dot{\a} \Dot{\b}}{}^{\g} + i\fracm15 T_{\Dot{\a} \Dot{\b}}{}^{\un c} (\s_{\un c})^{\g\d}(\nabla_{\d}L)  ~~~, \\
\d_S T_{\a \Dot{\b}}{}^{\g} \, =&~ \fracm12 L T_{\a \Dot{\b}}{}^{\g} + i\fracm15 T_{\a \Dot{\b}}{}^{\un{c}}(\s_{\un{c}})^{\d\g}(\nabla_{\d}L) + \fracm12 (\nabla_{\Dot{\b}}L)\d_{\a}{}^{\g} 
~ +\fracm{1}{20} (\s^{[2]})_{\Dot{\b}}{}^{\Dot{\d}}(\s_{[2]})_{\a}{}^{\g}(\nabla_{\Dot{\d}}L) ~~~, \\
\d_S T_{\a \Dot{\b}}{}^{\Dot{\g}} \, =&~ \fracm12 L T_{\a \Dot{\b}}{}^{\Dot{\g}} + i\fracm15 T_{\a \Dot{\b}}{}^{\un{c}}(\s_{\un{c}})^{\Dot{\d}\Dot{\g}}(\nabla_{\Dot{\d}}L) + \fracm12 (\nabla_{\a}L)\d_{\Dot{\b}}{}^{\Dot{\g}} 
~ +\fracm{1}{20} (\s^{[2]})_{\Dot{\b}}{}^{\Dot{\g}}(\s_{[2]})_{\a}{}^{\d}(\nabla_{\d}L) ~~~, \\
\d_S T_{\a \un{b}}{}^{\un c} \, =&~ \fracm12 L T_{\a \un b}{}^{\un c} + (\nabla_{\a}L)\d_{\un b}{}^{\un c} + \fracm15 (\s_{\un b}{}^{\un c})_{\a}{}^{\b}(\nabla_{\b}L) 
~+ i\fracm15(\s_{\un b})^{\b\d}(\nabla_{\d}L)T_{\a \b}{}^{\un c} \nonumber\\
&~ + i\fracm15(\s_{\un b})^{\Dot{\b}\Dot{\d}}(\nabla_{\Dot{\d}}L)T_{\a \Dot{\b}}{}^{\un c} ~~~, \\
\d_S T_{\Dot{\a} \un{b}}{}^{\un c} \, =&~ \fracm12 L T_{\Dot{\a} \un b}{}^{\un c} + (\nabla_{\Dot{\a}}L)\d_{\un b}{}^{\un c} + \fracm15 (\s_{\un b}{}^{\un c})_{\Dot{\a}}{}^{\Dot{\b}}(\nabla_{\Dot{\b}}L) 
~+ i\fracm15(\s_{\un b})^{\b\d}(\nabla_{\d}L)T_{\Dot{\a} \b}{}^{\un c}
\nonumber\\
& + i\fracm15(\s_{\un b})^{\Dot{\b}\Dot{\d}}(\nabla_{\Dot{\d}}L)T_{\Dot{\a} \Dot{\b}}{}^{\un c} ~~~, 
\end{align}

\begin{align}
\d_S T_{\a \un b}{}^{\g} \, =&~ LT_{\a \un b}{}^{\g} + i\fracm15 T_{\a \un b}{}^{\un c}(\s_{\un c})^{\g\d}(\nabla_{\d}L) - \fracm12 (\nabla_{\un b}L)\d_{\a}{}^{\g} -i\fracm15(\s_{\un b})^{\g\b}(\nabla_{\a}\nabla_{\b}L)  \nonumber\\
&~  + i\fracm15(\s_{\un b})^{\d\b}(\nabla_{\d}L)T_{\a \b}{}^{\g} + i\fracm15(\s_{\un b})^{\Dot{\d}\Dot{\b}}(\nabla_{\Dot{\d}}L)T_{\a \Dot{\b}}{}^{\g}  ~+ \fracm12 (\s_{\un b}{}^{\un c})_{\a}{}^{\g}(\nabla_{\un c}L) ~~~, \\
\d_S T_{\a \un{b}}{}^{\Dot{\g}} \, =&~ LT_{\a \un b}{}^{\Dot{\g}}  -i\fracm15(\s_{\un b})^{\Dot{\g}\Dot{\b}}(\nabla_{\a}\nabla_{\Dot{\b}}L) + i\fracm15(\s_{\un b})^{\d\b}(\nabla_{\d}L)T_{\a \b}{}^{\Dot{\g}} + i\fracm15(\s_{\un b})^{\Dot{\d}\Dot{\b}}(\nabla_{\Dot{\d}}L)T_{\a \Dot{\b}}{}^{\Dot{\g}} \nonumber\\
&~- i\fracm15 (\s_{\un c})^{\Dot{\d}\Dot{\g}}(\nabla_{\Dot{\d}}L)T_{\a \un{b}}{}^{\un{c}} ~~~, 
 \\
\d_S T_{\Dot{\a} \un{b}}{}^{\g} \, =&~  LT_{\Dot{\a} \un b}{}^{\g}  -i\fracm15(\s_{\un b})^{\g\b}(\nabla_{\Dot{\a}}\nabla_{\b}L) + i\fracm15(\s_{\un b})^{\d\b}(\nabla_{\d}L)T_{\Dot{\a} \b}{}^{\g} + i\fracm15(\s_{\un b})^{\Dot{\d}\Dot{\b}}(\nabla_{\Dot{\d}}L)T_{\Dot{\a} \Dot{\b}}{}^{\g} {~~~~~~~~~~~~~~~}  \nonumber\\
&~- i\fracm15 (\s_{\un c})^{\d\g}(\nabla_{\d}L)T_{\Dot{\a} \un{b}}{}^{\un{c}} ~~~, \\
\d_S T_{\Dot{\a} \un{b}}{}^{\Dot{\g}} \, =&~ LT_{\Dot{\a} \un b}{}^{\Dot{\g}} + i\fracm15T_{\Dot{\a} \un b}{}^{\un c}(\s_{\un c})^{\Dot{\g}\Dot{\d}}(\nabla_{\Dot{\d}}L) - \fracm12 (\nabla_{\un b}L)\d_{\Dot{\a}}{}^{\Dot{\g}} 
-i\fracm15(\s_{\un b})^{\Dot{\g}\Dot{\b}}(\nabla_{\Dot{\a}}\nabla_{\Dot{\b}}L)
\nonumber\\
&~  + i\fracm15(\s_{\un b})^{\d\b}(\nabla_{\d}L)T_{\Dot{\a}\b }{}^{\Dot{\g}} + i\fracm15(\s_{\un b})^{\Dot{\d}\Dot{\b}}(\nabla_{\Dot{\d}}L)T_{\Dot{\a} \Dot{\b}}{}^{\Dot{\g}} 
~+ \fracm12 (\s_{\un b}{}^{\un c})_{\Dot{\a}}{}^{\Dot{\g}}(\nabla_{\un c}L) ~~~, \\
\d_S T_{\un a \un b}{}^{\un c} \, =&~ L T_{\un a \un b}{}^{\un c} - i\fracm15 (\s_{[\un a})^{\a\b}(\nabla_{\a}L)  T_{\b| \un b]}{}^{\un c}- i\fracm15 (\s_{[\un a})^{\Dot{\a}\Dot{\b}}(\nabla_{\Dot{\a}}L)  T_{\Dot{\b}| \un b]}{}^{\un c}~~~, 
\end{align}

\begin{align}
\d_S R_{\a \b}{}^{\un{d}\un{e}} \, =&~ L R_{\a \b}{}^{\un{d}\un{e}} - T_{\a \b}{}^{[\un d}(\nabla^{\un e]}L) + \fracm15  T_{\a \b}{}^{\g}(\s^{\un{d}\un{e}})_{\g}{}^{\d}(\nabla_{\d}L) 
~+ \fracm15  T_{\a \b}{}^{\Dot{\g}}(\s^{\un{d}\un{e}})_{\Dot{\g}}{}^{\Dot{\d}}(\nabla_{\Dot{\d}}L)
{~~~~~~~~~~~~~~~~~~~~}
\nonumber\\
&  -\fracm15 (\s^{\un{d}\un{e}})_{(\a}{}^{\d}(\nabla_{\b)}\nabla_{\d}L)  ~~~, 
 \\
\d_S R_{\a \Dot{\b}}{}^{\un{d}\un{e}} \, =&~ L R_{\a \Dot{\b}}{}^{\un{d}\un{e}} - T_{\a \Dot{\b}}{}^{[\un d}(\nabla^{\un e]}L) + \fracm15  T_{\a \Dot{\b}}{}^{\g}(\s^{\un{d}\un{e}})_{\g}{}^{\d}(\nabla_{\d}L)
~+ \fracm15  T_{\a \Dot{\b}}{}^{\Dot{\g}}(\s^{\un{d}\un{e}})_{\Dot{\g}}{}^{\Dot{\d}}(\nabla_{\Dot{\d}}L)
 \nonumber\\
&  -\fracm15 (\s^{\un{d}\un{e}})_{\a}{}^{\d}(\nabla_{\Dot{\b}}\nabla_{\d}L)
 ~ -\fracm15 (\s^{\un{d}\un{e}})_{\Dot{\b}}{}^{\Dot{\g}}(\nabla_{\a}\nabla_{\Dot{\g}}L) ~~~,  \\
\d_S R_{\Dot{\a} \Dot{\b}}{}^{\un{d}\un{e}} \, =&~  L R_{\Dot{\a} \Dot{\b}}{}^{\un{d}\un{e}} - T_{\Dot{\a} \Dot{\b}}{}^{[\un d}(\nabla^{\un e]}L) + \fracm15  T_{\Dot{\a} \Dot{\b}}{}^{\g}(\s^{\un{d}\un{e}})_{\g}{}^{\d}(\nabla_{\d}L) 
~+ \fracm15  T_{\Dot{\a} \Dot{\b}}{}^{\Dot{\g}}(\s^{\un{d}\un{e}})_{\Dot{\g}}{}^{\Dot{\d}}(\nabla_{\Dot{\d}}L) 
\nonumber\\
& -\fracm15 (\s^{\un{d}\un{e}})_{(\Dot{\a}}{}^{\Dot{\d}}(\nabla_{\Dot{\b})}\nabla_{\Dot{\d}}L)  ~~~.
\end{align}

Similar to 11D, we define ${\cal J}$-tensors and use them to construct the Weyl tensor superfield. The ${\cal J}$-tensors constructed are the following,
\begin{equation}
\begin{split}
    {\cal J}_{\a}^{(1)} ~=&~ \fracm{1}{2} T_{\a\b}{}^{\b} ~~~,~~~ 
    {\cal J}_{\Dot{\a}}^{(1)} ~=~ \fracm{1}{2} T_{\Dot{\a}\Dot{\b}}{}^{\Dot{\b}} ~~~, \\
    {\cal J}_{\a}^{(2)} ~=&~ \fracm{1}{8} T_{\a\Dot{\b}}{}^{\Dot{\b}} ~~~,~~~ 
    {\cal J}_{\Dot{\a}}^{(2)} ~=~ \fracm{1}{8} T_{\Dot{\a}\b}{}^{\b} ~~~, \\
    {\cal J}_{\a}^{(3)} ~=&~ \fracm{1}{8} T_{\a\un{b}}{}^{\un{b}} ~~~~,~~~ 
    {\cal J}_{\Dot{\a}}^{(3)} ~=~ \fracm{1}{8} T_{\Dot{\a}\un{b}}{}^{\un{b}} ~~~.
\end{split}
\end{equation}
Thus we see
\begin{equation}
\begin{split}
    \d_{S} {\cal J}_{\a}^{(1)} ~=&~ \fracm{1}{2} L {\cal J}_{\a}^{(1)} ~+~ ( \nabla_{\a} L ) ~~~,~~~ 
    \d_{S} {\cal J}_{\Dot{\a}}^{(1)} ~=~ \fracm{1}{2} L {\cal J}_{\Dot{\a}}^{(1)} ~+~ ( \nabla_{\Dot{\a}} L ) ~~~, \\ 
    \d_{S} {\cal J}_{\a}^{(2)} ~=&~ \fracm{1}{2} L {\cal J}_{\a}^{(2)} ~+~ ( \nabla_{\a} L ) ~~~,~~~ 
    \d_{S} {\cal J}_{\Dot{\a}}^{(2)} ~=~ \fracm{1}{2} L {\cal J}_{\Dot{\a}}^{(2)} ~+~ ( \nabla_{\Dot{\a}} L ) ~~~, \\
    \d_{S} {\cal J}_{\a}^{(3)} ~=&~ \fracm{1}{2} L {\cal J}_{\a}^{(3)} ~+~ ( \nabla_{\a} L ) ~~~,~~~ 
    \d_{S} {\cal J}_{\Dot{\a}}^{(3)} ~=~ \fracm{1}{2} L {\cal J}_{\Dot{\a}}^{(3)} ~+~ ( \nabla_{\Dot{\a}} L ) ~~~.
\end{split}
\end{equation}
One can see there are only two independent objects (one from each copy of Grassmann coordinates) by changing basis for these superfields. Define
\begin{equation}
\begin{aligned}
    {\cal J}_{\a}^{(+)} ~=&~ \fracm{1}{3} \big( {\cal J}_{\a}^{(3)} ~+~ {\cal J}_{\a}^{(1)} ~+~ {\cal J}_{\a}^{(2)} \big) ~~~,~ &
    {\cal J}_{\Dot{\a}}^{(+)} ~=&~ \fracm{1}{3} \big( {\cal J}_{\Dot{\a}}^{(3)} ~+~ {\cal J}_{\Dot{\a}}^{(1)} ~+~ {\cal J}_{\Dot{\a}}^{(2)} \big) ~~~,  
    \\
    {\cal J}_{\a}^{(-1)} ~=&~ \fracm{1}{2} \big( {\cal J}_{\a}^{(3)} ~-~ {\cal J}_{\a}^{(1)} \big) ~~~,~ & 
    {\cal J}_{\Dot{\a}}^{(-1)} ~=&~ \fracm{1}{2} \big( {\cal J}_{\Dot{\a}}^{(3)} ~-~ {\cal J}_{\Dot{\a}}^{(1)} \big) ~~~, \\
    {\cal J}_{\a}^{(-2)} ~=&~ \fracm{1}{6} \big( {\cal J}_{\a}^{(3)} ~+~ {\cal J}_{\a}^{(1)} ~-~ 2 {\cal J}_{\a}^{(2)} \big) ~~~,~ &
    {\cal J}_{\Dot{\a}}^{(-2)} ~=&~ \fracm{1}{6} \big( {\cal J}_{\Dot{\a}}^{(3)} ~+~ {\cal J}_{\Dot{\a}}^{(1)} ~-~ 2 {\cal J}_{\Dot{\a}}^{(2)} \big) ~~~.
\end{aligned}
\end{equation}
The variations of these $\cal J$-tensors under the scale variation take the forms
\begin{equation}
\begin{aligned}
    \d_{S} {\cal J}_{\a}^{(+)} ~=&~ \fracm{1}{2} L {\cal J}_{\a}^{(+)} ~+~ ( \nabla_{\a} L ) ~~~,~ & 
    \d_{S} {\cal J}_{\Dot{\a}}^{(+)} ~=&~ \fracm{1}{2} L {\cal J}_{\Dot{\a}}^{(+)} ~+~ ( \nabla_{\Dot{\a}} L ) ~~~, \\
    \d_{S} {\cal J}_{\a}^{(-1)} ~=&~ \fracm{1}{2} L {\cal J}_{\a}^{(-1)} ~~~,~ &
    \d_{S} {\cal J}_{\Dot{\a}}^{(-1)} ~=&~ \fracm{1}{2} L {\cal J}_{\Dot{\a}}^{(-1)} ~~~, \\
    \d_{S} {\cal J}_{\a}^{(-2)} ~=&~ \fracm{1}{2} L {\cal J}_{\a}^{(-2)} ~~~,~ & 
    \d_{S} {\cal J}_{\Dot{\a}}^{(-2)} ~=&~ \fracm{1}{2} L {\cal J}_{\Dot{\a}}^{(-2)} ~~~.
\end{aligned}
\end{equation}
In this basis, only ${\cal J}_{\a}^{(+)}$ and ${\cal J}_{\Dot{\a}}^{(+)}$ serve as spinorial gauge connections. 
The Weyl tensor is then constructed using conformal weight 1 torsion supertensors and conformal weight $\fracm{1}{2}$ spinorial gauge connections.
By requiring $\d_{S} {\cal W}_{\un{a}\un{b}\un{c}} = L {\cal W}_{\un{a}\un{b}\un{c}}$, we obtain
\begin{equation}
\begin{split}
    {\cal W}_{\un{a}\un{b}\un{c}} ~=&~ \fracm{1}{32} \Big\{\, (\s^{\un d})_{\g\b}(\s_{\un{a}\un{b}\un{c}})^{\b\a}\, T_{\a\un d}{}^{\g} - i 2(\s_{\un{a}\un{b}\un{c}})^{\a\b}\, \big[\,  \nabla_{\a}\mathcal{J}_{\b}^{(+)} - \fracm{6}{5} \,\mathcal{J}_{\a}^{(+)} \mathcal{J}_{\b}^{(+)} \,\big] \\
    &~~~~~~ + (\s^{\un d})_{\Dot\g\Dot\b}(\s_{\un{a}\un{b}\un{c}})^{\Dot\b\Dot\a}\, T_{\Dot\a\un d}{}^{\Dot\g} - i 2(\s_{\un{a}\un{b}\un{c}})^{\Dot\a\Dot\b}\, \big[\, \nabla_{\Dot\a}\mathcal{J}_{\Dot\b}^{(+)} - \fracm{6}{5} \,\mathcal{J}_{\Dot\a}^{(+)} \mathcal{J}_{\Dot\b}^{(+)} \,\big] \,\Big\} ~~~.
\end{split}
\label{eq:WFS1}
\end{equation}

\subsection{10D, ${\cal N}~=~$IIB Supergravity Derivatives \& Scale Transformations}
\label{sec:analyticalB}

Now, the covariant derivative operators linear in the complex conformal compensator $\Psi$
(appropriate for the IIB theory) and necessary for a Nordstr\" om theory (Equations (8.92) - (8.94) in \cite{GHM1}) may be given by
\begin{align}
\nabla_{\a} ~=&~ {\rm D}_{\a} ~+~ \fracm{1}{2} \Psi {\rm D}_{\a} 
~+~ \fracm{1}{10} (\s^{\un{a}\un{b}})_{\a}{}^{\b} ({\rm D}_{\b} \Psi) {\cal M}_{\un{a}\un{b}} ~~~, 
\label{eq:N2B1} \\
\Bar{\nabla}_{\a} ~=&~ \Bar{\rm D}_{\a} ~+~ \fracm{1}{2} \Bar{\Psi} \Bar{\rm D}_{\a} 
~+~ \fracm{1}{10} (\s^{\un{a}\un{b}})_{\a}{}^{\b} (\Bar{\rm D}_{\b} \Bar{\Psi}) {\cal M}_{\un{a}\un{b}} ~~~, 
\label{eq:N2B2} \\
\begin{split}
\nabla_{\un{a}} ~=&~ \pa_{\un{a}} ~+~ \fracm{1}{2} \, (\, \Psi ~+~ \Bar{\Psi} \,)  \, \pa_{\un{a}} 
~-~ i \, \fracm{1}{32} \, (\s_{\un{a}})^{\a\b} \,  \left[ \, \Bar{\rm D}_{\a} ( \, \Psi  ~+~ \fracm{27}{5} \, \Bar{\Psi} \, )  \,  \right] \, {\rm D}_{\b}  \\
& ~-~ i \, \fracm{1}{32} \, (\s_{\un{a}})^{\a\b} \, \left[ \, {\rm D}_{\a}  ( \, \Bar{\Psi} ~+~ \fracm{27}{5} \, \Psi) \, \right] \, \Bar{\rm D}_{\b} 
~-~  \, \fracm{1}{2} \,  \left[ \, \pa_{\un{c}} (\, \Psi ~+~ \Bar{\Psi} \,)  \, \right] {\cal M}_{\un{a}}{}^{\un{c}} ~~~,
\end{split}
\label{eq:N2B3} 
\end{align}
here the ``bare'' algebra of the supersymmetry covariant derivatives and the
spacetime partial derivative takes the forms 
\be
\{ \, {\rm D}_{\a} ~,~ {\rm D}_{\b} \, \} ~=~ 0 ~~~,~~~
\{ \, \Bar{\rm D}_{\a} ~,~ \Bar{\rm D}_{\b} \, \} ~=~ 0  ~~~,~~~
\{ \, {\rm D}_{\a} ~,~ \Bar{\rm D}_{\b} \, \} ~=~  i \, (\s^{\un{a}})_{\a\b} \, \pa_{\un{a}} ~~~,
\label{eq:N2B4}
\ee
which indicate the non-vanishing torsion supertensors must be given
by
\begin{align}
T_{\a \b}{}^{\un{c}}~=~ 0 ~~~,~~~ T_{{\Bar \a}{\Bar \b}}{}^{\un{c}}~=~ 0 ~~~,~~~ 
T_{\a\Bar{\b}}{}^{\un{c}}~=~ i\, (\s^{\un{c}})_{\a\b} ~~~.
\label{eq:N2B5}
\end{align}
By repeating the steps used to obtain (\ref{eq:N2A1FL}), (\ref{eq:N2A2FL}),
and (\ref{eq:N2A3FL}) in the last section, here we find,
\begin{align}
\d_{S} \nabla_{\a} ~=&~ \fracm{1}{2} L {\nabla}_{\a} 
~+~ \fracm{1}{10}\, (\s^{\un{a}\un{b}})_{\a}{}^{\b} ({\nabla}_{\b} L) \, {\cal M}_{\un{a}\un{b}} ~~~,
\label{eq:N2B6} \\
\d_{S} \Bar{\nabla}_{\a} ~=&~ \fracm{1}{2} \Bar{L} \Bar{\nabla}_{\a}
~+~ \fracm {1}{10} \, (\s ^{\un{a}\un{b}})_{\a}{}^{\b} (\Bar{\nabla}_{\b} \, \Bar{L}){\cal M}_{\un{a}\un{b}} ~~~,
\label{eq:N2B7} \\
\begin{split}
\d_{S} \nabla_{\un{a}} ~=&~ \fracm{1}{2} \, (\, L  ~+~ \Bar{L} \,)  \, \nabla_{\un{a}} ~-~ i \, \fracm{1}{32} \, (\s_{\un{a}})^{\a\b} \,  \left[ \, \Bar{\nabla}_{\a} ( \, L  ~+~ \fracm{27}{5} \, \Bar{L} \, )  \,  \right] \, {\nabla}_{\b} \\
& ~-~ i \, \fracm{1}{32} \, (\s_{\un{a}})^{\a\b} \, \left[ \, {\nabla}_{\a} ( \, \Bar{L} ~+~ \fracm{27}{5} \, L) \, \right]
\, \Bar{\nabla}_{\b}  ~-~ \, \fracm{1}{2} \,  \left[ \, \nabla_{\un{c}} (\, L ~+~ \Bar{L} \,)  \, \right] {\cal M}_{\un{a}}{}^{\un{c}} ~~~.
\end{split}
\label{eq:N2B8} 
\end{align}
The super-Weyl parameter in (\ref{eq:N2A1FL}), (\ref{eq:N2A2FL}), and 
(\ref{eq:N2A3FL}) 
is distinguished from that in (\ref{eq:N2B6}), (\ref{eq:N2B7}), and 
(\ref{eq:N2B8}).  The former 
involve a real scaling parameter superfield $L$ = $\Bar L$, whereas the latter involve a complex  
scaling parameter superfield $L$ $\ne$ $\Bar L$.  This means the transformation laws in (\ref{eq:N2B6}), 
(\ref{eq:N2B7}), and (\ref{eq:N2B8}) imply both a real scaling parametrized by $\fracm12(L + \Bar L)$ and also 
a U(1) rotation parametrized by $i \fracm12(L - \Bar L)$.  The terms proportional to $\cal M$ in (\ref{eq:N2B6}), 
and (\ref{eq:N2B7}) as well as the ones proportional to the spinorial derivatives in (\ref{eq:N2B8})
depend on both the real superspace Weyl scaling parameter u $=~ \fracm12(L + \Bar L)$ and the real
U(1) rotation parameter v = $- i \fracm12(L - \Bar L)$.  To see this, one need only note
$L = ( {\rm u} + i {\rm v} )$ and  $\Bar L = ( {\rm u} - i {\rm v} )$ in the latter equations.

Upon using the definitions of torsion and curvature tensors 
\be \label{eq:TR1IIB}
\eqalign{
[ \,\nabla_{\a}~,~\nabla_{\b} \, \} ~=&~ T_{\a\b}{}^{\un{c}} \, \nabla_{\un{c}} ~+~ T_{\a\b}{}^{\g} \, \nabla_{\g}
~+~ T_{\a\b}{}^{\Bar \g} \, \Bar{\nabla}_{\g} ~+~ \fracm{1}{2}R_{\a\b \,\un{
d}}{}^{\un{e}}\, \mathcal{M}_{\un{e}}{}^{\un{d}}  ~~~, \cr
[ \,\nabla_{\a}~,~ \Bar{\nabla}_{\b} \, \}  ~=&~ T_{\a{\Bar \b}}{}^{\un{c}} \, \nabla_{\un{c}} 
~+~ T_{\a{\Bar \b}}{}^{\g} \, \nabla_{\g} ~+~ T_{\a{\Bar \b}}{}^{\Bar \g} \, \Bar{\nabla}_{\g}
~+~ \fracm{1}{2}R_{\a{\Bar \b} \,\un{
d}}{}^{\un{e}}\, \mathcal{M}_{\un{e}}{}^{\un{d}}  ~~~, \cr
[ \,\Bar{\nabla}_{\a}
~,~ \Bar{\nabla}_{\b} \, \} ~=&~ T_{{\Bar \a}{\Bar \b}}{}^{\un{c}} \, \nabla_{\un{c}} 
~+~ T_{{\Bar \a}{\Bar \b}}{}^{\g} \, \nabla_{\g} ~+~ T_{{\Bar \a}{\Bar \b}}{}^{\Bar \g} \, \Bar{\nabla}_{\g}
~+~ \fracm{1}{2}R_{{\Bar \a}{\Bar \b} \,\un{ d}}{}^{\un{e}}\, \mathcal{M}_{\un{e}}{}^{\un{d}}  ~~~, \cr
[ \, \nabla_{\a}~,~\nabla_{\un{b}} \, \} ~=&~ T_{\a\un{b}}{}^{\un{c}}\, \nabla_{\un{c}}\, 
~+~ T_{\a\un{b}}{}^{\g}\, \nabla_{\g} ~+~ T_{\a{\un b}}{}^{\Bar \g} \, \Bar{\nabla}_{\g}
 ~+~ \fracm{1}{2}R_{\a\un{b}\, \un{
d}}{}^{\un{e}} \, \mathcal{M}_{\un{e}}{}^{\un{d}}  ~~~~,  \cr
[ \, \Bar{\nabla}_{\a} ~,~\nabla_{\un{b}} \, \} ~=&~ T_{{\Bar \a}\un{b}}{}^{\un{c}}\, \nabla_{\un{c}}\, 
~+~ T_{{\Bar \a}\un{b}}{}^{\g}\, \nabla_{\g} ~+~ T_{{\Bar \a}{\un b}}{}^{\Bar \g} \, \Bar{\nabla}_{\g}
~+~ \fracm{1}{2}R_{{\Bar \a}\un{b}\, \un{
d}}{}^{\un{e}} \, \mathcal{M}_{\un{e}}{}^{\un{d}}    ~~~~,  \cr
[ \, \nabla_{\un{a}}~,~ \nabla_{\un{b}} \, \} ~=&~ T_{\un{a}\un{b}}{}^{\un{c}} \, \nabla_{\un{c}}\,\, 
~+~ T_{\un{a}\un{b}}{}^{\g} \, \nabla_{\g} ~+~ T_{{\un a}{\un b}}{}^{\Bar \g} \, \Bar{\nabla}_{\g}
~+~ \fracm{1}{2}R_{\un{a}\un{b} \, \un{
d}}{}^{\un{e}}\, \mathcal{M}_{\un{e}}{}^{\un{d}} ~~~~ , }
\ee
appropriate to the case of the 10D, $\cal N$ = IIB theory, we find the Weyl scaling properties of 
all the superspace torsion and curvature supertensors with weights of less than three-halves 
as below. Weyl scaling properties of all other superspace torsion and curvature supertensors 
with weights of equal and larger than three-halves will be presented in Appendix \ref{appen:Weyl-IIB}.
\begin{align}
\d_S T_{\a \b}{}^{\un c} \, =&~ \fracm12 (L-\overline{L}) T_{\a \b}{}^{\un c} ~~~, \label{equ:Talphabetac-IIB}\\
\d_S T_{\a \overline{\b}}{}^{\un c} \, =&~ 0 ~~~,  \\
\d_S T_{\overline{\a} \overline{\b}}{}^{\un c} \, =&~ -\, \fracm12 (L - \overline{L}) T_{\overline{\a} \overline{\b}}{}^{\un c}\label{equ:Talphabarbetabarc-IIB}  ~~~, 
{~~~~~~~~~~~~~~~~~~~~~~~~~~~~~~}
{~~~~~~~~~~~~~~~~~~~~~~~~~~~~~~}
\end{align}

\begin{align}
\d_S T_{\a \b}{}^{\overline{\g}} \, =&~ (L-\fracm12 \overline{L})T_{\a \b}{}^{\overline{\g}} + i T_{\a \b}{}^{\un c} (\s_{\un c})^{\d\g} (\nabla_{\d}(\fracm{1}{32}\overline{L} + \fracm{27}{160}L)) ~~~, \label{equ:Talphabetagammabar-IIB}\\
\d_S T_{\a \b}{}^{\g} \, =&~  \fracm12 LT_{\a \b}{}^{\g} + i T_{\a \b}{}^{\un c} (\s_{\un c})^{\g\d}( \Bar{\nabla}_{\d}(\fracm{1}{32}L + \fracm{27}{160}\overline{L}))\nonumber \\
&~ + \fracm12(\nabla_{(\a}L)\d_{\b)}{}^{\g}  +\fracm{1}{20}(\s^{[2]})_{(\a}{}^{\d}(\s_{[2]})_{\b)}{}^{\g}(\nabla_{\d}L) ~~~, \\
\d_S T_{\a \overline{\b}}{}^{\overline{\g}} \, =&~ \fracm12 L T_{\a \overline{\b}}{}^{\overline{\g}} + i T_{\a \overline{\b}}{}^{\un{c}}(\s_{\un{c}})^{\d\g}(\nabla_{\d}(\fracm{1}{32}\overline{L} + \fracm{27}{160}L))  + \fracm12 (\nabla_{\a}\overline{L})\d_{\b}{}^{\g} 
\nonumber\\
&~ +\fracm{1}{20} (\s^{[2]})_{\b}{}^{\g}(\s_{[2]})_{\a}{}^{\d}(\nabla_{\d}L) ~~~,  \\
\d_S T_{\a \un{b}}{}^{\un c} \, =&~ \fracm12 L T_{\a \un b}{}^{\un c} +\fracm12 (\nabla_{\a}(L+\overline{L}))\d_{\un b}{}^{\un c} + \fracm15 (\s_{\un b}{}^{\un c})_{\a}{}^{\b}(\nabla_{\b}L) \nonumber\\
&~+ i(\s_{\un b})^{\d\b}(\nabla_{\d}(\fracm{1}{32}\overline{L} + \fracm{27}{160}L))T_{\a \overline{\b}}{}^{\un c} + i(\s_{\un b})^{\d\b}(\Bar{\nabla}_{\d}(\fracm{1}{32}L + \fracm{27}{160}\overline{L}))T_{\a \b}{}^{\un c} ~~~, \\
\d_S T_{\overline{\a} \un{b}}{}^{\un c} \, =&~ \fracm12 \overline{L} T_{\overline{\a} \un b}{}^{\un c} + \fracm12(\Bar{\nabla}_{\a}(L+\overline{L}))\d_{\un b}{}^{\un c} + \fracm15 (\s_{\un b}{}^{\un c})_{\a}{}^{\b} (\Bar{\nabla}_{\b}\overline{L}) \nonumber\\
&~+ i(\s_{\un b})^{\b\d}(\Bar{\nabla}_{\d}(\fracm{1}{32}L + \fracm{27}{160}\overline{L}))T_{\overline{\a}\b}{}^{\un c} + i(\s_{\un b})^{\b\d}(\nabla_{\d}(\fracm{1}{32}\overline{L} + \fracm{27}{160}L))T_{\overline{\a} \overline{\b}}{}^{\un c} ~~~,  \\
\d_S T_{\a \overline{\b}}{}^{\g} \, =&~ \fracm12 \overline{L} T_{\a \overline{\b}}{}^{\g} + i T_{\a \overline{\b}}{}^{\un{c}}(\s_{\un{c}})^{\d\g}(\Bar{\nabla}_{\d}(\fracm{1}{32}L + \fracm{27}{160}\overline{L})) + \fracm12 (\Bar{\nabla}_{\b}L)\d_{\a}{}^{\g} \nonumber\\
&~ +\fracm{1}{20} (\s^{[2]})_{\b}{}^{\d}(\s_{[2]})_{\a}{}^{\g}(\Bar{\nabla}_{\d}\overline{L}) ~~~, \\
\d_S T_{\overline{\a} \overline{\b}}{}^{\overline{\g}} \, =&~ \fracm12 \overline{L}T_{\overline{\a} \overline{\b}}{}^{\overline{\g}} + i T_{\overline{\a} \overline{\b}}{}^{\un c} (\s_{\un c})^{\g \d}(\nabla_{\d}(\fracm{1}{32}\overline{L} + \fracm{27}{160}L)) \nonumber\\
&~+ \fracm12(\Bar{\nabla}_{(\a}\overline{L})\d_{\b)}{}^{\g}  +\fracm{1}{20}(\s^{[2]})_{(\a}{}^{\d}(\s_{[2]})_{\b)}{}^{\g} (\Bar{\nabla}_{\d}\overline{L}) ~~~, \\
\d_S T_{\overline{\a} \overline{\b}}{}^{\g} \, =&~ (\overline{L}- \fracm12 L)T_{\overline{\a} \overline{\b}}{}^{\g} + i T_{\overline{\a} \overline{\b}}{}^{\un c} (\s_{\un c})^{\g\d} (\Bar{\nabla}_{\d}(\fracm{1}{32}L + \fracm{27}{160}\overline{L}))\label{equ:Talphabarbetabargamma-IIB}  ~~~, 
\end{align}

\begin{align}
\d_S T_{\a \un{b}}{}^{\overline{\g}} \, =&~ LT_{\a \un b}{}^{\overline{\g}}  -i (\s_{\un b})^{\g\b}(\nabla_{\a}\nabla_{\b}(\fracm{1}{32}\overline{L} + \fracm{27}{160}L)) + i(\s_{\un b})^{\d\b}(\Bar{\nabla}_{\d}(\fracm{1}{32}L + \fracm{27}{160}\overline{L}))T_{\a \b}{}^{\overline{\g}} \nonumber\\
&~+ i(\s_{\un b})^{\d\b}(\nabla_{\d}(\fracm{1}{32}\overline{L} + \fracm{27}{160}L))T_{\a \overline{\b}}{}^{\overline{\g}} - i (\s_{\un c})^{\d\g}(\nabla_{\d}(\fracm{1}{32}\overline{L} + \fracm{27}{160}L))T_{\a \un{b}}{}^{\un{c}} ~~~, \label{equ:Talphabgammabar-IIB}\\
\d_S T_{\a \un b}{}^{\g} \, =&~ \fracm12(L+\overline{L})T_{\a \un b}{}^{\g} + iT_{\a \un b}{}^{\un c}(\s_{\un c})^{\g\d}(\Bar{\nabla}_{\d}(\fracm{1}{32}L + \fracm{27}{160}\overline{L})) - \fracm12 (\nabla_{\un b}L)\d_{\a}{}^{\g} \nonumber\\
&~ -i(\s_{\un b})^{\g\b}(\nabla_{\a}\Bar{\nabla}_{\b}(\fracm{1}{32}L + \fracm{27}{160}\overline{L})) + i(\s_{\un 
b})^{\d\b}(\nabla_{\d}(\fracm{1}{32}\overline{L} + \fracm{27}{160}L))T_{\a \overline{\b}}{}^{\g}\nonumber\\
&~ + i(\s_{\un b})^{\d\b}(\Bar{\nabla}_{\d}(\fracm{1}{32}L + \fracm{27}{160}\overline{L}))T_{\a \b}{}^{\g} + \fracm14 (\s_{\un b}{}^{\un c})_{\a}{}^{\g}(\nabla_{\un c}(L+\overline{L})) ~~~, \\
\d_S T_{\un a \un b}{}^{\un c} \, =&~ \fracm12(L+\overline{L}) T_{\un a \un b}{}^{\un c} - i (\s_{[\un a})^{\a\b}
\left[ \, (\Bar{\nabla}_{\a}(\fracm{1}{32}L + \fracm{27}{160}\overline{L}))  T_{\b| \un b]}{}^{\un c}
(\nabla_{\a}(\fracm{1}{32}\overline{L} + \fracm{27}{160}L))  T_{\overline{\b}| \un b]}{}^{\un c} \, \right] ~~~, \\
\d_S T_{\overline{\a} \un{b}}{}^{\overline{\g}} \, =&~ \fracm12(L+\overline{L})T_{\overline{\a} \un b}{}^{\overline{\g}} + iT_{\overline{\a} \un b}{}^{\un c}(\s_{\un c})^{\g\d}(\nabla_{\d}(\fracm{1}{32}\overline{L} + \fracm{27}{160}L)) - \fracm12 (\nabla_{\un b}\overline{L})\d_{\a}{}^{\g} \nonumber\\
&~ -i(\s_{\un b})^{\g\b}(\Bar{\nabla}_{\a}\nabla_{\b}(\fracm{1}{32}\overline{L} + \fracm{27}{160}L)) + i(\s_{\un b})^{\d\b}(\Bar{\nabla}_{\d}(\fracm{1}{32}L + \fracm{27}{160}\overline{L}))T_{\overline{\a}\b }{}^{\overline{\g}} \nonumber\\
&~+ i(\s_{\un b})^{\d\b}(\nabla_{\d}(\fracm{1}{32}\overline{L} + \fracm{27}{160}L))T_{\overline{\a} \overline{\b}}{}^{\overline{\g}} + \fracm14 (\s_{\un b}{}^{\un c})_{\a}{}^{\g}(\nabla_{\un c}(L+\overline{L})) ~~~, \\
\d_S T_{\overline{\a} \un{b}}{}^{\g} \, =&~  \overline{L}T_{\overline{\a} \un b}{}^{\g}  -i(\s_{\un b})^{\g\b}(\Bar{\nabla}_{\a}\Bar{\nabla}_{\b}(\fracm{1}{32}L + \fracm{27}{160}\overline{L})) + i(\s_{\un b})^{\d\b}(\Bar{\nabla}_{\d}(\fracm{1}{32}L + \fracm{27}{160}\overline{L}))T_{\overline{\a}\b}{}^{\g} \nonumber\\
&~ + i(\s_{\un b})^{\d\b}(\nabla_{\d}(\fracm{1}{32}\overline{L} + \fracm{27}{160}L))T_{\overline{\a} \overline{\b}}{}^{\g} - i (\s_{\un c})^{\d\g}(\Bar{\nabla}_{\d}(\fracm{1}{32}L + \fracm{27}{160}\overline{L}))T_{\overline{\a} \un{b}}{}^{\un{c}} ~~~,\label{equ:Talphabarbgamma-IIB}
\end{align}

\begin{align}
\d_S R_{\a \b}{}^{\un{d}\un{e}} \, =&~ L R_{\a \b}{}^{\un{d}\un{e}} - \fracm{1}{2}T_{\a \b}{}^{[\un d}(\nabla^{\un e]}(L+\overline{L})) + \fracm15  T_{\a \b}{}^{\g}(\s^{\un{d}\un{e}})_{\g}{}^{\d}(\nabla_{\d}L) \nonumber\\
&~+ \fracm15  T_{\a \b}{}^{\overline{\g}}(\s^{\un{d}\un{e}})_{\g}{}^{\d}(\Bar{\nabla}_{\d}\overline{L})  -\fracm15 (\s^{\un{d}\un{e}})_{(\a}{}^{\d}(\nabla_{\b)}\nabla_{\d}L)  ~~~, \label{equ:Ralphabeta-IIB}\\
\d_S R_{\a \overline{\b}}{}^{\un{d}\un{e}} \, =&~ \fracm12(L+\overline{L}) R_{\a \overline{\b}}{}^{\un{d}\un{e}} - \fracm12T_{\a \overline{\b}}{}^{[\un d}(\nabla^{\un e]}(L+\overline{L})) + \fracm15  T_{\a \overline{\b}}{}^{\g}(\s^{\un{d}\un{e}})_{\g}{}^{\d}(\nabla_{\d}L) \nonumber\\
&~+ \fracm15  T_{\a \overline{\b}}{}^{\overline{\g}}(\s^{\un{d}\un{e}})_{\g}{}^{\d}(\Bar{\nabla}_{\d}\overline{L})  - \fracm15 (\s^{\un{d}\un{e}})_{\a}{}^{\d}(\Bar{\nabla}_{\b}\nabla_{\d}L) - \fracm15 (\s^{\un{d}\un{e}})_{\b}{}^{\g}(\nabla_{\a}\Bar{\nabla}_{\g}\overline{L}) ~~~, \\
\d_S R_{\overline{\a} \overline{\b}}{}^{\un{d}\un{e}} \, =&~  \overline{L} R_{\overline{\a} \overline{\b}}{}^{\un{d}\un{e}} -\fracm12 T_{\overline{\a} \overline{\b}}{}^{[\un d}(\nabla^{\un e]}(L+\overline{L})) + \fracm15  T_{\overline{\a} \overline{\b}}{}^{\g}(\s^{\un{d}\un{e}})_{\g}{}^{\d}(\nabla_{\d}L) \nonumber\\
&~+ \fracm15  T_{\overline{\a} \overline{\b}}{}^{\overline{\g}}(\s^{\un{d}\un{e}})_{\g}{}^{\d}(\Bar{\nabla}_{\d}\overline{L})  -\fracm15 (\s^{\un{d}\un{e}})_{(\a}{}^{\d}(\Bar{\nabla}_{\b)}\Bar{\nabla}_{\d}\overline{L})\label{equ:Ralphabarbetabar-IIB}  ~~~.
\end{align}

The grouping of terms in (\ref{equ:Talphabetac-IIB}) - (\ref{equ:Ralphabarbetabar-IIB}) is according to the real scale weight and U(1) charge.  Their real scale weights
and U(1) charges can be read off by expressing all the $L$ and $\Bar L$ super parameters in terms of u and
v.  All this information is summarized in Table \ref{tab:IIBweights}.
\begin{table}[h!]
\setlength{\tabcolsep}{10pt}
\renewcommand{\arraystretch}{1.5}
\centering
\begin{tabular}{|c|c|c|c|} \hline
    Torsion / Curvature & Scaling Coefficient & Real Scale Weight (u) & U(1) Charge (v) \\ \hline\hline
    $T_{\a\b}{}^{\un{c}}$ & $\fracm{1}{2} (L - \Bar{L})$ & 0 & $+1$ \\ 
    $T_{\a\overline{\b}}{}^{\un{c}}$ & 0 & 0 & 0 \\ 
    $T_{\overline{\a}\overline{\b}}{}^{\un{c}}$ & $- \fracm{1}{2} (L - \Bar{L})$ & 0 & $-1$ \\ \hline
    $T_{\a\b}{}^{\overline{\g}}$ & $L - \fracm{1}{2} \Bar{L}$ & $\fracm{1}{2}$ & $+\fracm{3}{2}$ \\
    $T_{\a\b}{}^{\g}$ & $\fracm{1}{2} L$ & $\fracm{1}{2}$ & $+\fracm{1}{2}$ \\
    $T_{\a\overline{\b}}{}^{\overline{\g}}$ & $\fracm{1}{2} L$ & $\fracm{1}{2}$ & $+\fracm{1}{2}$ \\
    $T_{\a\un{b}}{}^{\un{c}}$ & $\fracm{1}{2} L$ & $\fracm{1}{2}$ & $+\fracm{1}{2}$ \\
    $T_{\overline{\a}\un{b}}{}^{\un{c}}$ & $\fracm{1}{2} \Bar{L}$ & $\fracm{1}{2}$ & $-\fracm{1}{2}$ \\
    $T_{\a\overline{\b}}{}^{\g}$ & $\fracm{1}{2} \Bar{L}$ & $\fracm{1}{2}$ & $-\fracm{1}{2}$ \\
    $T_{\overline{\a}\overline{\b}}{}^{\overline{\g}}$ & $\fracm{1}{2} \Bar{L}$ & $\fracm{1}{2}$ & $-\fracm{1}{2}$ \\
    $T_{\overline{\a}\overline{\b}}{}^{\g}$ & $\Bar{L} - \fracm{1}{2} L$ & $\fracm{1}{2}$ & $-\fracm{3}{2}$ \\ \hline
    $T_{\a\un{b}}{}^{\overline{\g}}$ & $L$ & 1 & $+1$ \\
    $T_{\a\un{b}}{}^{\g}$ & $\fracm{1}{2} ( L + \Bar{L} )$ & 1 & 0 \\
    $T_{\un{a}\un{b}}{}^{\un{c}}$ & $\fracm{1}{2} ( L + \Bar{L} )$ & 1 & 0 \\
    $T_{\overline{\a}\un{b}}{}^{\overline{\g}}$ & $\fracm{1}{2} ( L + \Bar{L} )$ & 1 & 0 \\
    $T_{\overline{\a}\un{b}}{}^{\g}$ & $\Bar{L}$ & 1 & $-1$ \\ \hline
    $R_{\a\b}{}^{\un{d}\un{e}}$ & $L$ & 1 & $+1$ \\
    $R_{\a\overline{\b}}{}^{\un{d}\un{e}}$ & $\fracm{1}{2} ( L + \Bar{L} )$ & 1 & 0 \\
    $R_{\overline{a}\overline{\b}}{}^{\un{d}\un{e}}$ & $\Bar{L}$ & 1 & $-1$ \\ \hline
\end{tabular}
\caption{Summary of Weyl complex scaling properties of 10D, ${\cal N}$ = 2B torsions and curvatures}
\label{tab:IIBweights}
\end{table}

Again, we can construct ${\cal J}$-tensors in Type IIB theory. There are two independent ones (and their complex conjugates). This is because our scaling parameter superfield is complex and thus we have two degrees of freedom. The ${\cal J}$-tensors are defined by
\begin{equation}
\begin{aligned}
    \mathcal{J}^{(1)}_{\a} ~=&~ \fracm{1}{4} T_{\a \b}{}^{\b} ~~~,~ &
    \overline{\mathcal{J}}^{(1)}_{\a} ~=&~ \fracm{1}{4} T_{\overline{\a} \overline{\b}}{}^{\overline{\b}} ~~~, \\
    \mathcal{J}^{(2)}_{\a} ~=&~ \fracm{5}{12}T_{\a \b}{}^{\b} + \fracm{1}{3} T_{\a \overline{\b}}{}^{\overline{\b}}
    -\fracm{1}{3}T_{\a \un{b}}{}^{\un{b}}   ~~~,~ & 
    \overline{\mathcal{J}}^{(2)}_{\a} ~=&~  \fracm{5}{12} T_{\overline{\a} \overline{\b}}{}^{\overline{\b}} + 
    \fracm{1}{3} T_{\overline{\a} \b}{}^{\b} -\fracm{1}{3} T_{\overline{\a} \un{b}}{}^{\un{b}} ~~~.
\end{aligned}
\end{equation}
Their scale transformations are
\begin{equation}
\begin{split}
    \d_S \mathcal{J}^{(1)}_{\a} ~=&~ \fracm{1}{2} L \, \mathcal{J}^{(1)}_{\a} + (\nabla_{\a}L) ~~~,~~~
    \d_S \overline{\mathcal{J}}^{(1)}_{\a} ~=~ \fracm{1}{2} \overline{L} \, \overline{\mathcal{J}}^{(1)}_{\a} +  (\overline{\nabla}_{\a}\overline{L}) ~~~, \\
    \d_S \mathcal{J}^{(2)}_{\a} ~=&~ \fracm{1}{2} L \, \mathcal{J}^{(2)}_{\a} + (\nabla_{\a}\overline{L}) ~~~,~~~
    \d_S \overline{\mathcal{J}}^{(2)}_{\a} ~=~ \fracm{1}{2} \overline{L} \, \overline{\mathcal{J}}^{(2)}_{\a} +  (\overline{\nabla}_{\a}L) ~~~.
\end{split}
\end{equation}
The Weyl tensor can now be constructed from various (u,v) = (1,0) torsions, and combinations of (u,v) = $(\fracm12,+\fracm12)$ and (u,v) = $(\fracm12,-\fracm12)$ spinorial covariant derivatives and spinorial gauge connections such that each term has scaling properties 
(u,v) = $(1,0)$. 
The final result is
\begin{equation}
\begin{split}
    {\cal W}_{\un{a}\un{b}\un{c}} ~=&~ \fracm{1}{32} \Big\{\, (\s^{\un d})_{\g\b}(\s_{\un{a}\un{b}\un{c}})^{\b\a} \, T_{\a\un d}{}^{\g} 
    + (\s^{\un d})_{\g\b}(\s_{\un{a}\un{b}\un{c}})^{\b\a}\, T_{\overline{\a}\un d}{}^{\overline{\g}} 
    - 4 T_{[\un{a}\un{b}\un{c}]}  \\
    &~~~~~~ -i (\s_{\un{a}\un{b}\un{c}})^{\a\b}\, \Big[\, 
    \fracm{27}{16} \big( \nabla_{\a}\overline{\mathcal{J}}^{(1)}_{\b} + \overline{\nabla}_{\a}\mathcal{J}^{(1)}_{\b} \big)
    + \fracm{5}{16} \big( \nabla_{\a}\overline{\mathcal{J}}^{(2)}_{\b} + \overline{\nabla}_{\a}\mathcal{J}^{(2)}_{\b} \big)  \\
    &~~~~~~~~~~~~~~~~~~~~~~~ + \fracm{1,863}{3,200} \mathcal{J}^{(1)}_{\a}\overline{\mathcal{J}}^{(1)}_{\b} 
    - \fracm{33}{128} \mathcal{J}^{(2)}_{\a}\overline{\mathcal{J}}^{(2)}_{\b} 
    - \fracm{411}{640} \big( \mathcal{J}^{(1)}_{\a}\overline{\mathcal{J}}^{(2)}_{\b} + \mathcal{J}^{(2)}_{\a}\overline{\mathcal{J}}^{(1)}_{\b} \big) \,\Big\} ~~~.
\end{split}
\label{eq:WFS2}
\end{equation}

\subsection{10D, ${\cal N}~=~1$ Supergravity Derivatives \& Scale Transformations}
\label{sec:analyticalC}

The covariant derivatives linear in the conformal compensator $\Psi$ are given by (Equations 
(8.27) - (8.28) in \cite{GHM1})
\begin{align}
    \nabla_{\alpha} ~=&~ {\rm D}_{\alpha}~+~ \fracm 12\Psi {\rm D}_{\alpha}
    ~+~  \fracm 1{10}(\sigma^{\un{a}\un{b}})_{\alpha}^{\ \beta}({\rm D}_{\beta}\Psi){\cal M}_{\un{a}\un{b}}  ~~~,
    \label{eq:N1a} \\
    \nabla_{\un{a}} ~=&~ \pa_{\un{a}} ~+~ \Psi\pa_{\un{a}} ~-~ i \, \fracm 25 \, (\sigma_{\un{a}})^{\alpha\beta} 
    ({\rm D}_{\alpha}\Psi)\, { {\rm D}}{}_{\beta} ~-~  (\partial_{\un{c}}\Psi){\cal M}_{\un{a}}^{\ \un{c}} ~~~,
    \label{eq:N1b}
\end{align}
where the $\cal N$ = 1 ``D-operators'' satisfy
\begin{equation}
\{ \, {\rm D}_{\alpha}~,~ {\rm D}_{\beta} \, \} ~=~ i \, (\sigma^{\un{a}})_{\alpha\beta} \, \pa_{\un{a}}
~~~.
\label{xx}
\end{equation}
As the reader has seen the argument that follows from (\ref{eq:N1a}) and (\ref{eq:N1b}), thus 
we find
\begin{align}
    \d{}_{S}\nabla_{\alpha} ~=&~ \fracm 12L {\nabla}_{\alpha}
    ~+~  \fracm 1{10}(\sigma^{\un{a}\un{b}})_{\alpha}^{\ \beta}({\nabla}_{\beta}L)\, {\cal M}_{\un{a}\un{b}} ~~~,
    \label{eq:N1c} \\
    \d{}_{S}\nabla_{\un{a}} ~=&~  L\nabla_{\un{a}} ~-~ i\, \fracm 25 \, (\sigma_{\un{a}})^{\alpha\beta} ({\nabla}_{\alpha}L){\nabla}_{\beta} ~-~  (\nabla_{\un{c}}L)\, {\cal M}_{\un{a}}^{\ \un{c}} ~~~.
    \label{eq:N1d}
\end{align}

These results in (\ref{eq:N1c}) and (\ref{eq:N1d}) may be compared with the results 
derived in \cite{SC5}.  The two sets of Weyl scaling laws are different.  However,
this is due to the different choices of constraints that were chosen.  Once this is
taken into account, the scaling laws agree.  The scaling laws used in \cite{SC5}
were presented as ansatz\" e, whereas the ones in (\ref{eq:N1c}) and (\ref{eq:N1d})
were derived from the Nordstr\" om supergravity results \cite{GHM1}.  

The definition of torsion and curvature tensors for the 10D, $\cal N$ = 1 theory
are identical in form to the equations given in (\ref{eq:TR}).  Thus, here we find
repeating the series of calculations for 10D, $\cal N$ = 1 superspace as was done for 
11D, $\cal N$ = 1 superspace yields the results that follow. Weyl scaling properties of all other superspace torsion and curvature supertensors with weights of equal and larger than three-halves will be presented in Appendix \ref{appen:Weyl-I}.
\begin{align}
\d_S T_{\a \b}{}^{\un c} \, =&~ 0 ~~~, 
{~~~~~~~~~~~~~~~~~~~~~~~~~~~~~~~~~~~~~~~~}
{~~~~~~~~~~~~~~~~~~~~~~~~~~~~~~~~~~~~~~~~}
\end{align}
\begin{align}
\d_S T_{\a \b}{}^{\g} \, =&~  \fracm12 L\, T_{\a \b}{}^{\g} ~+~ i\, \fracm25 \, T_{\a \b}{}^{\un c} (\s_{\un c})^{\g\d}(\nabla_{\d}L) ~+~ \fracm12\, (\nabla_{(\a}L)\d_{\b)}{}^{\g} 
{~~~~~~~~~~~~~~~~~~~~~~~~~~~~~~}  \nonumber\\
&~+~\fracm{1}{20}\, (\s^{[2]})_{(\a}{}^{\d}(\s_{[2]})_{\b)}{}^{\g}(\nabla_{\d}L) ~~~, \\
\d_S T_{\a \un b}{}^{\un c} \, =&~ \fracm12 L T_{\a \un b}{}^{\un c} ~+~ (\nabla_{\a}L)\d_{\un b}{}^{\un c} ~+~ \fracm15 \, (\s_{\un b}{}^{\un c})_{\a}{}^{\b}(\nabla_{\b}L) 
~+~ i\, \fracm25 \, (\s_{\un b})^{\b\d}(\nabla_{\d}L)T_{\a \b}{}^{\un c} ~~~, 
\end{align}
\begin{align}
\d_S T_{\a \un b}{}^{\g} \, =&~ L\, T_{\a \un b}{}^{\g} ~+~ i\, \fracm25 \, T_{\a \un b}{}^{\un c}(\s_{\un c})^{\g\d}(\nabla_{\d}L) ~-~ \fracm12 \, (\nabla_{\un b}L)\d_{\a}{}^{\g} \nonumber\\
&~ -i\, \fracm25\, (\s_{\un b})^{\g\b}(\nabla_{\a}\nabla_{\b}L) ~+~ i\, \fracm25 \, (\s_{\un b})^{\d\b}(\nabla_{\d}L)T_{\a \b}{}^{\g} ~+~ \fracm12 \, (\s_{\un b}{}^{\un c})_{\a}{}^{\g}(\nabla_{\un c}L) ~~~, \\
\d_S T_{\un a \un b}{}^{\un c} \, =&~ L \, T_{\un a \un b}{}^{\un c} ~-~ i\, \fracm25 \, (\s_{[\un a})^{\a\b}(\nabla_{\a}L) \, T_{\b| \un b]}{}^{\un c}~~~, \\
\d_S R_{\a \b}{}^{\un{d}\un{e}} \, =&~ L\,  R_{\a \b}{}^{\un{d}\un{e}} ~-~ T_{\a \b}{}^{[\un d}(\nabla^{\un e]}L) ~+~ \fracm15  \, T_{\a \b}{}^{\g}(\s^{\un{d}\un{e}})_{\g}{}^{\d}(\nabla_{\d}L) \nonumber\\
&~ -\fracm15 \, (\s^{\un{d}\un{e}})_{(\a}{}^{\d}(\nabla_{\b)}\nabla_{\d}L)  ~~~.
\end{align}

Here we define the ${\cal J}$-tensors
\begin{equation}
\begin{split}
{\cal J}_{\a}^{(+)} ~=&~ \fracm{1}{6} T_{\a\un{b}}{}^{\un{b}} ~~~,~~~
    {\cal J}_{\a}^{(-)} ~=~ T_{\a\b}{}^{\b} ~~~, 
\end{split}
\end{equation}
and they transform as
\begin{equation}
\begin{split}
    \d_{S} {\cal J}_{\a}^{(+)} ~=&~ \fracm{1}{2} L {\cal J}_{\a}^{(+)} ~+~ ( \nabla_{\a} L )  ~~~,~~~
    \d_{S} {\cal J}_{\a}^{(-)} ~=~ \fracm{1}{2} L {\cal J}_{\a}^{(-)}  ~~~. 
\end{split}
\end{equation}
Again, similar as in 11D, $\cal N$=1 case, ${\cal J}{}_{\a}^{(-)}$ transforms like a scale covariant tensor of weight $\fracm 12$,
the quantity ${\cal J}{}_{\a}^{(+)}$ transforms with a scale weight of $\fracm 12$ while
being a spinorial gauge connection under a scaling transformation!

The necessary engineering dimension
one superfield ${\cal W}_{{\un a} {\un b} {\un c} }$ with Weyl weight one can be defined by
\begin{equation}
    {\cal W}_{\un{a}\un{b}\un{c}} ~=~ \fracm{1}{16} \Big\{\, (\s^{\un d})_{\g\b}(\s_{\un{a}\un{b}\un{c}})^{\b\a}\, T_{\a\un d}{}^{\g} - i 4(\s_{\un{a}\un{b}\un{c}})^{\a\b} \,\big[\, \nabla_{\a}\mathcal{J}_{\b}^{(+)} - \fracm{22}{25}\mathcal{J}_{\a}^{(+)} \mathcal{J}_{\b}^{(+)} \,\big] \,\Big\} ~~~.
    \label{eq:WFS3}
\end{equation}

To our knowledge, the results presented in equations (\ref{eq:WFS1}), (\ref{eq:WFS2}), and (\ref{eq:WFS3})
mark the first time that off-shell definitions of 10D Weyl superfield supergravity field strength tensors
have been explicitly identified in the physics literature.

\newpage
\section{A New Methodology: Adynkras \& ADA Scans}
\label{sec:ADA}

Recently \cite{GHM2,GHM3}, we have established a breakthrough approach which substantially 
lowered computational costs of determining how to embed a set of component fields within a 
superfield.  Just as MRI (magnetic resonance imaging - based on the phenomenon of nuclear 
magnetic resonance) has brought amazing progress in creating high definition images of 
internal structures of bodies in biological domains, the 
approach in \cite{GHM2,GHM3} (which can be called ``Adyndra Digital Analysis'' or ``ADA'')
permits the rapid assay of the Lorentz spectrum of component fields within superfields.

There are a number of ways to define an ``adynkra\footnote{This is a word created 
from the concatenation of the works ``adinkra'' and ``Dynkin'' in an effort to denote the fusion 
of these two concepts.}.''  One such meaning is as the name for a certain class of graphs.  
However, for our purposes here, we will use a definition tailored to 10D, ${\cal N} = 1$ SUSY 
where an adynkra is a collection of sets of Dynkin Labels that can be broken into subsets 
associated with a ``Level number.'' The latter of these is an integer that takes on values from 
zero to sixteen.  There exists a ``generator'' for such lists that takes the form
\be{ \label{equ:calG} \eqalign{
\ytableausetup{boxsize=0.8em}
{\cal G} ~=~
&1 ~\oplus~ \ell \, \left\{  \left(\, \CMTred{\ydiagram{1}} \,\right) \, 
\times [a_1,b_1,c_1,d_1,e_1]  \right\} ~\oplus  
~ \bigoplus_{p = 2}^{16}
 \, {\fracm 1{p!}} \, (\ell){}^p  \left\{  \left( \,
\CMTred{\ydiagram{1}} ~ ( \wedge \, \CMTred{\ydiagram{1}} \, )^{p - 2}  \wedge
 \CMTred{\ydiagram{1}} \, \right)  \, \times
[a_p,b_p,c_p,d_p,e_p]  \right\}    ~~,
} }\ee
which can be expanded to,
\be{ \label{equ:calG1} \eqalign{
\ytableausetup{boxsize=0.8em}
{\cal G} ~=~
& \ell \, \left\{  \left(\, \CMTred{\ydiagram{1}} \,\right) \, 
\times [a_1,b_1,c_1,d_1,e_1]  \right\}        \cr
&{~\, \,}\oplus  \,
\bigoplus_{p = 1}^{7} \, {\fracm 1{p!}} \, (\ell){}^{2p + 1}  \left\{  \left( \,
\CMTred{\ydiagram{1}} ~ ( \wedge \, \CMTred{\ydiagram{1}} \, )^{2 p - 1} 
\wedge \CMTred{\ydiagram{1}} \, \right)  \, \times
[a_{2p + 1},b_{2p + 1},c_{2p + 1},d_{2p + 1},e_{2p + 1}]  \right\}      \cr
&{~\, \,}\oplus  \, 1 \,\oplus  \,
\bigoplus_{p = 1}^{8} \, {\fracm 1{(2p)!}} \, (\ell){}^{2p}  \left\{  \left( \,
\CMTred{\ydiagram{1}} ~ ( \wedge \, \CMTred{\ydiagram{1}} \, )^{2 (p - 1) }  
\wedge \CMTred{\ydiagram{1}} \, \right)  \, \times
[a_{2p},b_{2p},c_{2p},d_{2p},e_{2p}]  \right\}   ~~~,
} }\ee
where $ [a_p,b_p,c_p,d_p,e_p]$ (with $p$ = 1, $\dots$, 16) denote Dynkin Labels appropriate 
for the fields over the 10-dimensional manifold.  The ``Level number'' is the exponent seen 
associated with various powers of the level parameter $\ell$ in this expression.  Finally 
$\ytableausetup{boxsize=0.8em} \CMTred{\ydiagram{1}}$ is a spinorial Young Tableau and $\wedge$ 
denotes the ``wedge'' product of the tableau.  A more complete discussion of conventions 
and notation used in writing (\ref{equ:calG}) can be found in \cite{GHM3}.  All the terms on the 
first two lines of (\ref{equ:calG1}) are fermionic representations as the Young Tableau associated 
with $\ytableausetup{boxsize=0.8em} \CMTred{\ydiagram{1}}$ corresponds to the fundamental spinor representation of $\mathfrak{so}(1,9)$.   
As the final terms have only even powers of $\CMTred{\ydiagram{1}}$ and due to the identity
\be
{\CMTR{\ydiagram{1}}} \, \wedge \, {\CMTR{\ydiagram{1}}} ~=~ {\CMTR{\ydiagram{1,1}}} 
~=~  {\CMTB{\ydiagram{1,1,1}}} ~~~~~,
\label{F2B}
\ee
these can be expressed  {\it {solely}} in terms of YT's involving $\ytableausetup{boxsize=0.8em} \CMTB{\ydiagram{1}}$ (bosonic Young Tableaux), i.\ e.\ bosonic representations of $\mathfrak{so}(1,9)$.

The generator (\ref{equ:calG}) provides a basis for the creation of
algorithms that are extraordinarily efficient at encoding representations of component 
fields contained within superfields.  It is this efficiency that enabled unprecedented
clarity about the component field contents of superfields in ten and eleven dimensions
\cite{GHM2,GHM3,GHM4} and the computational tools we use are noted in \cite{Susyno,LiE,LieART}.

The discussion in \cite{GHM3} is devoted to using the concepts of Branching Rules, Dynkin 
Labels, Plethysm, and Young Tableaux to  calculate ${\cal G}$.  In \cite{GHM2,GHM3} the basic 
``scalar'' adynkra, denoted by $\cal V$, was calculated and described in detail.   It is the foundational 
quantity expressed in terms of YT's.  The computation expense to find all related quantities 
is substantially less than that needed to explicitly determine $\cal V$.  This follows as higher 
representations are simply found from multiplication of $\cal V$ by Dynkin Labels when the 
latter are represented by YT's. Thus, we have
\be \eqalign{
{\cal V}_{\CMTB {[0,0,1,0,0]}} ~&=~  {\CMTB {[0,0,1,0,0]}} \, \otimes \, {\cal V} ~~~, ~~~  
{\cal V}_{\CMTred {[1,0,1,0,1]}} ~=~ {\CMTred {[1,0,1,0,1]}}  \, \otimes \, {\cal V}~~~, \cr 
{\cal V}_{\CMTred {[3,0,0,0,1]}} ~&=~  {\CMTred {[3,0,0,0,1]}}  \, \otimes \, {\cal V}~~~, ~~~
{\cal V}_{\CMTB {[4,0,0,0,0]}} ~=~ {\CMTB {[4,0,0,0,0]}} \, \otimes \, {\cal V}~~~, 
} \label{equ:REPs}
\ee
and these multiplications are straightforward by the use of the tools indicated in the
references \cite{Susyno,LiE,LieART}.  For any bosonic irrep ${\CMTB {{\cal R}}}$ 
or fermionic irrep ${\CMTR {{\cal R}}}$, (\ref{equ:REPs}) imples
$
{ \rm {dim}}({\cal V}_{\CMTB {\cal R}}) =  { \rm {dim}}({\CMTB {\cal R}}) \, { \rm {dim}}({\cal V})$ or  
${ \rm {dim}}({\cal V}_{\CMTred {\cal R}}) = {\rm {dim}}({\CMTred {\cal R}}) \, {\rm {dim}}({\cal V}), 
$ where ${ \rm {dim}}({\cal V})$ = 65,536.
The Dynkin Label Library of the ${\cal V}$-superfield is explicitly shown below and similar results
are shown in the Appendix \ref{appen:Lib} for all superfields in (\ref{equ:REPs}).

\begin{itemize} \sloppy
\item Level-0: $\CMTB {[0,0,0,0,0]}$
\item Level-1: $\CMTred {[0,0,0,1,0]}$
\item Level-2: $\CMTB {[0,0,1,0,0]}$
\item Level-3: $\CMTred {[0,1,0,0,1]}$
\item Level-4: $\CMTB {[0,2,0,0,0]} \oplus \CMTB {[1,0,0,0,2]}$
\item Level-5: $\CMTred {[0,0,0,0,3]} \oplus \CMTred {[1,1,0,0,1]}$
\item Level-6: $\CMTB {[0,1,0,0,2]} \oplus \CMTB {[2,0,1,0,0]}$
\item Level-7: $\CMTred {[3,0,0,1,0]} \oplus \CMTred {[1,0,1,0,1]}$
\item Level-8: $\CMTB {[4,0,0,0,0]} \oplus \CMTB {[0,0,2,0,0]} \oplus \CMTB {[2,0,0,1,1]}$
\item Level-9: $\CMTred {[3,0,0,0,1]} \oplus \CMTred {[1,0,1,1,0]}$
\item Level-10: $\CMTB {[0,1,0,2,0]} \oplus \CMTB {[2,0,1,0,0]}$
\item Level-11: $\CMTred {[0,0,0,3,0]} \oplus \CMTred {[1,1,0,1,0]}$
\item Level-12: $\CMTB {[0,2,0,0,0]} \oplus \CMTB {[1,0,0,2,0]}$
\item Level-13: $\CMTred {[0,1,0,1,0]}$
\item Level-14: $\CMTB {[0,0,1,0,0]}$
\item Level-15: $\CMTred {[0,0,0,0,1]}$
\item Level-16: $\CMTB {[0,0,0,0,0]}$
\end{itemize}

The utility of such a library is straightforward.

For example, one could ask whether the component field corresponding to the bosonic irrep $\CMTB {[4,0,1,0,0]}$ occurs
within the $\theta$ expansion of $\cal V$?  Upon examination of the library above, it is seen
that the answer is, ``no.''  In a similar manner, one can ask if the pair of representations given
by $\CMTB {[0,2,0,0,0]}$ and $\CMTR {[3,0,0,0,1]}$ occur in the superfield $\cal V$ at 
adjacent levels with the fermionic one higher than the lower one?  Once more a quick consultation
of the library above returns the answer, ``no.''   Clearly, more complicated questions of this
nature can be pursued and we turn to this next.

\subsection{Component Dynkin Label Examples \& ADA Scans}
\label{subsec: CompX}

The component fermionic and bosonic representation content of the on-shell 10D, $\cal N$ = 1 
supergravity multiplet \cite{10Dsg1,10Dsg2} is given by
\be  \eqalign{
{\CMTR { \{ {\cal F} \} } } {}_{\rm {SGI}} ~&=~ (\, \chi{}_{ \a}, \, \psi{}_{\un a}{}^{ \a} \, ) ~\, 
~~~\, ~=~ ( \CMTred{[0,0,0,1,0]}, \, \CMTred{[1,0,0,0,1]} )    ~~~~~~~~~~~~~~~~~\,~~~~,   \cr
{\CMTB { \{ {\cal B} \} } } {}_{\rm {SGI}} ~&=~  
(\, \varphi ,\,  B{}_{{\un a} \, {\un b}}, \,{\rm e}{}_{{\un a}}{}^{{\un m}}  \, )
~=~ 
(\CMTB{[0,0,0,0,0]}, \,  {\CMTB{[0,1,0,0,0]}}, \,  {\CMTB{[2,0,0,0,0]}})  ~~~~~, 
}  \label{eq:I}
\ee
in its most common form.  However, as first emphasized in the work of \cite{d-Het1},
there is an alternate version where the replacement $B{}_{{\un a} \, {\un b}}$ $\to$
$M{}_{ {\un a}{}_1 \, \cdots \,  {\un a}{}_{6} }$ is compatible with the absence of anomalies
\cite{G-SM1,G-SM2} as a realization of the LEEA of the heterotic string \cite{Hstrng}.

To complete 10D, $\cal N$ = 1 supergravity into either the 10D, $\cal 
N$ = IIA or IIB supergravity multiplets, it is necessary to find two other 10D, $\cal 
N$ = 1 multiplets that contain the matter gravitino multiplets (MGM) discussed
in Appendix C of the work \cite{FX1} and that indicated the existence of two such 10D,
$\cal N$ = 1 MGM systems. We may call one of them the ``IIAMGM'' system and the
``IIBMGM'' system.  In terms of their component fermionic and bosonic representation
contents, these look as
\be  \eqalign{
{\CMTR { \{ {\cal F} \} } } {}_{\rm {IIAMGM}} ~&=~   
(\, \chi^{}_{\Dot \a}, \, \psi{}_{\un a}{}^{\Dot \a} \, ) ~\, 
~=~ 
( \CMTred{[0,0,0,0,1]}, \, \CMTred{[1,0,0,1,0]}  )
~~~~~,   \cr
{\CMTB { \{ {\cal B} \} } } {}_{\rm {IIAMGM}} ~&=~  (\, B{}_{\un a}, \,A{}_{{\un a} \, {\un b}\, {\un c}}  \, )  
~=~ (\CMTB{[1,0,0,0,0]}, \,  \CMTB{[0,0,1,0,0]})   ~~~~~, 
}  \label{eq:1IIA}
\ee
and
\be  \eqalign{
{~~~~~~~}
{\CMTR { \{ {\cal F} \} } } {}_{\rm {IIBMGM}} ~&=~  (\, \chi^{\prime}_{\a}, \, \psi^{\prime}_{\un a}{}^{\a} \, )  ~~~~~~~~~ ~=~  
(\CMTred{[0,0,0,1,0]}, \, \CMTred{[1,0,0,0,1]})   ~~~~~~~~~~~~~~~~~\,~~~~, \cr
{\CMTB { \{ {\cal B} \} } } {}_{\rm {IIBMGM}} ~&=~ (\, A,\,  B^{\prime}{}_{{\un a} \, {\un b}}, \,A{}_{{\un a} \, {\un b}\, {\un c}   \, {\un d} }  \, ) 
 ~=~ (\CMTB{[0,0,0,0,0]}, \,  \CMTB{[0,1,0,0,0]}
, \,  \CMTB{[0,0,0,1,1]})   ~~~~~,
}  \label{eq:2IIB}
\ee
respectively.  It can be seen that the fermionic representations are presented in a manner
where they are ``higher'' than the bosonic ones.  This is due to the fact that dynamical fermionic fields 
possess higher engineering dimensions than dynamical bosonic fields.  In terms of the ``Level''
numbers, the fermions are higher than the bosons.

Both the engineering dimensions and the Lorentz representations of all fields are key 
data inputs in the construction of Adynkra Digital Analysis (ADA) scans.

The adynkras for $ {\cal V}$,  ${\cal V}_{\CMTB {[0,0,1,0,0]}}$, ${\cal V}_{\CMTred {[1,0,1,0,1]}}$, 
${\cal V}_{\CMTred {[3,0,0,0,1]}}$, and ${\cal V}_{\CMTB {[4,0,0,0,0]}}$ can be regarded as 
a set of ``libraries.''  An ADA scan asks simple questions as any number of such queries can
be asked. One is, ``Given the data of Level difference and the Dynkin Labels demonstrated
in (\ref{eq:1IIA}) does such a data pattern occur in the $ {\cal V}$-library?''  Another might
be, ``How many times does such a pattern occur?''  

Going back to the work of \cite{G-1}, we have long asserted the interpretation of adinkras as 
being the analogs of genetic sequence content where superfields play the roles of biological bodies.  An alternative
would have been to analogize adinkras to quarks.  However, this analogy suffers when one
realizes the numbers of degrees of freedom of systems in ten and eleven dimensions (e.\ g.
11D SG possess 2,147,483,648 bosonic and the same number of fermionic degrees).  So the number
of possible targets in searchs are of more ``biological'' in order of magnitude than the numbers 
encountered in determing the quark content in hadronic spectroscopy problems.  

Thus, the process of querying adynkra libraries more closely favors the challenges encountered
 in DNA analysis than the analysis of the quark spectra of hadronic matter.  With modern 
 IT platforms, both hardware and software, it is a straightforward matter to meet 
the challenges of writing codes to query such libraries... after they have been constructed.
To focus this more accurately, the analogy is to regard the data (as given in (\ref{eq:I}), 
(\ref{eq:1IIA}), and (\ref{eq:2IIB})) as primary biological sequence content, (e.\ g.\ similar 
in spirit to nucleotides of DNA/RNA or protein amino-acids).  The adynkra ``libraries'' play 
the roles of genetic sequence data bases/libraries\footnote{One example of such a bioinfomatics 
IT tool is the ``basis local alignment search tool'' (BLAST) \cite{BLASTn} that serves this
function.}.

\newpage
\section{Toward the Rest of the Story}
\label{sec:story}

The efforts we have described in chapters  \ref{sec:primer} and \ref{sec:analytical0} follow the ``traditional'' 
routes to understand supergravity in superspace where attention is focused on the ``outside'' variables, 
i.\ e.\  $\nabla{}_{\un A}$, $T{}_{{\un A} \, {\un B}}{}^{{\un C}}$, and $R{}_{{\un A} \, {\un B}} 
{}_{\, \un c} {}^{\un d}$.  However, there is also a less trod pathway based on the study of the 
prepotentials, i.\ e.\ $ \Psi$, ${\cal A}^{{\un a} \, {\un b}} $, ${\cal N}{}_{\a}{}^{\b}$, and 
$ {{\rm  H}}{}_\b{}^{\un b}$.  Really, due to the presence of constraints and covariance, only 
$ \Psi$ and $ {{\rm  H}}{}_\b{}^{\un b}$ require deeper study.  Heretofore, mostly the literature 
\cite{HW1} -  \cite{M2}, has focused on the ``outside'' superfields rather than on the ``inside'' 
ones, $ \Psi$, and $ {{\rm  H}}{}_\b{}^{\un b}$.  The most obvious reason for this is the traditional 
approach requires a high computational price be paid to elucidate the $\theta$-expansion for component 
fields residing within the ``inside'' superfields.   

As we wish to treat the cases of the $\cal N$ = 1, $\cal N$ = 2A, and $\cal N$ = 2B uniformly,
there is a conceptual approach available as an effective enabling strategy.  We now turn to a 
discussion of this.

In the works of \cite{NF,Nuf0,Nuf1,Nuf2,Nuf3} there was initiated an approach where superfields
with a higher realization of supersymmetry were formulated in terms of superfields that provide
a lower realization of supersymmetry.  In these cases, mostly superfields with 4D, $\cal N$ = 2 SUSY
were expressed in terms of superfields with 4D, $\cal N$ = 1 SUSY.  More recently \cite{BB,TAMU1,TAMU2,TAMU3},
this approach has been implemented in the context of superfields with 11D, $\cal N$ = 1 SUSY
expressed in terms of superfields with 4D, $\cal N$ = 1 SUSY.  Since we are working in the arena
of 10D, $\cal N$ = 1, $\cal N$ = 2A, and $\cal N$ = 2B SUSY, the lower dimensional
superfields in which to conduct our investigations are ones that realize 10D, $\cal N$ = 1
SUSY.

To create a set of notational conventions that are graphically easy to follow in our subsequent
discussion, we will denote the coordinates of 10D, $\cal N$ = 1 superspace by $(x^{\un{a}}, \, 
\theta^{\a})$.  In order to describe the superfields with $\cal N$ = 2A, and $\cal N$ = 2B SUSY,
we introduce a second Grassmann coordinate of the form $\CMTG {\theta^{\Dot \a}}$ in the
case of the $\cal N$ = 2A superfields or of the form $\CMTG {\theta^{\a}}$ in the case of the $\cal 
N$ = 2B superfields.  Now we can continue to study the superfields ${\cal V} (x^{\un{a}}, \, 
\theta^{\a})$, ${\cal V}{}_{IIA}(x^{\un{a}}, \, \theta^{\a}, \, {\CMTG {\theta^{\dot{\a}}}})$, or ${\cal V} 
{}_{IIB}(x^{\un{a}}, \, \theta^{\a}, \, {\CMTG {\theta^{{\a}}}})$ in each respective case.  Next we treat 
these superfields, but in the latter two cases we use respective explicit expansions in 
terms of ``green $\theta$'s,''
\begin{equation}
{\cal V}{}_{IIA} (x^{\un{a}}, \, \theta^{\a}, \, {\CMTG {\theta^{\Dot \a}}}) ~=~ {\cal V}^{(0)} (x^{\un{a}}
, \, \theta^{\a}) ~+~ {\CMTG {\theta^{\Dot \a}}} \, {\cal V}^{(1)}_{\dot{\a}} (x^{\un{a}}, \, \theta^{\a}) 
~+~ {\CMTG {\theta^{\Dot \a}}} {\CMTG {\theta^{\Dot \b}}} \, {\cal V}^{(2)}_{\dot{\a}\dot{\b}}  (x^{\un{a}}, 
\, \theta^{\a}) ~+~ \dots   ~~~,
\label{eq:IIAx}
\end{equation}
\begin{equation}
{\cal V}{}_{IIB} (x^{\un{a}}, \, \theta^{\a}, \, {\CMTG {\theta^{\a}}}) ~=~ {\cal V}^{(0)} (x^{\un{a}}, \, 
\theta^{\a}) ~+~ {\CMTG {\theta^{ \a}}} \, {\cal V}^{(1)}_{{\a}} (x^{\un{a}}, \, \theta^{\a}) ~+~ {\CMTG 
{\theta^{ \a}}} {\CMTG {\theta^{ \b}}} \, {\cal V}^{(2)}_{{\a}{\b}}  (x^{\un{a}}, \, \theta^{\a}) ~+~  \dots  
~~~,
\label{eq:IIBx}
\end{equation}
and where these expansions terminate at the sixteenth order of the ``green $\theta$'s.''  Clearly,
these indicate respective sets of sixteen distinct 10D, $\cal N$ = 1 superfields within the 
two types of 10D, ${\cal N} = 2$ superfields.

Although the two expansions in (\ref{eq:IIAx}) and (\ref{eq:IIBx}) appear rather similar, an interesting dichotomy emerges when ADA algorithms are applied to them.  In the case of (\ref{eq:IIAx}), it is seen for all the fermionic superfields, {\it {none}} contributes the conformal graviton representation to level-16 of Type IIA and IIB scalar superfields.  In the case of (\ref{eq:IIBx}), it is seen for all of the fermionic superfields except for $\mathcal{V}_{\a}$, $\mathcal{V}^{\a}$, $\mathcal{V}_{\{\aone\bone\cone\done\eone\}{}^{+}}{}^{\a}$, and $\mathcal{V}_{\{\aone\bone\cone\done\eone\}{}^{-}}{}_{\a}$, {\it {all}} contain the conformal graviton representation.  This result suggests if we wish to find formulations of the Type-IIA and Type-IIB theories that possess a common truncation to the Type-I theory, we should eliminate the fermionic superfields in both expansions as subjects of additional study in this regard.

Now concentrating solely on the bosonic superfields in (\ref{eq:IIAx}) and (\ref{eq:IIBx}), we apply ADA 
algorithms to search within the bosonic 10D, $\cal N$ = 1 superfields for the number $b_{\CMTB{\{1\}}}$ 
of singlet irreps, the number $b_{\CMTB{\{45\}}}$ of $\CMTB {\{45\}}$ irreps, the number $b_{\CMTB{\{
54\}}}$ of $\CMTB {\{54\}}$ irreps, the number $b_{\CMTB{\{210\}}}$\footnote{The reason this irrep is 
considered in the search is due to the fact that it is known there is a duality ``flip'' \cite{CHM} that 
exchanges the $b_{\CMTB{\{54\}}}$ and $b_{\CMTB{\{210\}}}$ irreps in 10D, $\cal N$ = 1 supergravity 
and which is consistent with the anomaly-freedom conditions and Green-Schwarz $\s$-models associated 
with the low-energy heterotic string \cite{d-Het1,d-Het2} effective action.} of $\CMTB {\{210\}}$ irreps, the number $b_{\CMTred{\{\overline{16}\}}}$ of $\CMTR {\{\overline{16}\}}$ irreps, and 
the number $b_{\CMTred{\{\overline{144}\}}}$ of $\CMTR {\{\overline{144}\}}$ irreps.  Furthermore, the 
ADA scans for these irreps within the bosonic 10D, $\cal N$ = 1 superfields were designed additionally 
so that all bosonic irreps should occur at the same level in the superfield and also to fulfill the requirement 
that the ${\CMTred{\{\overline{16}\}}}$ irrep and the ${\CMTred{\{\overline{144}\}}}$ irrep should occur at one level higher than the bosonic irreps.  The 10D, ${\cal N}=2$A
results are shown in the Table \ref{tab:IIA_contribution_54_lv16} below.
\begin{table}[htp!]
\centering
\begin{tabular}{|c|c|c|c|c|c|c|c|c|}
\hline
Type-I Superfield & Dynkin Label & $b_{\CMTB{\{1\}}}$ & $b_{\CMTB{\{45\}}}$ & $b_{\CMTB{\{54\}}}$ &    
$b_{\CMTB{\{210\}}}$ &    $b_{\CMTred{\{\overline{16}\}}}$&    $b_{\CMTred{\{\overline{144}\}}}$ & \begin{tabular}[c]{@{}c@{}}Level of  
$\CMTB{\{54\}}$ in \\ Type-I Superfield\end{tabular}
\\ \hline \hline       
$\mathcal{V}_{\{ \aone \bone \cone \}}$   & $\CMTB{[0,0,1,0,0]}$ &  1 & 1 & 1 & 2  & 1 & 1 & 14  \\ \hline
$\mathcal{V}_{\{ \aone \bone, \atwo \btwo \}}$ & $\CMTB{[0,2,0,0,0]}$ &  1 & 1 & 1 & 2& 1  & 1 & 12  \\ \hline
$\mathcal{V}_{\{ \atwo | \aone \bone \cone \done \eone \}{}^{-}}$ & $\CMTB{[1,0,0,2,0]}$  &  0 & 0 & 1 & 1 & 0 & 1 & 12  \\ \hline
$\mathcal{V}_{\{ \atwo \btwo | \aone \bone \cone \done \eone \}{}^{-}}$& $\CMTB{[0,1,0,2,0]}$  &  0 & 0 & 1 & 1 & 0  & 1 & 10   \\ \hline
$\mathcal{V}_{\{ \atwo , \athree | \aone \bone \cone \}}$  & $\CMTB{[2,0,1,0,0]}$  &  1 & 2 & 2 & 4& 1 & 3 & 10  \\ \hline
$\mathcal{V}_{\{ \aone,  \atwo,  \athree, \afour \}}$  & $\CMTB{[4,0,0,0,0]}$  &  1 & 1 & 1 & 1& 1 & 1 & 8 \\ \hline
$\mathcal{V}_{\{ \aone \bone 
\cone , \atwo \btwo \ctwo \}}$ & $\CMTB{[0,0,2,0,0]}$  &  1 & 1 & 1 & 3& 1 & 2 & 8  \\ \hline
$\mathcal{V}_{\{ \atwo, \athree | \aone \bone \cone \done \}}$ & $\CMTB{[2,0,0,1,1]}$  &  1 & 3 & 2 & 6& 2 & 5 & 8  \\ \hline
$\mathcal{V}_{\{\atwo \btwo| \aone \bone \cone \done \eone \}{}^{+}}$ & $\CMTB{[0,1,0,0,2]}$  &  0 & 0 & 1 & 1& 0 & 1 & 6  \\ \hline
$\mathcal{V}_{\{ \atwo , \athree | \aone \bone \cone \}}$  & $\CMTB{[2,0,1,0,0]}$  &  1 & 2 & 2 & 4& 2 & 4 & 6  \\ \hline
$\mathcal{V}_{\{ \aone \bone, \atwo \btwo \}}$ & $\CMTB{[0,2,0,0,0]}$  &  1 & 1 & 1 & 2& 1 & 2 & 4  \\ \hline
$\mathcal{V}_{\{ \atwo | \aone \bone \cone \done \eone \}{}^{+}}$ & $\CMTB{[1,0,0,0,2]}$  &  0 & 0 & 1 & 1& 0 & 1 & 4 \\ \hline
$\mathcal{V}_{\{ \aone \bone \cone \}}$  & $\CMTB{[0,0,1,0,0]}$ &  1 & 1 & 1 & 2& 1 & 2 & 2  \\ \hline
\end{tabular}
\caption{Contributions of Each Type-I Superfield to the component bosonic and fermionic 
\newline $~~~~~~~~~~~~~~\,$representation
content of the 10D, $\cal N$ = 1 supergravity multiplet at Level-16/17 of 
\newline $~~~~~~~~~~~~~~\,$Type-IIA Scalar Superfield Decomposition}
\label{tab:IIA_contribution_54_lv16}
\end{table}

Looking at this table, the reader's attention is directed to the Dynkin Labels.  The distinct
ones are $\CMTB{[0,0,1,0,0]}$, $\CMTB{[0,2,0,0,0]}$, $\CMTB{[1,0,0,2,0]}$, $\CMTB{[0,1,0,2,0]}$, $\CMTB{[2,0,1,0,0]}$,
$\CMTB{[4,0,0,0,0]}$, $\CMTB{[0,0,2,0,0]}$, $\CMTB{[2,0,0,1,1]}$, $\CMTB{[0,1,0,0,2]}$, and $\CMTB{[1,0,0,0,2]}$.

In a similar manner ADA scans can be applied to the bosonic 10D, $\cal N$ = 1 superfields in 
the expansion shown in (\ref{eq:IIBx}).  In this case the sets of members of bosonic 10D, 
$\cal N$ = 1 superfields that are found is identical to the ones found by starting from the
expansion in (\ref{eq:IIAx}). The 10D, ${\cal N} = 2$B results are shown in Table \ref{tab:IIB_contribution_54_lv16}.
\begin{table}[htp!]
\centering
\begin{tabular}{|c|c|c|c|c|c|c|c|c|}
\hline
Type-I Superfield & Dynkin Label & $b_{\CMTB{\{1\}}}$ & $b_{\CMTB{\{45\}}}$ & $b_{\CMTB{\{54\}}}$ &    
$b_{\CMTB{\{210\}}}$ &    $b_{\CMTred{\{\overline{16}\}}}$&    $b_{\CMTred{\{\overline{144}\}}}$ & \begin{tabular}[c]{@{}c@{}}Level of  
$\CMTB{\{54\}}$ in \\  Type-I Superfield\end{tabular} \\ 
\hline \hline         
$\mathcal{V}_{\{ \aone \bone \cone \}}$   & $\CMTB{[0,0,1,0,0]}$  & 1   & 1 & 1 & 2 & 1 & 1 & 14\\ \hline
$\mathcal{V}_{\{ \aone \bone, \atwo \btwo \}}$  & $\CMTB{[0,2,0,0,0]}$  &  1   & 1 & 1 & 2& 1 & 1 & 12  \\ \hline
$\mathcal{V}_{\{ \atwo | \aone \bone \cone \done \eone \}{}^{+}}$ & $\CMTB{[1,0,0,0,2]}$  &  1  & 2 & 1 & 3 & 1& 2 & 12 \\ \hline
$\mathcal{V}_{\{\atwo \btwo| \aone \bone \cone \done \eone \}{}^{+}}$ & $\CMTB{[0,1,0,0,2]}$  &  1 & 2 & 1 & 4 & 2 & 3 & 10 \\ \hline
$\mathcal{V}_{\{ \atwo , \athree | \aone \bone \cone \}}$  & $\CMTB{[2,0,1,0,0]}$  &  1 & 2 & 2& 4& 1 & 3  & 10  \\ \hline
$\mathcal{V}_{\{ \aone,  \atwo,  \athree, \afour \}}$  & $\CMTB{[4,0,0,0,0]}$  &  1 & 1 & 1 & 1& 1  & 1 & 8  \\ \hline
$\mathcal{V}_{\{ \aone \bone 
\cone , \atwo \btwo \ctwo \}}$  & $\CMTB{[0,0,2,0,0]}$  &  1  & 1 & 1 & 3 & 1 & 2 & 8 \\ \hline
$\mathcal{V}_{\{ \atwo, \athree | \aone \bone \cone \done \}}$ & $\CMTB{[2,0,0,1,1]}$  &  1 & 3 & 2 & 6 & 2 & 5 & 8  \\ \hline
$\mathcal{V}_{\{ \atwo \btwo | \aone \bone \cone \done \eone \}{}^{-}}$ & $\CMTB{[0,1,0,2,0]}$  &  1 & 2 & 1 & 4 & 1 & 3 & 6  \\ \hline
$\mathcal{V}_{\{ \atwo , \athree | \aone \bone \cone \}}$ & $\CMTB{[2,0,1,0,0]}$   &  1 & 2 & 2 & 4 & 2 & 4 & 6  \\ \hline
$\mathcal{V}_{\{ \aone \bone , \atwo \btwo \}}$ & $\CMTB{[0,2,0,0,0]}$  &  1  & 1 & 1 & 2 & 1 & 2 & 4 \\ \hline
$\mathcal{V}_{\{ \atwo | 
\aone \bone \cone \done \eone \}{}^{-}}$ & $\CMTB{[1,0,0,2,0]}$  &  1 & 2 & 1 & 3 & 2 & 3 & 4  \\ \hline
$\mathcal{V}_{\{ \aone \bone \cone \}}$  & $\CMTB{[0,0,1,0,0]}$ &  1  & 1 & 1 & 2 & 1& 2 & 2 \\ \hline
\end{tabular}
\caption{Contributions of Each Type-I Superfield to the component bosonic and fermionic 
\newline $~~~~~~~~~~~~~~\,$representation
content of the 10D, $\cal N$ = 1 supergravity multiplet at Level-16/17 of 
\newline $~~~~~~~~~~~~~~\,$Type-IIB Scalar Superfield Decomposition}
\label{tab:IIB_contribution_54_lv16}
\end{table}

For all Type-I superfields shown in the two tables above, if there is $\CMTB{\{54\}}$ 
appears in level-$n$, there exits $\CMTB{\{120\}}$ irrep in level-$(n+2)$, which is the 
three(seven)-form (i.e. the field strength for either the ${\CMTB{\{45\}}}$ or ${\CMTB{\{210\}}}$
irreps).  Also upon comparing the two tables, it is clear that the same two sets of 10D, 
$\cal N$ = 1 superfields appear in both.  The only differences that occur take place 
when a prepotential involves a 5-form index.  In these cases, the 5-form indices are 
``flipped'' between the 10D, $\cal N$ = IIA and 10D, $\cal N$ = IIB theories.

The last column of the preceding two tables gives the height of the conformal
graviton irrep in the Type-I superfield.  This simultaneously gives the location
of the Type-I superfield in the ${\CMTG {\theta}}$-expansion.  So for example
the conformal graviton irrep indicated in the first row of either of the preceding
tables is found at Level number 14.  This means that the type-I superfield
containing those conformal graviton irreps are located at quadratic order
in the ${\CMTG {\theta}}$-expansion.

We must exercise a note of caution in regards to our proposal, however.  There is a 
long standing alternate proposal \cite{HNvP} for the 10D, $\cal N$ = 1 superspace 
supergravity prepotential.  This work has suggested that the ${\CMTB{\{120\}}}$ irrep
(the ${\CMTB{ [0,0,1,0,0]}}$ irrep or the superfield described by $\mathcal{V}_{\{ \aone \bone 
\cone \}}$), should play this role.  One argument that mitigates against this concern
is the fact that the ${\CMTB{ [0,0,1,0,0]}}$ irrep in the 10D theory does not connect 
to the unique conformal graviton seen in the 11D scalar supermultiplet.

In this chapter, we have looked at the issue of what superfield could serve as the 
10D, $\cal N$ = 1 supergravity prepotential.  Our investigation began from the
scalar superfield for both the Type-IIA and Type-IIB superspaces. The choice of
the scalar superfield is motivated by our study of the 11D, $\cal N$ = 1 superspace
gravity theory \cite{GHM3} where the scalar superfield is the simplest superfield
that contains the conformal graviton representation.  In this context, it was found
that the conformal graviton occurs exactly at the middle level of the superfield.
A truncation of 11D, $\cal N$ = 1 superspace to 10D, $\cal N$ = IIA superspace
must also then yield a scalar superfield as the supergravity prepotential.

Next in this line of reasoning, we may consider a truncation of 10D, $\cal N$ = IIA 
superspace to 10D, $\cal N$ = 1 superspace.  As our discussions in this chapter
showed, there are choices to explore as the putative supergravity prepotential.
The ADA algorithms inform us there are only a few options to explore, at least 
under the set of assumptions we are using.  These include the ten possibilities
given by
$\mathcal{V}_{\{ \aone  \atwo  \athree \}}$, 
$\mathcal{V}_{\{ \aone \bone, \atwo \btwo \}}$,
$\mathcal{V}_{\{ \aone, \atwo, \athree, \afour \}}$,
$\mathcal{V}_{\{ \atwo,\athree | \aone \bone \cone \}}$,
$\mathcal{V}_{\{ \aone \bone \cone, \atwo, \btwo \ctwo \}}$,
$\mathcal{V}_{\{ \atwo \athree | \aone \bone \cone \done \}}$,
$\mathcal{V}_{\{ \atwo| \aone \bone \cone \done \eone \}{}^+}$,
$\mathcal{V}_{\{ \atwo | \aone \bone \cone \done \eone \}{}^-}$,
$\mathcal{V}_{\{ \atwo \btwo | \aone \bone \cone \done \eone \}{}^+}$, and
$\mathcal{V}_{\{ \atwo \btwo | \aone \bone \cone \done \eone \}{}^-}$.

However, among the ten choices indicated by the two tables in this chapter, there 
is one that is extremely interesting as it contains a single copy of each of the 
component irreps that were the parameters of our ADA scan.  This representation is 
the superfield  $\mathcal{V}_{\{ \aone, \atwo, \athree, \afour \}}$, with Dynkin
Label $\CMTB{[4,0,0,0,0]}$ which corresponds to a 10D, totally symmetical, completely traceless  
fourth rank tensor superfield
\be
{\cal V}{}_{  {\CMTB{[4,0,0,0,0]}}}
(x^{\un{a}}, \, \theta^{\a} ) ~=~
{\cal V}{}_{\{{\un a}_1 , {\un a}_2 , {\un a}_3 , {\un a}_4\}}
(x^{\un{a}}, \, \theta^{\a} )  ~~~.
\label{eq:SG10Pre}
\ee
It also possesses the property that the conformal graviton irrep occurs in the middle 
level of the superfield.  These two properties suggest to us this should be the 
superfield that is the most likely candidate for the 10D, $\cal N$ = 1 supergravity 
prepotential. 

The next task becomes one of finding if there are Type-I superfields (SF's) in the expansions
shown in (\ref{eq:IIAx}) and (\ref{eq:IIBx}) that will allow the patterns of height assignments 
and irreps seen in (\ref{eq:1IIA}) and (\ref{eq:2IIB}) to be found within them?  So we set up an 
ADA scan in (\ref{eq:IIAx}) looking for the first pattern and one in (\ref{eq:IIBx}) looking for the 
second pattern at levels compatible with the 10D, ${\cal N} = 1$ SG prepotential identified in 
(\ref{eq:SG10Pre}).  The task is simplified by an observation on how the non-manifest supercharges 
act relative to the identification of the ${\CMTB{ [4,0,0,0,0]}}$ supergravity prepotential which occurs
at $\CMTG {\theta}$-Level eight.  The second non-manifest supercharge must connect quantities
at different $\CMTG {\theta}$-Levels.  The simplest example of this phenomenon is seen in the 
4D, $\cal N$ = 1 chiral supermultiplet. The spin-0 fields are at the lowest ${\theta}$-Level and a 
supercharge acting on these spin-0 fields connect them to the spin-1/2 fields at one 
${\theta}$-Level higher.  Translating this lesson for the present consideration, the ${\CMTB{ [4,0,0,0,0]}}$ 
supergravity prepotential should be connected to a 10D, ${\cal N} = 1$ superfield at $\CMTG {\theta}$-Level 
nine. So the present scan can be restricted to a search at this level.  The results of these scans are 
shown in Tables \ref{tab:IIA_MGM} and \ref{tab:IIB_MGM}.

\begin{table}[htp!]
\centering
\begin{tabular}{|c|c|c|c|c|c|c|c|c|}
\hline
Type-I SF & Dynkin Label & $b_{\CMTB{\{10\}}}$ & $b_{\CMTB{\{120\}}}$ & $b_{\CMTred{\{16\}}}$ &
$b_{\CMTred{\{144\}}}$ & \begin{tabular}[c]{@{}c@{}} $\CMTG {\theta}$-Level   \\ 
\end{tabular}
& \begin{tabular}[c]{@{}c@{}}Boson  \\ Height\end{tabular} & \begin{tabular}[c]{@{}c@{}}Fermion  \\ Height
\end{tabular}\\ \hline \hline
$\Psi_{\{ \aone , \atwo , \athree \}}{}_{\a}$ & $\CMTred{[3,0,0,1,0]}$ & 1& 3 & 2 & 3 & 9 & 7 & 8 \\ \hline 
$\Psi_{\{ \atwo | \aone 
\bone \cone \}}{}^{\a}$ & $\CMTred{[1,0,1,0,1]}$ & 1& 4 & 2 & 5  & 9 & 7 & 8\\ \hline 
\end{tabular}
\caption{Type-I Superfield Contributions to Bosons at Level-16 \& Fermions at Level-17
 in Type-IIA 
$~~~~~~~~~~~~~$ Scalar Superfield Decomposition}
    \label{tab:IIA_MGM}
\end{table}         

\begin{table}[htp!]
\centering
\begin{tabular}{|c|c|c|c|c|c|c|c|c|c|}
\hline
Type-I SF & Dynkin Label & $b_{\CMTB{\{1\}}}$ & $b_{\CMTB{\{45\}}}$ & $b_{\CMTB{\{210\}}}$ &  
$b_{\CMTred{\{\overline{16}\}}}$ &  $b_{\CMTred{\{\overline{144}\}}}$ & 
$\CMTG {\theta}$-Level
& \begin{tabular}[c]{@{}c@{}} Boson \\ Height \end{tabular} & \begin{tabular}[c]{@{}c@{}}Fermion  \\ 
Height \end{tabular}\\ \hline \hline
$\Psi_{\{ \aone , \atwo , \athree \}}{}^{\a}$ & $\CMTred{[3,0,0,0,1]}$ & 1& 2 & 3 & 2 & 3 & 9 & 7 & 8  \\ \hline
$\Psi_{\{ \atwo | \aone 
\bone \cone \}}{}_{\a}$ & $\CMTred{[1,0,1,1,0]}$ & 1& 3 & 6 & 2 & 5 & 9 & 7 & 8  \\ \hline
\end{tabular}
\caption{Type-I Superfield Contributions to Bosons at Level-16 \& Fermions at Level-17  
in Type-IIB 
$~~~~~~~~~~~~~$ Scalar Superfield Decomposition}
\label{tab:IIB_MGM}
\end{table}

Beyond this point, ADA scans provide no guidance or other means will be required for 
further progress.  But these scans have provided valuable insights that have ``shone a 
spotlight'' within these systems containing up to 2,147,483,648 bosonic degrees of 
freedom and 2,147,483,648 fermionic degrees of freedom.  The results derived indicate
where future investigations might show the greatest return on investment of time and
energy.

$$~~$$
\newpage
$$~~$$
\newpage

\section{Conclusion}
\label{sec:CONcL}

In chapters \ref{sec:primer} and \ref{sec:analytical0}, we have demonstrated the prepotential
basis for the argument that imposing a sufficient number of constraints on the superframe
fields to express them in terms of $\Psi$ and ${\rm H}{}_{\a}{}^{\un a}$ as their only 
independent superfield variables necessarily leads to Poincar\' e systems that nevertheless
possess Weyl covariance.  This observation was used to derive a set of infinitesimal Weyl 
scaling laws for ten and eleven supergravity.  On this basis the transformation laws for
the superspace torsion and curvature supertensors were derived and the super Weyl covariant
supertensors containing the on-shell degrees of freedom were thus identified within components
of the torsion supertensors.  In the case of the $\cal N$ = 1 theories in 10D and 11D, the 
infinitesimal Weyl scaling laws confirm results derived previously in the work of \cite{SC5} and
\cite{10DConFRM}.  For the cases 10D, $\cal N$ = 2A and $\cal N$ = 2B, the infinitesimal Weyl 
scaling laws derived in chapter \ref{sec:analytical0} have not been presented previously.  In 
chapter \ref{sec:ADA}, the conceptual basis provided by the concept of ``adynkras'' \cite{GHM2,GHM3} 
was reviewed in a way that is relevant to the discussion of supergravity theories in ten and 
eleven dimensions.  Finally, in chapter \ref{sec:story}, the presentations and results of the
previous chapters were used to identify possible supergravity prepotentials candidates.

In the discussion in \cite{GHM3} of the 11D, $\cal N$ = 1 theory, the primary putative supergravity 
superfield was a scalar superfield also.  To distinguish it from the superfields discussed in 
(\ref{eq:IIAx}) and (\ref{eq:IIBx}), we can denote the eleven dimensional scalar superfield 
by ${\cal V}{}_{11} (x^{\un{a}}, \, \theta^{\a})$ which together with ${\cal V}{}_{IIA} (x^{\un{a}}, 
\, \theta^{\a}, \, {\CMTG {\theta^{\Dot \a}}})$ and ${\cal V}{}_{IIB} (x^{\un{a}}, \, \theta^{\a}, \, 
{\CMTG {\theta^\a}})$\footnote{Of course, the range of the spinor index for the 11D 
theory goes from one to thirty-two while the range for the 10D theories is half of this.} form 
a triplet of superfields that hold the possibility of being the fundamental supergravity 
prepotentials of 11D, $\cal N$ = 1, 10D, $\cal N$ = IIA, and 10D, $\cal N$ = IIB theories.  
This raises the intriguing possibility this triplet is related to duality properties among these 
high dimensional models.

\vspace{.05in}
 \begin{center}
\parbox{4in}{{\it ``The best way to have a good idea is to have a lot of  \\ $~~$
ideas.'' \\ ${~}$
\\ ${~}$ }\,\,-\,\, Linus Pauling}
 \parbox{4in}{
 $~~$} 
 \end{center}
 \noindent
{\bf {Acknowledgements}}\\[.1in] \indent

We wish to acknowledge discussions with B.\ de Wit, S.\ J.\ Gates, III, S.\ Kuzenko, W.\ Linch, W.\ Siegel, 
M.\ Ro\v cek, and E. Witten.  The research of S.\ J.\ G., Y.\ Hu, and S.-N.\ Mak is supported 
in part by the endowment of the Ford Foundation Professorship of Physics at Brown 
University and they gratefully acknowledge the support of the Brown Theoretical Physics 
Center.

\newpage
\appendix
\section{Appendix on 10D Weyl Scaling Properties of  Weight ${\bm {\ge \fracm32}}$ Super Tensors \label{appen:Weyl}}

In this appendix, we record the Weyl scaling transformation laws for the super torsion and super curvature
tensors of scale weight equal or greater than three-halves in the respective cases of the 
$\cal N$ = 2A, $\cal N$ = 2B and $\cal N$ = 1 theories.

\subsection{10D, $\mathcal{N} = 2$A\label{appen:Weyl-IIA}}

\begin{align}
\d_S T_{\un a \un b}{}^{\g} \,  =&~ \fracm32 L T_{\un a \un b}{}^{\g} + i\fracm15 (\s_{[\un a})^{\a\g}(\nabla_{\un b]}\nabla_{\a}L) - i\fracm15 (\s_{[\un a})^{\a\b}(\nabla_{\a}L)T_{\b| \un b]}{}^{\g} 
{~~~~~~~~~~~~~~~~~~~~~~~~~~~~~~~~~~~~~~}  \nonumber\\
&~- i\fracm15 (\s_{[\un a})^{\Dot{\a}\Dot{\b}}(\nabla_{\Dot{\a}}L)T_{\Dot{\b}| \un b]}{}^{\g} + i\fracm15 (\s_{\un c})^{\g\d}(\nabla_{\d}L)T_{\un a \un b}{}^{\un c} ~~~, \\
\d_S T_{\un{a} \un{b}}{}^{\Dot{\g}} \, =&~ \fracm32 L T_{\un a \un b}{}^{\Dot{\g}} + i\fracm15 (\s_{[\un a})^{\Dot{\a}\Dot{\g}}(\nabla_{\un b]}\nabla_{\Dot{\a}}L) - i\fracm15 (\s_{[\un a})^{\a\b}(\nabla_{\a}L)T_{\b| \un b]}{}^{\Dot{\g}} \nonumber\\&~- i\fracm15 (\s_{[\un a})^{\Dot{\a}\Dot{\b}}(\nabla_{\Dot{\a}}L)T_{\Dot{\b}| \un b]}{}^{\Dot{\g}} + i\fracm15 (\s_{\un c})^{\Dot{\g}\Dot{\d}}(\nabla_{\Dot{\d}}L)T_{\un a \un b}{}^{\un c} ~~~,
\end{align}

\begin{align}
    \d_S R_{\a \un{b}}{}^{\un{d}\un{e}} \, =&~ \fracm32 L R_{\a \un b}{}^{\un{d}\un{e}} - T_{\a \un b}{}^{[\un d}(\nabla^{\un e]}L) + \fracm15 T_{\a \un b}{}^{\g}(\s^{\un{d}\un{e}})_{\g}{}^{\d}(\nabla_{\d}L)+ \fracm15 T_{\a \un b}{}^{\Dot{\g}}(\s^{\un{d}\un{e}})_{\Dot{\g}}{}^{\Dot{\d}}(\nabla_{\Dot{\d}}L)
{~~~~~~~~~~~~~~~~~~~~}    
     \nonumber\\
    &~ +\fracm15(\s^{\un{d}\un{e}})_{\a}{}^{\b}(\nabla_{\un b}\nabla_{\b}L) + i\fracm15(\s_{\un b})^{\d\b}(\nabla_{\d}L)R_{\a \b}{}^{\un{d}\un{e}} + i\fracm15(\s_{\un b})^{\Dot{\d}\Dot{\b}}(\nabla_{\Dot{\d}}L)R_{\a \Dot{\b}}{}^{\un{d}\un{e}} \nonumber\\
    &~-(\nabla_{\a}\nabla^{[\un{d}}L)\d_{\un b}{}^{\un e]}  ~~~, \\
    \d_S R_{\Dot{\a} \un{b}}{}^{\un{d}\un{e}} \, =&~ \fracm32 L R_{\Dot{\a} \un b}{}^{\un{d}\un{e}} - T_{\Dot{\a} \un b}{}^{[\un d}(\nabla^{\un e]}L) + \fracm15 T_{\Dot{\a} \un b}{}^{\g}(\s^{\un{d}\un{e}})_{\g}{}^{\d}(\nabla_{\d}L)+ \fracm15 T_{\Dot{\a} \un b}{}^{\Dot{\g}}(\s^{\un{d}\un{e}})_{\Dot{\g}}{}^{\Dot{\d}}(\nabla_{\Dot{\d}}L) \nonumber\\
    &~ +\fracm15(\s^{\un{d}\un{e}})_{\Dot{\a}}{}^{\Dot{\b}}(\nabla_{\un b}\nabla_{\Dot{\b}}L) + i\fracm15(\s_{\un b})^{\d\b}(\nabla_{\d}L)R_{\Dot{\a}\b }{}^{\un{d}\un{e}} + i\fracm15(\s_{\un b})^{\Dot{\d}\Dot{\b}}(\nabla_{\Dot{\d}}L)R_{\Dot{\a} \Dot{\b}}{}^{\un{d}\un{e}} \nonumber\\
    &~-(\nabla_{\Dot{\a}}\nabla^{[\un{d}}L)\d_{\un b}{}^{\un e]}\ ~~~,
\end{align}

\begin{align}
\d_S R_{\un a \un b}{}^{\un{d}\un{e}} \, =&~ 2L R_{\un a \un b}{}^{\un{d}\un{e}} - T_{\un a \un b}{}^{[\un d}(\nabla^{\un e]}L) + \fracm15T_{\un a \un b}{}^{\g}(\s^{\un d\un e})_{\g}{}^{\d}(\nabla_{\d}L)+ \fracm15T_{\un a \un b}{}^{\Dot{\g}}(\s^{\un d\un e})_{\Dot{\g}}{}^{\Dot{\d}}(\nabla_{\Dot{\d}}L) \nonumber\\
&~ -i\fracm15 (\s_{[\un a})^{\a\b}(\nabla_{\a}L)R_{\b| \un b]}{}^{\un{d}\un{e}}-i\fracm15 (\s_{[\un a})^{\Dot{\a}\Dot{\b}}(\nabla_{\Dot{\a}}L)R_{\Dot{\b}| \un b]}{}^{\un{d}\un{e}} - (\nabla_{[\un a}\nabla^{[\un d}L)\d_{\un b]}{}^{\un e]}    ~~~.  {~~~~~~~~~}
\end{align}

\subsection{10D, $\mathcal{N} = 2$B\label{appen:Weyl-IIB}}

\begin{align}
\d_S T_{\un a \un b}{}^{\g} \,  =&~ (\fracm12 L + \overline{L}) T_{\un a \un b}{}^{\g} + i (\s_{[\un a})^{\a\g}(\nabla_{\un b]}\Bar{\nabla}_{\a}(\fracm{1}{32}L + \fracm{27}{160}\overline{L})) - i (\s_{[\un a})^{\a\b}(\Bar{\nabla}_{\a}(\fracm{1}{32}L + \fracm{27}{160}\overline{L}))T_{\b| \un b]}{}^{\g} \nonumber\\
&~- i (\s_{[\un a})^{\a\b}(\nabla_{\a}(\fracm{1}{32}\overline{L} + \fracm{27}{160}L))T_{\overline{\b}| \un b]}{}^{\g} + i (\s_{\un c})^{\g\d}(\Bar{\nabla}_{\d}(\fracm{1}{32}L + \fracm{27}{160}\overline{L}))T_{\un a \un b}{}^{\un c} ~~~, \\
\d_S T_{\un{a} \un{b}}{}^{\overline{\g}} \, =&~ (\fracm12\overline{L} + L) T_{\un a \un b}{}^{\overline{\g}} + i (\s_{[\un a})^{\a\g}(\nabla_{\un b]}\nabla_{\a}(\fracm{1}{32}\overline{L} + \fracm{27}{160}L)) - i (\s_{[\un a})^{\a\b}(\Bar{\nabla}_{\a}(\fracm{1}{32}L + \fracm{27}{160}\overline{L}))T_{\b| \un b]}{}^{\overline{\g}} \nonumber\\
&~- i (\s_{[\un a})^{\a\b}(\nabla_{\a}(\fracm{1}{32}\overline{L} + \fracm{27}{160}L))T_{\overline{\b}| \un b]}{}^{\overline{\g}} + i (\s_{\un c})^{\g\d}(\nabla_{\d}(\fracm{1}{32}\overline{L} + \fracm{27}{160}L))T_{\un a \un b}{}^{\un c} ~~~.
\end{align}

\begin{align}
\d_S R_{\a \un{b}}{}^{\un{d}\un{e}} \, =&~ (\fracm12\overline{L} + L) R_{\a \un b}{}^{\un{d}\un{e}} -\fracm12 T_{\a \un b}{}^{[\un d}(\nabla^{\un e]}(L+\overline{L})) + \fracm15 T_{\a \un b}{}^{\g}(\s^{\un{d}\un{e}})_{\g}{}^{\d}(\nabla_{\d}L)+ \fracm15 T_{\a \un b}{}^{\overline{\g}}(\s^{\un{d}\un{e}})_{\g}{}^{\d}(\Bar{\nabla}_{\d}\overline{L}) \nonumber\\
&~ +\fracm15(\s^{\un{d}\un{e}})_{\a}{}^{\b}(\nabla_{\un b}\nabla_{\b}L) + i(\s_{\un b})^{\d\b}(\Bar{\nabla}_{\d}(\fracm{1}{32}L + \fracm{27}{160}\overline{L}))R_{\a \b}{}^{\un{d}\un{e}}  \nonumber\\
&~+ i(\s_{\un b})^{\d\b}(\nabla_{\d}(\fracm{1}{32}\overline{L} + \fracm{27}{160}L))R_{\a \overline{\b}}{}^{\un{d}\un{e}}-\fracm12(\nabla_{\a}\nabla^{[\un{d}}(L+\overline{L}))\d_{\un b}{}^{\un e]}  ~~~, \\
\d_S R_{\overline{\a} \un{b}}{}^{\un{d}\un{e}} \, =&~ (\fracm12 L +\overline{L} ) R_{\overline{\a} \un b}{}^{\un{d}\un{e}} -\fracm12 T_{\overline{\a} \un b}{}^{[\un d}(\nabla^{\un e]}(L+\overline{L})) + \fracm15 T_{\overline{\a} \un b}{}^{\g}(\s^{\un{d}\un{e}})_{\g}{}^{\d}(\nabla_{\d}L)+ \fracm15 T_{\overline{\a} \un b}{}^{\overline{\g}}(\s^{\un{d}\un{e}})_{\g}{}^{\d}(\Bar{\nabla}_{\d}\overline{L}) \nonumber\\
&~ +\fracm15(\s^{\un{d}\un{e}})_{\a}{}^{\b}(\nabla_{\un b}\Bar{\nabla}_{\b}\overline{L}) + i(\s_{\un b})^{\d\b}(\Bar{\nabla}_{\d}(\fracm{1}{32}L + \fracm{27}{160}\overline{L}))R_{\overline{\a}\b }{}^{\un{d}\un{e}} + i(\s_{\un b})^{\d\b}(\nabla_{\d}(\fracm{1}{32}\overline{L} + \fracm{27}{160}L))R_{\overline{\a} \overline{\b}}{}^{\un{d}\un{e}} \nonumber\\
&~-\fracm12(\Bar{\nabla}_{\a}\nabla^{[\un{d}}(L+\overline{L}))\d_{\un b}{}^{\un e]}\ ~~~, \\
\d_S R_{\un a \un b}{}^{\un{d}\un{e}} \, =&~ (L+\overline{L}) R_{\un a \un b}{}^{\un{d}\un{e}} - \fracm12 T_{\un a \un b}{}^{[\un d}(\nabla^{\un e]}(L+\overline{L})) + \fracm15T_{\un a \un b}{}^{\g}(\s^{\un d\un e})_{\g}{}^{\d}(\nabla_{\d}L)+ \fracm15 T_{\un a \un b}{}^{\overline{\g}} (\s^{\un d\un e})_{\g}{}^{\d}(\Bar{\nabla}_{\d}\overline{L}) \nonumber\\
&~ -i (\s_{[\un a})^{\a\b}(\Bar{\nabla}_{\a}(\fracm{1}{32}L + \fracm{27}{160}\overline{L}))R_{\b| \un b]}{}^{\un{d}\un{e}}-i (\s_{[\un a})^{\a\b}(\nabla_{\a}(\fracm{1}{32}\overline{L} + \fracm{27}{160}L))R_{\overline{\b}| \un b]}{}^{\un{d}\un{e}} \nonumber\\
&~ - \fracm12(\nabla_{[\un a}\nabla^{[\un d}(L+\overline{L}))\d_{\un b]}{}^{\un e]}
~~~.
\end{align}

\subsection{10D, $\mathcal{N} = 1$\label{appen:Weyl-I}}

\begin{align}
\d_S T_{\un a \un b}{}^{\g} \,  =&~ \fracm32 L \, T_{\un a \un b}{}^{\g} ~+~ i\, \fracm25 (\s_{[\un a})^{\a\g}(\nabla_{\un b]}\nabla_{\a}L) ~-~ i\, \fracm25 \, (\s_{[\un a})^{\a\b}(\nabla_{\a}L)\, T_{\b| \un b]}{}^{\g} \nonumber\\
&~+~ i\, \fracm25 \, (\s_{\un c})^{\g\d}(\nabla_{\d}L)\, T_{\un a \un b}{}^{\un c} ~~~, \\
\d_S R_{\a \un b}{}^{\un{d}\un{e}} \,  =&~ \fracm32 L \, R_{\a \un b}{}^{\un{d}\un{e}} ~-~ T_{\a \un b}{}^{[\un d}(\nabla^{\un e]}L) ~+~ \fracm15 \, T_{\a \un b}{}^{\g}\, (\s^{\un{d}\un{e}})_{\g}{}^{\d}\, (\nabla_{\d}L) \nonumber\\
&~+~\fracm15 \, (\s^{\un{d}\un{e}})_{\a}{}^{\b}(\nabla_{\un b}\nabla_{\b}L) ~+~ i\, \fracm25(\s_{\un b})^{\d\b}(\nabla_{\d}L)R_{\a \b}{}^{\un{d}\un{e}} ~-~ (\nabla_{\a}\nabla^{[\un{d}}L)\d_{\un b}{}^{\un e]}  ~~~,
\end{align}
\begin{align}
\d_S R_{\un a \un b}{}^{\un{d}\un{e}} \, =&~ 2L \, R_{\un a \un b}{}^{\un{d}\un{e}} ~-~ T_{\un a \un b}{}^{[\un d}(\nabla^{\un e]}L) ~+~ \fracm15T_{\un a \un b}{}^{\g}(\s^{\un d\un e})_{\g}{}^{\d}(\nabla_{\d}L) \nonumber\\
&~-~i\, \fracm25 (\s_{[\un a})^{\a\b}(\nabla_{\a}L)R_{\b| \un b]}{}^{\un{d}\un{e}} ~-~ (\nabla_{[\un a}\nabla^{[\un d}L)\d_{\un b]}{}^{\un e]}
    ~~~.  {~~~~~~~~~~~~~~~~~~~~~~~~~~~} 
\end{align}

\newpage

\section{Appendix on Adynkra Libraries \label{appen:Lib}}

In this appendix, we explicitly present the adynkra libraries of all superfields that 
appear in equation (\ref{equ:REPs}) in the text.

{\subsection{Dynkin Label Library of ${\cal V}_{\CMTB{[0,0,1,0,0]}}$   \label{appen:L1}}
\begin{itemize} \sloppy
\item Level-0: $\CMTB {[0,0,1,0,0]}$
\item Level-1: $\CMTred {[0,0,0,0,1]} \oplus \CMTred {[1,0,0,1,0]} \oplus \CMTred {[0,1,0,0,1]} \oplus \CMTred {[0,0,1,1,0]}$
\item Level-2: $\CMTB {[0,0,0,0,0]} \oplus \CMTB {[0,1,0,0,0]} \oplus \CMTB {[2,0,0,0,0]} \oplus (2) \CMTB {[0,0,0,1,1]} \oplus \CMTB {[0,2,0,0,0]} \oplus \CMTB {[1,0,1,0,0]} \oplus \CMTB {[1,0,0,2,0]} \oplus \CMTB {[1,0,0,0,2]} \oplus \CMTB {[0,0,2,0,0]} \oplus \CMTB {[0,1,0,1,1]}$
\item Level-3: $\CMTred {[0,0,0,1,0]} \oplus (2) \CMTred {[1,0,0,0,1]} \oplus (2) \CMTred {[0,1,0,1,0]} \oplus \CMTred {[0,0,0,0,3]} \oplus \CMTred {[2,0,0,1,0]} \oplus (2) \CMTred {[0,0,1,0,1]} \oplus \CMTred {[0,0,0,2,1]} \oplus (2) \CMTred {[1,1,0,0,1]} \oplus \CMTred {[0,2,0,1,0]} \oplus \CMTred {[1,0,1,1,0]} \oplus \CMTred {[1,0,0,1,2]} \oplus \CMTred {[0,1,1,0,1]}$
\item Level-4: $(2) \CMTB {[0,0,1,0,0]} \oplus \CMTB {[0,0,0,0,2]} \oplus (2) \CMTB {[1,1,0,0,0]} \oplus (3) \CMTB {[1,0,0,1,1]} \oplus (2) \CMTB {[0,1,1,0,0]} \oplus \CMTB {[0,1,0,2,0]} \oplus (3) \CMTB {[0,1,0,0,2]} \oplus (2) \CMTB {[2,0,1,0,0]} \oplus \CMTB {[1,2,0,0,0]} \oplus \CMTB {[2,0,0,0,2]} \oplus \CMTB {[0,0,0,1,3]} \oplus \CMTB {[0,0,1,1,1]} \oplus \CMTB {[0,2,1,0,0]} \oplus (2) \CMTB {[1,1,0,1,1]} \oplus \CMTB {[1,0,1,0,2]}$
\item Level-5: $\CMTred {[1,0,0,1,0]} \oplus (3) \CMTred {[0,1,0,0,1]} \oplus (2) \CMTred {[2,0,0,0,1]} \oplus \CMTred {[0,0,1,1,0]} \oplus (2) \CMTred {[0,0,0,1,2]} \oplus \CMTred {[3,0,0,1,0]} \oplus (3) \CMTred {[1,1,0,1,0]} \oplus (2) \CMTred {[1,0,0,0,3]} \oplus (2) \CMTred {[0,2,0,0,1]} \oplus (4) \CMTred {[1,0,1,0,1]} \oplus \CMTred {[1,0,0,2,1]} \oplus (2) \CMTred {[2,1,0,0,1]} \oplus \CMTred {[0,1,1,1,0]} \oplus \CMTred {[0,0,1,0,3]} \oplus (2) \CMTred {[0,1,0,1,2]} \oplus \CMTred {[2,0,1,1,0]} \oplus \CMTred {[1,2,0,1,0]} \oplus \CMTred {[2,0,0,1,2]} \oplus \CMTred {[1,1,1,0,1]}$
\item Level-6: $\CMTB {[2,0,0,0,0]} \oplus \CMTB {[0,0,0,1,1]} \oplus \CMTB {[4,0,0,0,0]} \oplus (2) \CMTB {[0,2,0,0,0]} \oplus (2) \CMTB {[1,0,1,0,0]} \oplus \CMTB {[1,0,0,2,0]} \oplus (3) \CMTB {[1,0,0,0,2]} \oplus (2) \CMTB {[2,1,0,0,0]} \oplus \CMTB {[0,0,0,0,4]} \oplus (2) \CMTB {[0,0,2,0,0]} \oplus (3) \CMTB {[0,1,0,1,1]} \oplus (2) \CMTB {[0,0,1,0,2]} \oplus (4) \CMTB {[2,0,0,1,1]} \oplus \CMTB {[0,0,0,2,2]} \oplus \CMTB {[3,0,1,0,0]} \oplus \CMTB {[2,2,0,0,0]} \oplus \CMTB {[3,0,0,2,0]} \oplus \CMTB {[3,0,0,0,2]} \oplus (3) \CMTB {[1,1,1,0,0]} \oplus \CMTB {[1,1,0,2,0]} \oplus (3) \CMTB {[1,1,0,0,2]} \oplus \CMTB {[1,0,0,1,3]} \oplus (2) \CMTB {[1,0,1,1,1]} \oplus \CMTB {[0,2,0,1,1]} \oplus \CMTB {[2,0,2,0,0]} \oplus \CMTB {[0,1,1,0,2]} \oplus \CMTB {[2,1,0,1,1]}$
\item Level-7: $\CMTred {[1,0,0,0,1]} \oplus \CMTred {[0,1,0,1,0]} \oplus \CMTred {[0,0,0,0,3]} \oplus (2) \CMTred {[2,0,0,1,0]} \oplus (2) \CMTred {[0,0,1,0,1]} \oplus \CMTred {[0,0,0,2,1]} \oplus (3) \CMTred {[3,0,0,0,1]} \oplus (4) \CMTred {[1,1,0,0,1]} \oplus \CMTred {[4,0,0,1,0]} \oplus \CMTred {[0,2,0,1,0]} \oplus (4) \CMTred {[1,0,1,1,0]} \oplus (3) \CMTred {[1,0,0,1,2]} \oplus (3) \CMTred {[2,1,0,1,0]} \oplus \CMTred {[0,1,0,0,3]} \oplus \CMTred {[2,0,0,0,3]} \oplus (3) \CMTred {[0,1,1,0,1]} \oplus \CMTred {[0,0,2,1,0]} \oplus \CMTred {[0,1,0,2,1]} \oplus (3) \CMTred {[2,0,1,0,1]} \oplus \CMTred {[1,2,0,0,1]} \oplus \CMTred {[3,1,0,0,1]} \oplus (2) \CMTred {[2,0,0,2,1]} \oplus \CMTred {[0,0,1,1,2]} \oplus \CMTred {[3,0,1,1,0]} \oplus \CMTred {[1,1,1,1,0]} \oplus \CMTred {[1,0,2,0,1]} \oplus \CMTred {[1,1,0,1,2]}$
\item Level-8: $\CMTB {[0,0,1,0,0]} \oplus \CMTB {[3,0,0,0,0]} \oplus \CMTB {[1,1,0,0,0]} \oplus (3) \CMTB {[1,0,0,1,1]} \oplus (2) \CMTB {[0,1,1,0,0]} \oplus (2) \CMTB {[0,1,0,2,0]} \oplus (2) \CMTB {[0,1,0,0,2]} \oplus (5) \CMTB {[2,0,1,0,0]} \oplus \CMTB {[1,2,0,0,0]} \oplus (2) \CMTB {[3,1,0,0,0]} \oplus (2) \CMTB {[2,0,0,2,0]} \oplus (2) \CMTB {[2,0,0,0,2]} \oplus (3) \CMTB {[0,0,1,1,1]} \oplus (2) \CMTB {[1,0,2,0,0]} \oplus (3) \CMTB {[3,0,0,1,1]} \oplus \CMTB {[0,2,1,0,0]} \oplus (4) \CMTB {[1,1,0,1,1]} \oplus \CMTB {[4,0,1,0,0]} \oplus (2) \CMTB {[1,0,1,2,0]} \oplus (2) \CMTB {[1,0,1,0,2]} \oplus \CMTB {[1,0,0,2,2]} \oplus \CMTB {[2,1,1,0,0]} \oplus \CMTB {[0,0,3,0,0]} \oplus \CMTB {[2,1,0,2,0]} \oplus \CMTB {[2,1,0,0,2]} \oplus \CMTB {[0,1,1,1,1]} \oplus \CMTB {[2,0,1,1,1]}$
\item Level-9: $\CMTred {[1,0,0,1,0]} \oplus \CMTred {[0,1,0,0,1]} \oplus \CMTred {[0,0,0,3,0]} \oplus (2) \CMTred {[2,0,0,0,1]} \oplus (2) \CMTred {[0,0,1,1,0]} \oplus \CMTred {[0,0,0,1,2]} \oplus (3) \CMTred {[3,0,0,1,0]} \oplus (4) \CMTred {[1,1,0,1,0]} \oplus \CMTred {[4,0,0,0,1]} \oplus \CMTred {[0,2,0,0,1]} \oplus (4) \CMTred {[1,0,1,0,1]} \oplus (3) \CMTred {[1,0,0,2,1]} \oplus (3) \CMTred {[2,1,0,0,1]} \oplus \CMTred {[0,1,0,3,0]} \oplus \CMTred {[2,0,0,3,0]} \oplus (3) \CMTred {[0,1,1,1,0]} \oplus \CMTred {[0,0,2,0,1]} \oplus \CMTred {[0,1,0,1,2]} \oplus (3) \CMTred {[2,0,1,1,0]} \oplus \CMTred {[1,2,0,1,0]} \oplus \CMTred {[3,1,0,1,0]} \oplus (2) \CMTred {[2,0,0,1,2]} \oplus \CMTred {[0,0,1,2,1]} \oplus \CMTred {[3,0,1,0,1]} \oplus \CMTred {[1,1,1,0,1]} \oplus \CMTred {[1,0,2,1,0]} \oplus \CMTred {[1,1,0,2,1]}$
\end{itemize}

\begin{itemize} \sloppy
\item Level-10: $\CMTB {[2,0,0,0,0]} \oplus \CMTB {[0,0,0,1,1]} \oplus \CMTB {[4,0,0,0,0]} \oplus (2) \CMTB {[0,2,0,0,0]} \oplus (2) \CMTB {[1,0,1,0,0]} \oplus (3) \CMTB {[1,0,0,2,0]} \oplus \CMTB {[1,0,0,0,2]} \oplus (2) \CMTB {[2,1,0,0,0]} \oplus \CMTB {[0,0,0,4,0]} \oplus (2) \CMTB {[0,0,2,0,0]} \oplus (3) \CMTB {[0,1,0,1,1]} \oplus (2) \CMTB {[0,0,1,2,0]} \oplus (4) \CMTB {[2,0,0,1,1]} \oplus \CMTB {[0,0,0,2,2]} \oplus \CMTB {[3,0,1,0,0]} \oplus \CMTB {[2,2,0,0,0]} \oplus \CMTB {[3,0,0,2,0]} \oplus \CMTB {[3,0,0,0,2]} \oplus (3) \CMTB {[1,1,1,0,0]} \oplus (3) \CMTB {[1,1,0,2,0]} \oplus \CMTB {[1,1,0,0,2]} \oplus \CMTB {[1,0,0,3,1]} \oplus (2) \CMTB {[1,0,1,1,1]} \oplus \CMTB {[0,2,0,1,1]} \oplus \CMTB {[2,0,2,0,0]} \oplus \CMTB {[0,1,1,2,0]} \oplus \CMTB {[2,1,0,1,1]}$
\item Level-11: $\CMTred {[1,0,0,0,1]} \oplus (3) \CMTred {[0,1,0,1,0]} \oplus (2) \CMTred {[2,0,0,1,0]} \oplus \CMTred {[0,0,1,0,1]} \oplus (2) \CMTred {[0,0,0,2,1]} \oplus \CMTred {[3,0,0,0,1]} \oplus (3) \CMTred {[1,1,0,0,1]} \oplus (2) \CMTred {[1,0,0,3,0]} \oplus (2) \CMTred {[0,2,0,1,0]} \oplus (4) \CMTred {[1,0,1,1,0]} \oplus \CMTred {[1,0,0,1,2]} \oplus (2) \CMTred {[2,1,0,1,0]} \oplus \CMTred {[0,1,1,0,1]} \oplus \CMTred {[0,0,1,3,0]} \oplus (2) \CMTred {[0,1,0,2,1]} \oplus \CMTred {[2,0,1,0,1]} \oplus \CMTred {[1,2,0,0,1]} \oplus \CMTred {[2,0,0,2,1]} \oplus \CMTred {[1,1,1,1,0]}$
\item Level-12: $(2) \CMTB {[0,0,1,0,0]} \oplus \CMTB {[0,0,0,2,0]} \oplus (2) \CMTB {[1,1,0,0,0]} \oplus (3) \CMTB {[1,0,0,1,1]} \oplus (2) \CMTB {[0,1,1,0,0]} \oplus (3) \CMTB {[0,1,0,2,0]} \oplus \CMTB {[0,1,0,0,2]} \oplus (2) \CMTB {[2,0,1,0,0]} \oplus \CMTB {[1,2,0,0,0]} \oplus \CMTB {[2,0,0,2,0]} \oplus \CMTB {[0,0,0,3,1]} \oplus \CMTB {[0,0,1,1,1]} \oplus \CMTB {[0,2,1,0,0]} \oplus (2) \CMTB {[1,1,0,1,1]} \oplus \CMTB {[1,0,1,2,0]}$
\item Level-13: $\CMTred {[0,0,0,0,1]} \oplus (2) \CMTred {[1,0,0,1,0]} \oplus (2) \CMTred {[0,1,0,0,1]} \oplus \CMTred {[0,0,0,3,0]} \oplus \CMTred {[2,0,0,0,1]} \oplus (2) \CMTred {[0,0,1,1,0]} \oplus \CMTred {[0,0,0,1,2]} \oplus (2) \CMTred {[1,1,0,1,0]} \oplus \CMTred {[0,2,0,0,1]} \oplus \CMTred {[1,0,1,0,1]} \oplus \CMTred {[1,0,0,2,1]} \oplus \CMTred {[0,1,1,1,0]}$
\item Level-14: $\CMTB {[0,0,0,0,0]} \oplus \CMTB {[0,1,0,0,0]} \oplus \CMTB {[2,0,0,0,0]} \oplus (2) \CMTB {[0,0,0,1,1]} \oplus \CMTB {[0,2,0,0,0]} \oplus \CMTB {[1,0,1,0,0]} \oplus \CMTB {[1,0,0,2,0]} \oplus \CMTB {[1,0,0,0,2]} \oplus \CMTB {[0,0,2,0,0]} \oplus \CMTB {[0,1,0,1,1]}$
\item Level-15: $\CMTred {[0,0,0,1,0]} \oplus \CMTred {[1,0,0,0,1]} \oplus \CMTred {[0,1,0,1,0]} \oplus \CMTred {[0,0,1,0,1]}$
\item Level-16: $\CMTB {[0,0,1,0,0]}$
\end{itemize}

\newpage
{\subsection{Dynkin Label Library of ${\cal V}_{\CMTB{[1,0,1,0,1]}}$   \label{appen:L2}}
\begin{itemize} \sloppy
\item Level-0: $\CMTred {[1,0,1,0,1]}$
\item Level-1: $\CMTB {[1,0,1,0,0]} \oplus \CMTB {[1,0,0,0,2]} \oplus \CMTB {[0,0,2,0,0]} \oplus \CMTB {[0,1,0,1,1]} \oplus \CMTB {[0,0,1,0,2]} \oplus \CMTB {[2,0,0,1,1]} \oplus \CMTB {[1,1,1,0,0]} \oplus \CMTB {[1,1,0,0,2]} \oplus \CMTB {[1,0,1,1,1]}$
\item Level-2: $\CMTred {[1,0,0,0,1]} \oplus \CMTred {[0,1,0,1,0]} \oplus \CMTred {[0,0,0,0,3]} \oplus \CMTred {[2,0,0,1,0]} \oplus (2) \CMTred {[0,0,1,0,1]} \oplus \CMTred {[0,0,0,2,1]} \oplus \CMTred {[3,0,0,0,1]} \oplus (3) \CMTred {[1,1,0,0,1]} \oplus \CMTred {[0,2,0,1,0]} \oplus (3) \CMTred {[1,0,1,1,0]} \oplus (3) \CMTred {[1,0,0,1,2]} \oplus \CMTred {[2,1,0,1,0]} \oplus \CMTred {[0,1,0,0,3]} \oplus \CMTred {[2,0,0,0,3]} \oplus (3) \CMTred {[0,1,1,0,1]} \oplus \CMTred {[0,0,2,1,0]} \oplus \CMTred {[0,1,0,2,1]} \oplus (2) \CMTred {[2,0,1,0,1]} \oplus \CMTred {[1,2,0,0,1]} \oplus \CMTred {[2,0,0,2,1]} \oplus \CMTred {[0,0,1,1,2]} \oplus \CMTred {[1,1,1,1,0]} \oplus \CMTred {[1,0,2,0,1]} \oplus \CMTred {[1,1,0,1,2]}$
\item Level-3: $\CMTB {[0,0,1,0,0]} \oplus \CMTB {[0,0,0,2,0]} \oplus \CMTB {[0,0,0,0,2]} \oplus \CMTB {[3,0,0,0,0]} \oplus (2) \CMTB {[1,1,0,0,0]} \oplus (5) \CMTB {[1,0,0,1,1]} \oplus (4) \CMTB {[0,1,1,0,0]} \oplus (3) \CMTB {[0,1,0,2,0]} \oplus (4) \CMTB {[0,1,0,0,2]} \oplus (4) \CMTB {[2,0,1,0,0]} \oplus (2) \CMTB {[1,2,0,0,0]} \oplus \CMTB {[3,1,0,0,0]} \oplus (2) \CMTB {[2,0,0,2,0]} \oplus (4) \CMTB {[2,0,0,0,2]} \oplus \CMTB {[0,0,0,3,1]} \oplus (2) \CMTB {[0,0,0,1,3]} \oplus (5) \CMTB {[0,0,1,1,1]} \oplus \CMTB {[1,0,0,0,4]} \oplus (4) \CMTB {[1,0,2,0,0]} \oplus (2) \CMTB {[3,0,0,1,1]} \oplus (2) \CMTB {[0,2,1,0,0]} \oplus (7) \CMTB {[1,1,0,1,1]} \oplus \CMTB {[0,2,0,2,0]} \oplus (2) \CMTB {[0,2,0,0,2]} \oplus (2) \CMTB {[1,0,1,2,0]} \oplus (5) \CMTB {[1,0,1,0,2]} \oplus (2) \CMTB {[1,0,0,2,2]} \oplus (2) \CMTB {[2,1,1,0,0]} \oplus \CMTB {[0,0,3,0,0]} \oplus \CMTB {[2,1,0,2,0]} \oplus (2) \CMTB {[2,1,0,0,2]} \oplus \CMTB {[0,0,2,0,2]} \oplus \CMTB {[0,1,0,1,3]} \oplus (3) \CMTB {[0,1,1,1,1]} \oplus \CMTB {[2,0,0,1,3]} \oplus (2) \CMTB {[2,0,1,1,1]} \oplus \CMTB {[1,1,2,0,0]} \oplus \CMTB {[1,2,0,1,1]} \oplus \CMTB {[1,1,1,0,2]}$
\item Level-4: $(2) \CMTred {[1,0,0,1,0]} \oplus (4) \CMTred {[0,1,0,0,1]} \oplus (2) \CMTred {[0,0,0,3,0]} \oplus (4) \CMTred {[2,0,0,0,1]} \oplus (5) \CMTred {[0,0,1,1,0]} \oplus (4) \CMTred {[0,0,0,1,2]} \oplus (3) \CMTred {[3,0,0,1,0]} \oplus (8) \CMTred {[1,1,0,1,0]} \oplus (4) \CMTred {[1,0,0,0,3]} \oplus \CMTred {[4,0,0,0,1]} \oplus (5) \CMTred {[0,2,0,0,1]} \oplus (11) \CMTred {[1,0,1,0,1]} \oplus (7) \CMTred {[1,0,0,2,1]} \oplus (7) \CMTred {[2,1,0,0,1]} \oplus (2) \CMTred {[0,1,0,3,0]} \oplus \CMTred {[2,0,0,3,0]} \oplus (8) \CMTred {[0,1,1,1,0]} \oplus (3) \CMTred {[0,0,1,0,3]} \oplus (5) \CMTred {[0,0,2,0,1]} \oplus (8) \CMTred {[0,1,0,1,2]} \oplus (7) \CMTred {[2,0,1,1,0]} \oplus \CMTred {[0,0,0,2,3]} \oplus (4) \CMTred {[1,2,0,1,0]} \oplus (2) \CMTred {[3,1,0,1,0]} \oplus (7) \CMTred {[2,0,0,1,2]} \oplus (3) \CMTred {[0,0,1,2,1]} \oplus \CMTred {[0,3,0,0,1]} \oplus \CMTred {[3,0,0,0,3]} \oplus (4) \CMTred {[1,1,0,0,3]} \oplus (3) \CMTred {[3,0,1,0,1]} \oplus (9) \CMTred {[1,1,1,0,1]} \oplus \CMTred {[2,2,0,0,1]} \oplus \CMTred {[3,0,0,2,1]} \oplus \CMTred {[1,0,0,1,4]} \oplus (3) \CMTred {[1,0,2,1,0]} \oplus (4) \CMTred {[1,1,0,2,1]} \oplus (2) \CMTred {[0,2,1,1,0]} \oplus (4) \CMTred {[1,0,1,1,2]} \oplus (2) \CMTred {[0,2,0,1,2]} \oplus (2) \CMTred {[0,1,2,0,1]} \oplus \CMTred {[0,1,1,0,3]} \oplus (2) \CMTred {[2,1,1,1,0]} \oplus \CMTred {[2,0,2,0,1]} \oplus (2) \CMTred {[2,1,0,1,2]} \oplus \CMTred {[2,0,1,0,3]} \oplus \CMTred {[1,2,1,0,1]}$
\item Level-5: $\CMTB {[0,1,0,0,0]} \oplus \CMTB {[2,0,0,0,0]} \oplus (3) \CMTB {[0,0,0,1,1]} \oplus \CMTB {[4,0,0,0,0]} \oplus (4) \CMTB {[0,2,0,0,0]} \oplus (7) \CMTB {[1,0,1,0,0]} \oplus (6) \CMTB {[1,0,0,2,0]} \oplus (5) \CMTB {[1,0,0,0,2]} \oplus (5) \CMTB {[2,1,0,0,0]} \oplus \CMTB {[0,0,0,4,0]} \oplus \CMTB {[0,0,0,0,4]} \oplus (6) \CMTB {[0,0,2,0,0]} \oplus (13) \CMTB {[0,1,0,1,1]} \oplus (6) \CMTB {[0,0,1,2,0]} \oplus (7) \CMTB {[0,0,1,0,2]} \oplus (2) \CMTB {[0,3,0,0,0]} \oplus (12) \CMTB {[2,0,0,1,1]} \oplus (5) \CMTB {[0,0,0,2,2]} \oplus \CMTB {[4,1,0,0,0]} \oplus (6) \CMTB {[3,0,1,0,0]} \oplus (3) \CMTB {[2,2,0,0,0]} \oplus (3) \CMTB {[3,0,0,2,0]} \oplus (5) \CMTB {[3,0,0,0,2]} \oplus (13) \CMTB {[1,1,1,0,0]} \oplus (9) \CMTB {[1,1,0,2,0]} \oplus (12) \CMTB {[1,1,0,0,2]} \oplus (3) \CMTB {[1,0,0,3,1]} \oplus (7) \CMTB {[1,0,0,1,3]} \oplus (2) \CMTB {[0,1,0,0,4]} \oplus (6) \CMTB {[0,1,2,0,0]} \oplus (16) \CMTB {[1,0,1,1,1]} \oplus (9) \CMTB {[0,2,0,1,1]} \oplus (2) \CMTB {[4,0,0,1,1]} \oplus (2) \CMTB {[2,0,0,0,4]} \oplus (6) \CMTB {[2,0,2,0,0]} \oplus (4) \CMTB {[0,1,1,2,0]} \oplus (8) \CMTB {[0,1,1,0,2]} \oplus (11) \CMTB {[2,1,0,1,1]} \oplus (4) \CMTB {[1,2,1,0,0]} \oplus (4) \CMTB {[0,1,0,2,2]} \oplus (3) \CMTB {[2,0,1,2,0]} \oplus (8) \CMTB {[2,0,1,0,2]} \oplus (2) \CMTB {[3,1,1,0,0]} \oplus (3) \CMTB {[0,0,2,1,1]} \oplus (2) \CMTB {[0,0,1,1,3]} \oplus (2) \CMTB {[1,2,0,2,0]} \oplus (4) \CMTB {[1,2,0,0,2]} \oplus (3) \CMTB {[2,0,0,2,2]} \oplus \CMTB {[3,1,0,2,0]} \oplus (2) \CMTB {[3,1,0,0,2]} \oplus \CMTB {[1,0,3,0,0]} \oplus \CMTB {[0,3,0,1,1]} \oplus \CMTB {[0,2,2,0,0]} \oplus \CMTB {[1,0,1,0,4]} \oplus \CMTB {[3,0,0,1,3]} \oplus (3) \CMTB {[1,1,0,1,3]} \oplus (2) \CMTB {[1,0,2,0,2]} \oplus (2) \CMTB {[3,0,1,1,1]} \oplus (6) \CMTB {[1,1,1,1,1]} \oplus \CMTB {[0,2,1,0,2]} \oplus \CMTB {[2,2,0,1,1]} \oplus \CMTB {[2,1,2,0,0]} \oplus \CMTB {[2,1,1,0,2]}$
\item Level-6: $\CMTred {[0,0,0,1,0]} \oplus (3) \CMTred {[1,0,0,0,1]} \oplus (7) \CMTred {[0,1,0,1,0]} \oplus (2) \CMTred {[0,0,0,0,3]} \oplus (6) \CMTred {[2,0,0,1,0]} \oplus (7) \CMTred {[0,0,1,0,1]} \oplus (7) \CMTred {[0,0,0,2,1]} \oplus (6) \CMTred {[3,0,0,0,1]} \oplus (13) \CMTred {[1,1,0,0,1]} \oplus (5) \CMTred {[1,0,0,3,0]} \oplus (3) \CMTred {[4,0,0,1,0]} \oplus (10) \CMTred {[0,2,0,1,0]} \oplus (17) \CMTred {[1,0,1,1,0]} \oplus (14) \CMTred {[1,0,0,1,2]} \oplus (12) \CMTred {[2,1,0,1,0]} \oplus (6) \CMTred {[0,1,0,0,3]} \oplus \CMTred {[5,0,0,0,1]} \oplus (7) \CMTred {[2,0,0,0,3]} \oplus (16) \CMTred {[0,1,1,0,1]} \oplus (2) \CMTred {[0,0,0,1,4]} \oplus (2) \CMTred {[0,0,1,3,0]} \oplus (7) \CMTred {[0,0,2,1,0]} \oplus (13) \CMTred {[0,1,0,2,1]} \oplus (17) \CMTred {[2,0,1,0,1]} \oplus (2) \CMTred {[0,0,0,3,2]} \oplus (11) \CMTred {[1,2,0,0,1]} \oplus (7) \CMTred {[3,1,0,0,1]} \oplus (11) \CMTred {[2,0,0,2,1]} \oplus (9) \CMTred {[0,0,1,1,2]} \oplus \CMTred {[1,0,0,0,5]} \oplus (3) \CMTred {[0,3,0,1,0]} \oplus \CMTred {[3,0,0,3,0]} \oplus (3) \CMTred {[1,1,0,3,0]} \oplus (7) \CMTred {[3,0,1,1,0]} \oplus \CMTred {[4,1,0,1,0]} \oplus (16) \CMTred {[1,1,1,1,0]} \oplus (4) \CMTred {[2,2,0,1,0]} \oplus (7) \CMTred {[3,0,0,1,2]} \oplus (10) \CMTred {[1,0,2,0,1]} \oplus (6) \CMTred {[1,0,1,0,3]} \oplus (3) \CMTred {[0,2,0,0,3]} \oplus (16) \CMTred {[1,1,0,1,2]} \oplus \CMTred {[4,0,0,0,3]} \oplus (6) \CMTred {[0,2,1,0,1]} \oplus (3) \CMTred {[1,0,0,2,3]} \oplus \CMTred {[1,3,0,0,1]} \oplus (2) \CMTred {[4,0,1,0,1]} \oplus (6) \CMTred {[1,0,1,2,1]} \oplus (4) \CMTred {[0,2,0,2,1]} \oplus (4) \CMTred {[2,1,0,0,3]} \oplus \CMTred {[0,0,3,0,1]} \oplus \CMTred {[3,2,0,0,1]} \oplus \CMTred {[4,0,0,2,1]} \oplus (3) \CMTred {[0,1,2,1,0]} \oplus \CMTred {[0,1,0,1,4]} \oplus \CMTred {[0,0,2,0,3]} \oplus (9) \CMTred {[2,1,1,0,1]} \oplus (3) \CMTred {[2,0,2,1,0]} \oplus \CMTred {[2,0,0,1,4]} \oplus (4) \CMTred {[2,1,0,2,1]} \oplus (4) \CMTred {[0,1,1,1,2]} \oplus (2) \CMTred {[1,2,1,1,0]} \oplus (4) \CMTred {[2,0,1,1,2]} \oplus \CMTred {[3,1,1,1,0]} \oplus (2) \CMTred {[1,2,0,1,2]} \oplus \CMTred {[3,1,0,1,2]} \oplus \CMTred {[3,0,2,0,1]} \oplus (2) \CMTred {[1,1,2,0,1]} \oplus \CMTred {[1,1,1,0,3]}$
\item Level-7: $\CMTB {[1,0,0,0,0]} \oplus (4) \CMTB {[0,0,1,0,0]} \oplus (3) \CMTB {[0,0,0,2,0]} \oplus (2) \CMTB {[0,0,0,0,2]} \oplus (2) \CMTB {[3,0,0,0,0]} \oplus (5) \CMTB {[1,1,0,0,0]} \oplus (13) \CMTB {[1,0,0,1,1]} \oplus \CMTB {[5,0,0,0,0]} \oplus (12) \CMTB {[0,1,1,0,0]} \oplus (11) \CMTB {[0,1,0,2,0]} \oplus (9) \CMTB {[0,1,0,0,2]} \oplus (12) \CMTB {[2,0,1,0,0]} \oplus (8) \CMTB {[1,2,0,0,0]} \oplus (5) \CMTB {[3,1,0,0,0]} \oplus (9) \CMTB {[2,0,0,2,0]} \oplus (9) \CMTB {[2,0,0,0,2]} \oplus (5) \CMTB {[0,0,0,3,1]} \oplus (5) \CMTB {[0,0,0,1,3]} \oplus (15) \CMTB {[0,0,1,1,1]} \oplus \CMTB {[1,0,0,4,0]} \oplus (3) \CMTB {[1,0,0,0,4]} \oplus (13) \CMTB {[1,0,2,0,0]} \oplus (12) \CMTB {[3,0,0,1,1]} \oplus (10) \CMTB {[0,2,1,0,0]} \oplus (28) \CMTB {[1,1,0,1,1]} \oplus (3) \CMTB {[1,3,0,0,0]} \oplus (4) \CMTB {[4,0,1,0,0]} \oplus (8) \CMTB {[0,2,0,2,0]} \oplus (9) \CMTB {[0,2,0,0,2]} \oplus (13) \CMTB {[1,0,1,2,0]} \oplus (15) \CMTB {[1,0,1,0,2]} \oplus (2) \CMTB {[3,2,0,0,0]} \oplus (2) \CMTB {[4,0,0,2,0]} \oplus (4) \CMTB {[4,0,0,0,2]} \oplus (12) \CMTB {[1,0,0,2,2]} \oplus (14) \CMTB {[2,1,1,0,0]} \oplus (3) \CMTB {[0,0,3,0,0]} \oplus (9) \CMTB {[2,1,0,2,0]} \oplus (13) \CMTB {[2,1,0,0,2]} \oplus \CMTB {[0,0,1,0,4]} \oplus \CMTB {[0,0,0,2,4]} \oplus (2) \CMTB {[0,0,2,2,0]} \oplus (4) \CMTB {[0,0,2,0,2]} \oplus (4) \CMTB {[0,1,0,3,1]} \oplus (7) \CMTB {[0,1,0,1,3]} \oplus (16) \CMTB {[0,1,1,1,1]} \oplus \CMTB {[5,0,0,1,1]} \oplus (3) \CMTB {[2,0,0,3,1]} \oplus (7) \CMTB {[2,0,0,1,3]} \oplus \CMTB {[0,3,1,0,0]} \oplus (3) \CMTB {[0,0,1,2,2]} \oplus \CMTB {[3,0,0,0,4]} \oplus (17) \CMTB {[2,0,1,1,1]} \oplus (4) \CMTB {[3,0,2,0,0]} \oplus (2) \CMTB {[1,1,0,0,4]} \oplus \CMTB {[0,3,0,2,0]} \oplus \CMTB {[0,3,0,0,2]} \oplus (7) \CMTB {[1,1,2,0,0]} \oplus (11) \CMTB {[1,2,0,1,1]} \oplus (7) \CMTB {[3,1,0,1,1]} \oplus \CMTB {[4,1,1,0,0]} \oplus (2) \CMTB {[2,2,1,0,0]} \oplus (2) \CMTB {[3,0,1,2,0]} \oplus (5) \CMTB {[3,0,1,0,2]} \oplus (5) \CMTB {[1,1,1,2,0]} \oplus (9) \CMTB {[1,1,1,0,2]} \oplus \CMTB {[4,1,0,0,2]} \oplus \CMTB {[2,2,0,2,0]} \oplus (2) \CMTB {[2,2,0,0,2]} \oplus (2) \CMTB {[3,0,0,2,2]} \oplus \CMTB {[0,1,3,0,0]} \oplus (5) \CMTB {[1,1,0,2,2]} \oplus (4) \CMTB {[1,0,2,1,1]} \oplus (2) \CMTB {[1,0,1,1,3]} \oplus \CMTB {[0,2,0,1,3]} \oplus \CMTB {[2,0,3,0,0]} \oplus (2) \CMTB {[0,2,1,1,1]} \oplus \CMTB {[0,1,2,0,2]} \oplus \CMTB {[4,0,1,1,1]} \oplus \CMTB {[2,1,0,1,3]} \oplus \CMTB {[2,0,2,0,2]} \oplus (3) \CMTB {[2,1,1,1,1]}$
\item Level-8: $(2) \CMTred {[0,0,0,0,1]} \oplus (5) \CMTred {[1,0,0,1,0]} \oplus (8) \CMTred {[0,1,0,0,1]} \oplus (3) \CMTred {[0,0,0,3,0]} \oplus (6) \CMTred {[2,0,0,0,1]} \oplus (10) \CMTred {[0,0,1,1,0]} \oplus (7) \CMTred {[0,0,0,1,2]} \oplus (6) \CMTred {[3,0,0,1,0]} \oplus (16) \CMTred {[1,1,0,1,0]} \oplus (5) \CMTred {[1,0,0,0,3]} \oplus (4) \CMTred {[4,0,0,0,1]} \oplus (12) \CMTred {[0,2,0,0,1]} \oplus (19) \CMTred {[1,0,1,0,1]} \oplus (17) \CMTred {[1,0,0,2,1]} \oplus (15) \CMTred {[2,1,0,0,1]} \oplus (6) \CMTred {[0,1,0,3,0]} \oplus \CMTred {[5,0,0,1,0]} \oplus (5) \CMTred {[2,0,0,3,0]} \oplus (19) \CMTred {[0,1,1,1,0]} \oplus \CMTred {[0,0,0,4,1]} \oplus (4) \CMTred {[0,0,1,0,3]} \oplus (8) \CMTred {[0,0,2,0,1]} \oplus (17) \CMTred {[0,1,0,1,2]} \oplus (19) \CMTred {[2,0,1,1,0]} \oplus (4) \CMTred {[0,0,0,2,3]} \oplus (14) \CMTred {[1,2,0,1,0]} \oplus (8) \CMTred {[3,1,0,1,0]} \oplus (16) \CMTred {[2,0,0,1,2]} \oplus (10) \CMTred {[0,0,1,2,1]} \oplus (5) \CMTred {[0,3,0,0,1]} \oplus (4) \CMTred {[3,0,0,0,3]} \oplus (8) \CMTred {[1,1,0,0,3]} \oplus (12) \CMTred {[3,0,1,0,1]} \oplus (3) \CMTred {[4,1,0,0,1]} \oplus (22) \CMTred {[1,1,1,0,1]} \oplus (7) \CMTred {[2,2,0,0,1]} \oplus (7) \CMTred {[3,0,0,2,1]} \oplus (2) \CMTred {[1,0,0,1,4]} \oplus (10) \CMTred {[1,0,2,1,0]} \oplus (3) \CMTred {[1,0,1,3,0]} \oplus (2) \CMTred {[0,2,0,3,0]} \oplus (18) \CMTred {[1,1,0,2,1]} \oplus (7) \CMTred {[0,2,1,1,0]} \oplus (3) \CMTred {[1,0,0,3,2]} \oplus (2) \CMTred {[1,3,0,1,0]} \oplus (3) \CMTred {[4,0,1,1,0]} \oplus (12) \CMTred {[1,0,1,1,2]} \oplus (7) \CMTred {[0,2,0,1,2]} \oplus (2) \CMTred {[2,1,0,3,0]} \oplus \CMTred {[0,0,3,1,0]} \oplus \CMTred {[3,2,0,1,0]} \oplus (3) \CMTred {[4,0,0,1,2]} \oplus (5) \CMTred {[0,1,2,0,1]} \oplus (2) \CMTred {[0,1,1,0,3]} \oplus (10) \CMTred {[2,1,1,1,0]} \oplus (2) \CMTred {[0,1,0,2,3]} \oplus (6) \CMTred {[2,0,2,0,1]} \oplus (10) \CMTred {[2,1,0,1,2]} \oplus (3) \CMTred {[2,0,1,0,3]} \oplus \CMTred {[5,0,1,0,1]} \oplus (4) \CMTred {[0,1,1,2,1]} \oplus \CMTred {[0,0,2,1,2]} \oplus \CMTred {[1,2,0,0,3]} \oplus \CMTred {[2,0,0,2,3]} \oplus \CMTred {[3,1,0,0,3]} \oplus (3) \CMTred {[1,2,1,0,1]} \oplus (4) \CMTred {[2,0,1,2,1]} \oplus (3) \CMTred {[3,1,1,0,1]} \oplus (2) \CMTred {[1,2,0,2,1]} \oplus \CMTred {[3,1,0,2,1]} \oplus \CMTred {[3,0,2,1,0]} \oplus (2) \CMTred {[1,1,2,1,0]} \oplus \CMTred {[1,0,3,0,1]} \oplus \CMTred {[3,0,1,1,2]} \oplus (2) \CMTred {[1,1,1,1,2]}$
\item Level-9: $\CMTB {[0,0,0,0,0]} \oplus (3) \CMTB {[0,1,0,0,0]} \oplus (2) \CMTB {[2,0,0,0,0]} \oplus (6) \CMTB {[0,0,0,1,1]} \oplus \CMTB {[4,0,0,0,0]} \oplus (6) \CMTB {[0,2,0,0,0]} \oplus (10) \CMTB {[1,0,1,0,0]} \oplus (8) \CMTB {[1,0,0,2,0]} \oplus (7) \CMTB {[1,0,0,0,2]} \oplus (6) \CMTB {[2,1,0,0,0]} \oplus \CMTB {[0,0,0,4,0]} \oplus \CMTB {[0,0,0,0,4]} \oplus (8) \CMTB {[0,0,2,0,0]} \oplus (19) \CMTB {[0,1,0,1,1]} \oplus (9) \CMTB {[0,0,1,2,0]} \oplus (8) \CMTB {[0,0,1,0,2]} \oplus (5) \CMTB {[0,3,0,0,0]} \oplus (16) \CMTB {[2,0,0,1,1]} \oplus (8) \CMTB {[0,0,0,2,2]} \oplus (2) \CMTB {[4,1,0,0,0]} \oplus (8) \CMTB {[3,0,1,0,0]} \oplus (6) \CMTB {[2,2,0,0,0]} \oplus (6) \CMTB {[3,0,0,2,0]} \oplus (6) \CMTB {[3,0,0,0,2]} \oplus (19) \CMTB {[1,1,1,0,0]} \oplus (16) \CMTB {[1,1,0,2,0]} \oplus (15) \CMTB {[1,1,0,0,2]} \oplus (8) \CMTB {[1,0,0,3,1]} \oplus (8) \CMTB {[1,0,0,1,3]} \oplus \CMTB {[0,4,0,0,0]} \oplus \CMTB {[0,1,0,4,0]} \oplus \CMTB {[0,1,0,0,4]} \oplus (8) \CMTB {[0,1,2,0,0]} \oplus (24) \CMTB {[1,0,1,1,1]} \oplus (16) \CMTB {[0,2,0,1,1]} \oplus (5) \CMTB {[4,0,0,1,1]} \oplus \CMTB {[2,0,0,4,0]} \oplus \CMTB {[2,0,0,0,4]} \oplus \CMTB {[5,0,1,0,0]} \oplus (9) \CMTB {[2,0,2,0,0]} \oplus \CMTB {[2,3,0,0,0]} \oplus \CMTB {[5,0,0,0,2]} \oplus (9) \CMTB {[0,1,1,2,0]} \oplus (9) \CMTB {[0,1,1,0,2]} \oplus (20) \CMTB {[2,1,0,1,1]} \oplus \CMTB {[0,0,0,3,3]} \oplus (8) \CMTB {[1,2,1,0,0]} \oplus (9) \CMTB {[0,1,0,2,2]} \oplus (9) \CMTB {[2,0,1,2,0]} \oplus (10) \CMTB {[2,0,1,0,2]} \oplus (5) \CMTB {[3,1,1,0,0]} \oplus (5) \CMTB {[0,0,2,1,1]} \oplus (2) \CMTB {[0,0,1,3,1]} \oplus (2) \CMTB {[0,0,1,1,3]} \oplus (6) \CMTB {[1,2,0,2,0]} \oplus (7) \CMTB {[1,2,0,0,2]} \oplus (8) \CMTB {[2,0,0,2,2]} \oplus (3) \CMTB {[3,1,0,2,0]} \oplus (5) \CMTB {[3,1,0,0,2]} \oplus (2) \CMTB {[1,0,3,0,0]} \oplus (2) \CMTB {[0,3,0,1,1]} \oplus \CMTB {[0,2,2,0,0]} \oplus \CMTB {[3,0,0,3,1]} \oplus (2) \CMTB {[3,0,0,1,3]} \oplus (3) \CMTB {[1,1,0,3,1]} \oplus (4) \CMTB {[1,1,0,1,3]} \oplus (2) \CMTB {[1,0,2,2,0]} \oplus (2) \CMTB {[1,0,2,0,2]} \oplus \CMTB {[4,0,2,0,0]} \oplus (6) \CMTB {[3,0,1,1,1]} \oplus (12) \CMTB {[1,1,1,1,1]} \oplus \CMTB {[4,1,0,1,1]} \oplus \CMTB {[0,2,1,2,0]} \oplus \CMTB {[0,2,1,0,2]} \oplus (3) \CMTB {[2,2,0,1,1]} \oplus (2) \CMTB {[2,1,2,0,0]} \oplus \CMTB {[4,0,1,0,2]} \oplus (2) \CMTB {[1,0,1,2,2]} \oplus \CMTB {[0,2,0,2,2]} \oplus \CMTB {[2,1,1,2,0]} \oplus (2) \CMTB {[2,1,1,0,2]} \oplus \CMTB {[0,1,2,1,1]} \oplus \CMTB {[2,1,0,2,2]} \oplus \CMTB {[2,0,2,1,1]}$
\item Level-10: $(2) \CMTred {[0,0,0,1,0]} \oplus (5) \CMTred {[1,0,0,0,1]} \oplus (8) \CMTred {[0,1,0,1,0]} \oplus (3) \CMTred {[0,0,0,0,3]} \oplus (6) \CMTred {[2,0,0,1,0]} \oplus (9) \CMTred {[0,0,1,0,1]} \oplus (7) \CMTred {[0,0,0,2,1]} \oplus (4) \CMTred {[3,0,0,0,1]} \oplus (14) \CMTred {[1,1,0,0,1]} \oplus (5) \CMTred {[1,0,0,3,0]} \oplus (2) \CMTred {[4,0,0,1,0]} \oplus (11) \CMTred {[0,2,0,1,0]} \oplus (17) \CMTred {[1,0,1,1,0]} \oplus (14) \CMTred {[1,0,0,1,2]} \oplus (12) \CMTred {[2,1,0,1,0]} \oplus (5) \CMTred {[0,1,0,0,3]} \oplus \CMTred {[5,0,0,0,1]} \oplus (4) \CMTred {[2,0,0,0,3]} \oplus (16) \CMTred {[0,1,1,0,1]} \oplus \CMTred {[0,0,0,1,4]} \oplus (3) \CMTred {[0,0,1,3,0]} \oplus (7) \CMTred {[0,0,2,1,0]} \oplus (14) \CMTred {[0,1,0,2,1]} \oplus (14) \CMTred {[2,0,1,0,1]} \oplus (3) \CMTred {[0,0,0,3,2]} \oplus (12) \CMTred {[1,2,0,0,1]} \oplus (6) \CMTred {[3,1,0,0,1]} \oplus (13) \CMTred {[2,0,0,2,1]} \oplus (8) \CMTred {[0,0,1,1,2]} \oplus (4) \CMTred {[0,3,0,1,0]} \oplus (2) \CMTred {[3,0,0,3,0]} \oplus (5) \CMTred {[1,1,0,3,0]} \oplus (7) \CMTred {[3,0,1,1,0]} \oplus \CMTred {[4,1,0,1,0]} \oplus (17) \CMTred {[1,1,1,1,0]} \oplus (5) \CMTred {[2,2,0,1,0]} \oplus (6) \CMTred {[3,0,0,1,2]} \oplus \CMTred {[1,0,0,4,1]} \oplus (7) \CMTred {[1,0,2,0,1]} \oplus (2) \CMTred {[1,0,1,0,3]} \oplus (2) \CMTred {[0,2,0,0,3]} \oplus (15) \CMTred {[1,1,0,1,2]} \oplus \CMTred {[4,0,0,0,3]} \oplus (6) \CMTred {[0,2,1,0,1]} \oplus (3) \CMTred {[1,0,0,2,3]} \oplus (2) \CMTred {[1,3,0,0,1]} \oplus (2) \CMTred {[4,0,1,0,1]} \oplus (9) \CMTred {[1,0,1,2,1]} \oplus (5) \CMTred {[0,2,0,2,1]} \oplus (2) \CMTred {[2,1,0,0,3]} \oplus \CMTred {[3,2,0,0,1]} \oplus \CMTred {[4,0,0,2,1]} \oplus (3) \CMTred {[0,1,2,1,0]} \oplus \CMTred {[0,1,1,3,0]} \oplus (8) \CMTred {[2,1,1,0,1]} \oplus \CMTred {[0,1,0,3,2]} \oplus (3) \CMTred {[2,0,2,1,0]} \oplus (6) \CMTred {[2,1,0,2,1]} \oplus \CMTred {[2,0,1,3,0]} \oplus (3) \CMTred {[0,1,1,1,2]} \oplus \CMTred {[0,0,2,2,1]} \oplus \CMTred {[2,0,0,3,2]} \oplus (2) \CMTred {[1,2,1,1,0]} \oplus (3) \CMTred {[2,0,1,1,2]} \oplus \CMTred {[3,1,1,1,0]} \oplus (2) \CMTred {[1,2,0,1,2]} \oplus \CMTred {[3,1,0,1,2]} \oplus \CMTred {[3,0,2,0,1]} \oplus \CMTred {[1,1,2,0,1]} \oplus \CMTred {[1,1,1,2,1]}$
\item Level-11: $\CMTB {[1,0,0,0,0]} \oplus (4) \CMTB {[0,0,1,0,0]} \oplus (2) \CMTB {[0,0,0,2,0]} \oplus (3) \CMTB {[0,0,0,0,2]} \oplus \CMTB {[3,0,0,0,0]} \oplus (5) \CMTB {[1,1,0,0,0]} \oplus (11) \CMTB {[1,0,0,1,1]} \oplus (10) \CMTB {[0,1,1,0,0]} \oplus (7) \CMTB {[0,1,0,2,0]} \oplus (9) \CMTB {[0,1,0,0,2]} \oplus (8) \CMTB {[2,0,1,0,0]} \oplus (6) \CMTB {[1,2,0,0,0]} \oplus (2) \CMTB {[3,1,0,0,0]} \oplus (6) \CMTB {[2,0,0,2,0]} \oplus (6) \CMTB {[2,0,0,0,2]} \oplus (3) \CMTB {[0,0,0,3,1]} \oplus (4) \CMTB {[0,0,0,1,3]} \oplus (11) \CMTB {[0,0,1,1,1]} \oplus \CMTB {[1,0,0,4,0]} \oplus \CMTB {[1,0,0,0,4]} \oplus (8) \CMTB {[1,0,2,0,0]} \oplus (6) \CMTB {[3,0,0,1,1]} \oplus (7) \CMTB {[0,2,1,0,0]} \oplus (19) \CMTB {[1,1,0,1,1]} \oplus (2) \CMTB {[1,3,0,0,0]} \oplus \CMTB {[4,0,1,0,0]} \oplus (5) \CMTB {[0,2,0,2,0]} \oplus (6) \CMTB {[0,2,0,0,2]} \oplus (9) \CMTB {[1,0,1,2,0]} \oplus (9) \CMTB {[1,0,1,0,2]} \oplus \CMTB {[3,2,0,0,0]} \oplus \CMTB {[4,0,0,2,0]} \oplus \CMTB {[4,0,0,0,2]} \oplus (8) \CMTB {[1,0,0,2,2]} \oplus (8) \CMTB {[2,1,1,0,0]} \oplus \CMTB {[0,0,3,0,0]} \oplus (6) \CMTB {[2,1,0,2,0]} \oplus (6) \CMTB {[2,1,0,0,2]} \oplus (2) \CMTB {[0,0,2,2,0]} \oplus \CMTB {[0,0,2,0,2]} \oplus (3) \CMTB {[0,1,0,3,1]} \oplus (3) \CMTB {[0,1,0,1,3]} \oplus (10) \CMTB {[0,1,1,1,1]} \oplus (3) \CMTB {[2,0,0,3,1]} \oplus (2) \CMTB {[2,0,0,1,3]} \oplus \CMTB {[0,3,1,0,0]} \oplus (2) \CMTB {[0,0,1,2,2]} \oplus (9) \CMTB {[2,0,1,1,1]} \oplus \CMTB {[3,0,2,0,0]} \oplus \CMTB {[0,3,0,0,2]} \oplus (3) \CMTB {[1,1,2,0,0]} \oplus (7) \CMTB {[1,2,0,1,1]} \oplus (3) \CMTB {[3,1,0,1,1]} \oplus \CMTB {[2,2,1,0,0]} \oplus \CMTB {[3,0,1,2,0]} \oplus \CMTB {[3,0,1,0,2]} \oplus (3) \CMTB {[1,1,1,2,0]} \oplus (3) \CMTB {[1,1,1,0,2]} \oplus \CMTB {[2,2,0,0,2]} \oplus \CMTB {[3,0,0,2,2]} \oplus (3) \CMTB {[1,1,0,2,2]} \oplus \CMTB {[1,0,2,1,1]} \oplus \CMTB {[1,0,1,3,1]} \oplus \CMTB {[0,2,1,1,1]} \oplus \CMTB {[2,1,1,1,1]}$
\item Level-12: $\CMTred {[0,0,0,0,1]} \oplus (3) \CMTred {[1,0,0,1,0]} \oplus (6) \CMTred {[0,1,0,0,1]} \oplus \CMTred {[0,0,0,3,0]} \oplus (4) \CMTred {[2,0,0,0,1]} \oplus (5) \CMTred {[0,0,1,1,0]} \oplus (5) \CMTred {[0,0,0,1,2]} \oplus (2) \CMTred {[3,0,0,1,0]} \oplus (8) \CMTred {[1,1,0,1,0]} \oplus (4) \CMTred {[1,0,0,0,3]} \oplus (6) \CMTred {[0,2,0,0,1]} \oplus (11) \CMTred {[1,0,1,0,1]} \oplus (7) \CMTred {[1,0,0,2,1]} \oplus (6) \CMTred {[2,1,0,0,1]} \oplus (2) \CMTred {[0,1,0,3,0]} \oplus (2) \CMTred {[2,0,0,3,0]} \oplus (8) \CMTred {[0,1,1,1,0]} \oplus (2) \CMTred {[0,0,1,0,3]} \oplus (4) \CMTred {[0,0,2,0,1]} \oplus (8) \CMTred {[0,1,0,1,2]} \oplus (7) \CMTred {[2,0,1,1,0]} \oplus \CMTred {[0,0,0,2,3]} \oplus (5) \CMTred {[1,2,0,1,0]} \oplus (2) \CMTred {[3,1,0,1,0]} \oplus (6) \CMTred {[2,0,0,1,2]} \oplus (4) \CMTred {[0,0,1,2,1]} \oplus (2) \CMTred {[0,3,0,0,1]} \oplus (2) \CMTred {[1,1,0,0,3]} \oplus (2) \CMTred {[3,0,1,0,1]} \oplus (8) \CMTred {[1,1,1,0,1]} \oplus (2) \CMTred {[2,2,0,0,1]} \oplus (2) \CMTred {[3,0,0,2,1]} \oplus (3) \CMTred {[1,0,2,1,0]} \oplus \CMTred {[1,0,1,3,0]} \oplus (6) \CMTred {[1,1,0,2,1]} \oplus (2) \CMTred {[0,2,1,1,0]} \oplus \CMTred {[1,0,0,3,2]} \oplus (3) \CMTred {[1,0,1,1,2]} \oplus (2) \CMTred {[0,2,0,1,2]} \oplus \CMTred {[0,1,2,0,1]} \oplus (2) \CMTred {[2,1,1,1,0]} \oplus (2) \CMTred {[2,1,0,1,2]} \oplus \CMTred {[0,1,1,2,1]} \oplus \CMTred {[1,2,1,0,1]} \oplus \CMTred {[2,0,1,2,1]}$
\item Level-13: $\CMTB {[0,1,0,0,0]} \oplus \CMTB {[2,0,0,0,0]} \oplus (2) \CMTB {[0,0,0,1,1]} \oplus (2) \CMTB {[0,2,0,0,0]} \oplus (4) \CMTB {[1,0,1,0,0]} \oplus (2) \CMTB {[1,0,0,2,0]} \oplus (4) \CMTB {[1,0,0,0,2]} \oplus (2) \CMTB {[2,1,0,0,0]} \oplus \CMTB {[0,0,0,0,4]} \oplus (3) \CMTB {[0,0,2,0,0]} \oplus (6) \CMTB {[0,1,0,1,1]} \oplus (2) \CMTB {[0,0,1,2,0]} \oplus (4) \CMTB {[0,0,1,0,2]} \oplus \CMTB {[0,3,0,0,0]} \oplus (5) \CMTB {[2,0,0,1,1]} \oplus (2) \CMTB {[0,0,0,2,2]} \oplus \CMTB {[3,0,1,0,0]} \oplus \CMTB {[2,2,0,0,0]} \oplus \CMTB {[3,0,0,2,0]} \oplus \CMTB {[3,0,0,0,2]} \oplus (5) \CMTB {[1,1,1,0,0]} \oplus (3) \CMTB {[1,1,0,2,0]} \oplus (5) \CMTB {[1,1,0,0,2]} \oplus \CMTB {[1,0,0,3,1]} \oplus (2) \CMTB {[1,0,0,1,3]} \oplus (2) \CMTB {[0,1,2,0,0]} \oplus (6) \CMTB {[1,0,1,1,1]} \oplus (3) \CMTB {[0,2,0,1,1]} \oplus \CMTB {[2,0,2,0,0]} \oplus \CMTB {[0,1,1,2,0]} \oplus (2) \CMTB {[0,1,1,0,2]} \oplus (3) \CMTB {[2,1,0,1,1]} \oplus \CMTB {[1,2,1,0,0]} \oplus \CMTB {[0,1,0,2,2]} \oplus \CMTB {[2,0,1,2,0]} \oplus \CMTB {[2,0,1,0,2]} \oplus \CMTB {[0,0,2,1,1]} \oplus \CMTB {[1,2,0,0,2]} \oplus \CMTB {[2,0,0,2,2]} \oplus \CMTB {[1,1,1,1,1]}$
\item Level-14: $\CMTred {[1,0,0,0,1]} \oplus \CMTred {[0,1,0,1,0]} \oplus \CMTred {[0,0,0,0,3]} \oplus \CMTred {[2,0,0,1,0]} \oplus (2) \CMTred {[0,0,1,0,1]} \oplus \CMTred {[0,0,0,2,1]} \oplus \CMTred {[3,0,0,0,1]} \oplus (3) \CMTred {[1,1,0,0,1]} \oplus \CMTred {[0,2,0,1,0]} \oplus (3) \CMTred {[1,0,1,1,0]} \oplus (3) \CMTred {[1,0,0,1,2]} \oplus \CMTred {[2,1,0,1,0]} \oplus \CMTred {[0,1,0,0,3]} \oplus \CMTred {[2,0,0,0,3]} \oplus (3) \CMTred {[0,1,1,0,1]} \oplus \CMTred {[0,0,2,1,0]} \oplus \CMTred {[0,1,0,2,1]} \oplus (2) \CMTred {[2,0,1,0,1]} \oplus \CMTred {[1,2,0,0,1]} \oplus \CMTred {[2,0,0,2,1]} \oplus \CMTred {[0,0,1,1,2]} \oplus \CMTred {[1,1,1,1,0]} \oplus \CMTred {[1,0,2,0,1]} \oplus \CMTred {[1,1,0,1,2]}$
\item Level-15: $\CMTB {[1,0,0,1,1]} \oplus \CMTB {[0,1,1,0,0]} \oplus \CMTB {[0,1,0,0,2]} \oplus \CMTB {[2,0,1,0,0]} \oplus \CMTB {[2,0,0,0,2]} \oplus \CMTB {[0,0,1,1,1]} \oplus \CMTB {[1,0,2,0,0]} \oplus \CMTB {[1,1,0,1,1]} \oplus \CMTB {[1,0,1,0,2]}$
\item Level-16: $\CMTred {[1,0,1,0,1]}$
\end{itemize}

\newpage
\subsection{Dynkin Label Library of ${\cal V}_{\CMTB{[3,0,0,0,1]}}$} \label{appen:L3}

\begin{itemize} \sloppy
\item Level-0: $\CMTred {[3,0,0,0,1]}$
\item Level-1: $\CMTB {[3,0,0,0,0]} \oplus \CMTB {[2,0,1,0,0]} \oplus \CMTB {[3,1,0,0,0]} \oplus \CMTB {[2,0,0,0,2]} \oplus \CMTB {[3,0,0,1,1]}$
\item Level-2: $\CMTred {[2,0,0,0,1]} \oplus (2) \CMTred {[3,0,0,1,0]} \oplus \CMTred {[1,1,0,1,0]} \oplus \CMTred {[4,0,0,0,1]} \oplus \CMTred {[1,0,1,0,1]} \oplus (2) \CMTred {[2,1,0,0,1]} \oplus \CMTred {[2,0,1,1,0]} \oplus \CMTred {[3,1,0,1,0]} \oplus \CMTred {[2,0,0,1,2]} \oplus \CMTred {[3,0,1,0,1]}$
\item Level-3: $\CMTB {[4,0,0,0,0]} \oplus \CMTB {[0,2,0,0,0]} \oplus \CMTB {[1,0,1,0,0]} \oplus \CMTB {[1,0,0,2,0]} \oplus (2) \CMTB {[2,1,0,0,0]} \oplus \CMTB {[0,1,0,1,1]} \oplus (3) \CMTB {[2,0,0,1,1]} \oplus \CMTB {[4,1,0,0,0]} \oplus (3) \CMTB {[3,0,1,0,0]} \oplus \CMTB {[2,2,0,0,0]} \oplus \CMTB {[3,0,0,2,0]} \oplus (2) \CMTB {[3,0,0,0,2]} \oplus (2) \CMTB {[1,1,1,0,0]} \oplus \CMTB {[1,1,0,2,0]} \oplus \CMTB {[1,1,0,0,2]} \oplus \CMTB {[1,0,1,1,1]} \oplus \CMTB {[4,0,0,1,1]} \oplus \CMTB {[2,0,2,0,0]} \oplus (2) \CMTB {[2,1,0,1,1]} \oplus \CMTB {[2,0,1,0,2]} \oplus \CMTB {[3,1,1,0,0]} \oplus \CMTB {[3,1,0,0,2]}$
\item Level-4: $(2) \CMTred {[0,1,0,1,0]} \oplus (2) \CMTred {[2,0,0,1,0]} \oplus \CMTred {[0,0,0,2,1]} \oplus (3) \CMTred {[3,0,0,0,1]} \oplus (3) \CMTred {[1,1,0,0,1]} \oplus \CMTred {[1,0,0,3,0]} \oplus (2) \CMTred {[4,0,0,1,0]} \oplus (2) \CMTred {[0,2,0,1,0]} \oplus (3) \CMTred {[1,0,1,1,0]} \oplus \CMTred {[1,0,0,1,2]} \oplus (4) \CMTred {[2,1,0,1,0]} \oplus \CMTred {[5,0,0,0,1]} \oplus \CMTred {[2,0,0,0,3]} \oplus \CMTred {[0,1,1,0,1]} \oplus \CMTred {[0,1,0,2,1]} \oplus (4) \CMTred {[2,0,1,0,1]} \oplus (2) \CMTred {[1,2,0,0,1]} \oplus (4) \CMTred {[3,1,0,0,1]} \oplus (2) \CMTred {[2,0,0,2,1]} \oplus (2) \CMTred {[3,0,1,1,0]} \oplus \CMTred {[4,1,0,1,0]} \oplus (2) \CMTred {[1,1,1,1,0]} \oplus \CMTred {[2,2,0,1,0]} \oplus (2) \CMTred {[3,0,0,1,2]} \oplus \CMTred {[1,0,2,0,1]} \oplus \CMTred {[1,1,0,1,2]} \oplus \CMTred {[4,0,0,0,3]} \oplus \CMTred {[4,0,1,0,1]} \oplus \CMTred {[2,1,0,0,3]} \oplus \CMTred {[3,2,0,0,1]} \oplus (2) \CMTred {[2,1,1,0,1]}$
\item Level-5: $\CMTB {[0,0,1,0,0]} \oplus \CMTB {[0,0,0,2,0]} \oplus \CMTB {[3,0,0,0,0]} \oplus (2) \CMTB {[1,1,0,0,0]} \oplus (3) \CMTB {[1,0,0,1,1]} \oplus \CMTB {[5,0,0,0,0]} \oplus (3) \CMTB {[0,1,1,0,0]} \oplus (3) \CMTB {[0,1,0,2,0]} \oplus \CMTB {[0,1,0,0,2]} \oplus (4) \CMTB {[2,0,1,0,0]} \oplus (3) \CMTB {[1,2,0,0,0]} \oplus (3) \CMTB {[3,1,0,0,0]} \oplus (3) \CMTB {[2,0,0,2,0]} \oplus (2) \CMTB {[2,0,0,0,2]} \oplus \CMTB {[0,0,0,3,1]} \oplus \CMTB {[0,0,1,1,1]} \oplus (2) \CMTB {[1,0,2,0,0]} \oplus (5) \CMTB {[3,0,0,1,1]} \oplus \CMTB {[5,1,0,0,0]} \oplus (2) \CMTB {[0,2,1,0,0]} \oplus (6) \CMTB {[1,1,0,1,1]} \oplus \CMTB {[1,3,0,0,0]} \oplus (3) \CMTB {[4,0,1,0,0]} \oplus \CMTB {[0,2,0,2,0]} \oplus \CMTB {[0,2,0,0,2]} \oplus (2) \CMTB {[1,0,1,2,0]} \oplus \CMTB {[1,0,1,0,2]} \oplus (2) \CMTB {[3,2,0,0,0]} \oplus \CMTB {[4,0,0,2,0]} \oplus (3) \CMTB {[4,0,0,0,2]} \oplus \CMTB {[1,0,0,2,2]} \oplus (5) \CMTB {[2,1,1,0,0]} \oplus (2) \CMTB {[2,1,0,2,0]} \oplus (4) \CMTB {[2,1,0,0,2]} \oplus \CMTB {[0,1,1,1,1]} \oplus \CMTB {[5,0,0,1,1]} \oplus \CMTB {[2,0,0,1,3]} \oplus \CMTB {[3,0,0,0,4]} \oplus (3) \CMTB {[2,0,1,1,1]} \oplus \CMTB {[3,0,2,0,0]} \oplus \CMTB {[1,1,2,0,0]} \oplus (2) \CMTB {[1,2,0,1,1]} \oplus (3) \CMTB {[3,1,0,1,1]} \oplus \CMTB {[4,1,1,0,0]} \oplus \CMTB {[2,2,1,0,0]} \oplus (2) \CMTB {[3,0,1,0,2]} \oplus \CMTB {[1,1,1,0,2]} \oplus \CMTB {[4,1,0,0,2]} \oplus \CMTB {[2,2,0,0,2]}$
\item Level-6: $\CMTred {[0,0,0,0,1]} \oplus (2) \CMTred {[1,0,0,1,0]} \oplus (3) \CMTred {[0,1,0,0,1]} \oplus \CMTred {[0,0,0,3,0]} \oplus (2) \CMTred {[2,0,0,0,1]} \oplus (3) \CMTred {[0,0,1,1,0]} \oplus \CMTred {[0,0,0,1,2]} \oplus (3) \CMTred {[3,0,0,1,0]} \oplus (6) \CMTred {[1,1,0,1,0]} \oplus (3) \CMTred {[4,0,0,0,1]} \oplus (4) \CMTred {[0,2,0,0,1]} \oplus (4) \CMTred {[1,0,1,0,1]} \oplus (4) \CMTred {[1,0,0,2,1]} \oplus (6) \CMTred {[2,1,0,0,1]} \oplus \CMTred {[0,1,0,3,0]} \oplus (2) \CMTred {[5,0,0,1,0]} \oplus \CMTred {[2,0,0,3,0]} \oplus (4) \CMTred {[0,1,1,1,0]} \oplus (2) \CMTred {[0,1,0,1,2]} \oplus (6) \CMTred {[2,0,1,1,0]} \oplus (5) \CMTred {[1,2,0,1,0]} \oplus (5) \CMTred {[3,1,0,1,0]} \oplus \CMTred {[6,0,0,0,1]} \oplus (4) \CMTred {[2,0,0,1,2]} \oplus \CMTred {[0,0,1,2,1]} \oplus (2) \CMTred {[0,3,0,0,1]} \oplus (2) \CMTred {[3,0,0,0,3]} \oplus \CMTred {[1,1,0,0,3]} \oplus (6) \CMTred {[3,0,1,0,1]} \oplus (4) \CMTred {[4,1,0,0,1]} \oplus (5) \CMTred {[1,1,1,0,1]} \oplus (4) \CMTred {[2,2,0,0,1]} \oplus (2) \CMTred {[3,0,0,2,1]} \oplus \CMTred {[1,0,2,1,0]} \oplus (3) \CMTred {[1,1,0,2,1]} \oplus \CMTred {[0,2,1,1,0]} \oplus \CMTred {[5,1,0,1,0]} \oplus \CMTred {[1,3,0,1,0]} \oplus (2) \CMTred {[4,0,1,1,0]} \oplus \CMTred {[1,0,1,1,2]} \oplus \CMTred {[0,2,0,1,2]} \oplus \CMTred {[3,2,0,1,0]} \oplus (2) \CMTred {[4,0,0,1,2]} \oplus (3) \CMTred {[2,1,1,1,0]} \oplus \CMTred {[2,0,2,0,1]} \oplus (3) \CMTred {[2,1,0,1,2]} \oplus \CMTred {[2,0,1,0,3]} \oplus \CMTred {[5,0,1,0,1]} \oplus \CMTred {[3,1,0,0,3]} \oplus \CMTred {[1,2,1,0,1]} \oplus (2) \CMTred {[3,1,1,0,1]}$
\item Level-7: $\CMTB {[0,0,0,0,0]} \oplus (2) \CMTB {[0,1,0,0,0]} \oplus \CMTB {[2,0,0,0,0]} \oplus (3) \CMTB {[0,0,0,1,1]} \oplus \CMTB {[4,0,0,0,0]} \oplus (3) \CMTB {[0,2,0,0,0]} \oplus (4) \CMTB {[1,0,1,0,0]} \oplus (3) \CMTB {[1,0,0,2,0]} \oplus (2) \CMTB {[1,0,0,0,2]} \oplus (3) \CMTB {[2,1,0,0,0]} \oplus (2) \CMTB {[0,0,2,0,0]} \oplus \CMTB {[6,0,0,0,0]} \oplus (6) \CMTB {[0,1,0,1,1]} \oplus (2) \CMTB {[0,0,1,2,0]} \oplus \CMTB {[0,0,1,0,2]} \oplus (3) \CMTB {[0,3,0,0,0]} \oplus (6) \CMTB {[2,0,0,1,1]} \oplus \CMTB {[0,0,0,2,2]} \oplus (3) \CMTB {[4,1,0,0,0]} \oplus (5) \CMTB {[3,0,1,0,0]} \oplus (4) \CMTB {[2,2,0,0,0]} \oplus (3) \CMTB {[3,0,0,2,0]} \oplus (3) \CMTB {[3,0,0,0,2]} \oplus (7) \CMTB {[1,1,1,0,0]} \oplus (5) \CMTB {[1,1,0,2,0]} \oplus (4) \CMTB {[1,1,0,0,2]} \oplus \CMTB {[1,0,0,3,1]} \oplus \CMTB {[1,0,0,1,3]} \oplus \CMTB {[0,4,0,0,0]} \oplus \CMTB {[0,1,2,0,0]} \oplus \CMTB {[6,1,0,0,0]} \oplus (5) \CMTB {[1,0,1,1,1]} \oplus (5) \CMTB {[0,2,0,1,1]} \oplus (5) \CMTB {[4,0,0,1,1]} \oplus (3) \CMTB {[5,0,1,0,0]} \oplus (3) \CMTB {[2,0,2,0,0]} \oplus \CMTB {[4,2,0,0,0]} \oplus \CMTB {[2,3,0,0,0]} \oplus \CMTB {[5,0,0,2,0]} \oplus (2) \CMTB {[5,0,0,0,2]} \oplus \CMTB {[0,1,1,2,0]} \oplus \CMTB {[0,1,1,0,2]} \oplus (9) \CMTB {[2,1,0,1,1]} \oplus (4) \CMTB {[1,2,1,0,0]} \oplus \CMTB {[0,1,0,2,2]} \oplus (2) \CMTB {[2,0,1,2,0]} \oplus (3) \CMTB {[2,0,1,0,2]} \oplus (5) \CMTB {[3,1,1,0,0]} \oplus (2) \CMTB {[1,2,0,2,0]} \oplus (3) \CMTB {[1,2,0,0,2]} \oplus (2) \CMTB {[2,0,0,2,2]} \oplus (2) \CMTB {[3,1,0,2,0]} \oplus (4) \CMTB {[3,1,0,0,2]} \oplus \CMTB {[6,0,0,1,1]} \oplus \CMTB {[0,3,0,1,1]} \oplus \CMTB {[3,0,0,1,3]} \oplus \CMTB {[1,1,0,1,3]} \oplus \CMTB {[4,0,2,0,0]} \oplus (3) \CMTB {[3,0,1,1,1]} \oplus (2) \CMTB {[1,1,1,1,1]} \oplus (2) \CMTB {[4,1,0,1,1]} \oplus (2) \CMTB {[2,2,0,1,1]} \oplus \CMTB {[2,1,2,0,0]} \oplus \CMTB {[4,0,1,0,2]} \oplus \CMTB {[2,1,1,0,2]}$
\item Level-8: $(2) \CMTred {[0,0,0,1,0]} \oplus (3) \CMTred {[1,0,0,0,1]} \oplus (4) \CMTred {[0,1,0,1,0]} \oplus \CMTred {[0,0,0,0,3]} \oplus (3) \CMTred {[2,0,0,1,0]} \oplus (4) \CMTred {[0,0,1,0,1]} \oplus (2) \CMTred {[0,0,0,2,1]} \oplus (3) \CMTred {[3,0,0,0,1]} \oplus (6) \CMTred {[1,1,0,0,1]} \oplus \CMTred {[1,0,0,3,0]} \oplus (3) \CMTred {[4,0,0,1,0]} \oplus (5) \CMTred {[0,2,0,1,0]} \oplus (6) \CMTred {[1,0,1,1,0]} \oplus (4) \CMTred {[1,0,0,1,2]} \oplus (7) \CMTred {[2,1,0,1,0]} \oplus \CMTred {[0,1,0,0,3]} \oplus (3) \CMTred {[5,0,0,0,1]} \oplus \CMTred {[2,0,0,0,3]} \oplus (5) \CMTred {[0,1,1,0,1]} \oplus \CMTred {[0,0,2,1,0]} \oplus (3) \CMTred {[0,1,0,2,1]} \oplus (6) \CMTred {[2,0,1,0,1]} \oplus (7) \CMTred {[1,2,0,0,1]} \oplus (6) \CMTred {[3,1,0,0,1]} \oplus (2) \CMTred {[6,0,0,1,0]} \oplus (5) \CMTred {[2,0,0,2,1]} \oplus \CMTred {[0,0,1,1,2]} \oplus (3) \CMTred {[0,3,0,1,0]} \oplus \CMTred {[3,0,0,3,0]} \oplus \CMTred {[1,1,0,3,0]} \oplus \CMTred {[7,0,0,0,1]} \oplus (6) \CMTred {[3,0,1,1,0]} \oplus (4) \CMTred {[4,1,0,1,0]} \oplus (6) \CMTred {[1,1,1,1,0]} \oplus (5) \CMTred {[2,2,0,1,0]} \oplus (4) \CMTred {[3,0,0,1,2]} \oplus \CMTred {[1,0,2,0,1]} \oplus \CMTred {[0,2,0,0,3]} \oplus (5) \CMTred {[1,1,0,1,2]} \oplus \CMTred {[4,0,0,0,3]} \oplus (2) \CMTred {[0,2,1,0,1]} \oplus \CMTred {[1,0,0,2,3]} \oplus (2) \CMTred {[5,1,0,0,1]} \oplus (2) \CMTred {[1,3,0,0,1]} \oplus (4) \CMTred {[4,0,1,0,1]} \oplus \CMTred {[1,0,1,2,1]} \oplus \CMTred {[0,2,0,2,1]} \oplus \CMTred {[2,1,0,0,3]} \oplus (2) \CMTred {[3,2,0,0,1]} \oplus (2) \CMTred {[4,0,0,2,1]} \oplus (5) \CMTred {[2,1,1,0,1]} \oplus \CMTred {[2,0,2,1,0]} \oplus (3) \CMTred {[2,1,0,2,1]} \oplus \CMTred {[5,0,1,1,0]} \oplus \CMTred {[5,0,0,1,2]} \oplus \CMTred {[1,2,1,1,0]} \oplus \CMTred {[2,0,1,1,2]} \oplus (2) \CMTred {[3,1,1,1,0]} \oplus \CMTred {[1,2,0,1,2]} \oplus \CMTred {[3,1,0,1,2]} \oplus \CMTred {[3,0,2,0,1]}$
\item Level-9: $\CMTB {[1,0,0,0,0]} \oplus (3) \CMTB {[0,0,1,0,0]} \oplus \CMTB {[0,0,0,2,0]} \oplus (2) \CMTB {[0,0,0,0,2]} \oplus \CMTB {[3,0,0,0,0]} \oplus (3) \CMTB {[1,1,0,0,0]} \oplus (5) \CMTB {[1,0,0,1,1]} \oplus \CMTB {[5,0,0,0,0]} \oplus (5) \CMTB {[0,1,1,0,0]} \oplus (2) \CMTB {[0,1,0,2,0]} \oplus (4) \CMTB {[0,1,0,0,2]} \oplus (5) \CMTB {[2,0,1,0,0]} \oplus (4) \CMTB {[1,2,0,0,0]} \oplus (3) \CMTB {[3,1,0,0,0]} \oplus (3) \CMTB {[2,0,0,2,0]} \oplus (3) \CMTB {[2,0,0,0,2]} \oplus \CMTB {[0,0,0,1,3]} \oplus \CMTB {[7,0,0,0,0]} \oplus (3) \CMTB {[0,0,1,1,1]} \oplus (3) \CMTB {[1,0,2,0,0]} \oplus (6) \CMTB {[3,0,0,1,1]} \oplus (2) \CMTB {[5,1,0,0,0]} \oplus (4) \CMTB {[0,2,1,0,0]} \oplus (9) \CMTB {[1,1,0,1,1]} \oplus (3) \CMTB {[1,3,0,0,0]} \oplus (4) \CMTB {[4,0,1,0,0]} \oplus (2) \CMTB {[0,2,0,2,0]} \oplus (3) \CMTB {[0,2,0,0,2]} \oplus (2) \CMTB {[1,0,1,2,0]} \oplus (3) \CMTB {[1,0,1,0,2]} \oplus (3) \CMTB {[3,2,0,0,0]} \oplus (3) \CMTB {[4,0,0,2,0]} \oplus (2) \CMTB {[4,0,0,0,2]} \oplus (2) \CMTB {[1,0,0,2,2]} \oplus (7) \CMTB {[2,1,1,0,0]} \oplus (5) \CMTB {[2,1,0,2,0]} \oplus (4) \CMTB {[2,1,0,0,2]} \oplus \CMTB {[0,1,0,1,3]} \oplus (2) \CMTB {[0,1,1,1,1]} \oplus (3) \CMTB {[5,0,0,1,1]} \oplus \CMTB {[2,0,0,3,1]} \oplus \CMTB {[2,0,0,1,3]} \oplus \CMTB {[6,0,1,0,0]} \oplus \CMTB {[0,3,1,0,0]} \oplus \CMTB {[6,0,0,0,2]} \oplus (5) \CMTB {[2,0,1,1,1]} \oplus (2) \CMTB {[3,0,2,0,0]} \oplus \CMTB {[0,3,0,0,2]} \oplus \CMTB {[1,1,2,0,0]} \oplus (5) \CMTB {[1,2,0,1,1]} \oplus (6) \CMTB {[3,1,0,1,1]} \oplus (2) \CMTB {[4,1,1,0,0]} \oplus (2) \CMTB {[2,2,1,0,0]} \oplus (2) \CMTB {[3,0,1,2,0]} \oplus \CMTB {[3,0,1,0,2]} \oplus \CMTB {[1,1,1,2,0]} \oplus \CMTB {[1,1,1,0,2]} \oplus \CMTB {[4,1,0,2,0]} \oplus \CMTB {[4,1,0,0,2]} \oplus \CMTB {[2,2,0,2,0]} \oplus \CMTB {[2,2,0,0,2]} \oplus \CMTB {[3,0,0,2,2]} \oplus \CMTB {[1,1,0,2,2]} \oplus \CMTB {[4,0,1,1,1]} \oplus \CMTB {[2,1,1,1,1]}$
\item Level-10: $\CMTred {[0,0,0,0,1]} \oplus (2) \CMTred {[1,0,0,1,0]} \oplus (4) \CMTred {[0,1,0,0,1]} \oplus (3) \CMTred {[2,0,0,0,1]} \oplus (2) \CMTred {[0,0,1,1,0]} \oplus (2) \CMTred {[0,0,0,1,2]} \oplus (3) \CMTred {[3,0,0,1,0]} \oplus (5) \CMTred {[1,1,0,1,0]} \oplus (2) \CMTred {[1,0,0,0,3]} \oplus (2) \CMTred {[4,0,0,0,1]} \oplus (4) \CMTred {[0,2,0,0,1]} \oplus (6) \CMTred {[1,0,1,0,1]} \oplus (2) \CMTred {[1,0,0,2,1]} \oplus (6) \CMTred {[2,1,0,0,1]} \oplus (2) \CMTred {[5,0,0,1,0]} \oplus \CMTred {[2,0,0,3,0]} \oplus (3) \CMTred {[0,1,1,1,0]} \oplus \CMTred {[0,0,1,0,3]} \oplus \CMTred {[0,0,2,0,1]} \oplus (3) \CMTred {[0,1,0,1,2]} \oplus (6) \CMTred {[2,0,1,1,0]} \oplus (5) \CMTred {[1,2,0,1,0]} \oplus (6) \CMTred {[3,1,0,1,0]} \oplus \CMTred {[6,0,0,0,1]} \oplus (4) \CMTred {[2,0,0,1,2]} \oplus (2) \CMTred {[0,3,0,0,1]} \oplus \CMTred {[1,1,0,0,3]} \oplus (4) \CMTred {[3,0,1,0,1]} \oplus (3) \CMTred {[4,1,0,0,1]} \oplus (5) \CMTred {[1,1,1,0,1]} \oplus (4) \CMTred {[2,2,0,0,1]} \oplus (4) \CMTred {[3,0,0,2,1]} \oplus \CMTred {[1,0,2,1,0]} \oplus (3) \CMTred {[1,1,0,2,1]} \oplus \CMTred {[4,0,0,3,0]} \oplus \CMTred {[0,2,1,1,0]} \oplus \CMTred {[5,1,0,1,0]} \oplus \CMTred {[1,3,0,1,0]} \oplus (3) \CMTred {[4,0,1,1,0]} \oplus \CMTred {[1,0,1,1,2]} \oplus \CMTred {[0,2,0,1,2]} \oplus \CMTred {[2,1,0,3,0]} \oplus (2) \CMTred {[3,2,0,1,0]} \oplus \CMTred {[4,0,0,1,2]} \oplus (4) \CMTred {[2,1,1,1,0]} \oplus (2) \CMTred {[2,1,0,1,2]} \oplus \CMTred {[5,0,1,0,1]} \oplus \CMTred {[1,2,1,0,1]} \oplus \CMTred {[2,0,1,2,1]} \oplus \CMTred {[3,1,1,0,1]} \oplus \CMTred {[3,1,0,2,1]}$
\item Level-11: $\CMTB {[0,1,0,0,0]} \oplus \CMTB {[2,0,0,0,0]} \oplus \CMTB {[0,0,0,1,1]} \oplus \CMTB {[4,0,0,0,0]} \oplus (2) \CMTB {[0,2,0,0,0]} \oplus (3) \CMTB {[1,0,1,0,0]} \oplus \CMTB {[1,0,0,2,0]} \oplus (3) \CMTB {[1,0,0,0,2]} \oplus (3) \CMTB {[2,1,0,0,0]} \oplus \CMTB {[0,0,0,0,4]} \oplus \CMTB {[0,0,2,0,0]} \oplus (3) \CMTB {[0,1,0,1,1]} \oplus (2) \CMTB {[0,0,1,0,2]} \oplus \CMTB {[0,3,0,0,0]} \oplus (5) \CMTB {[2,0,0,1,1]} \oplus (2) \CMTB {[4,1,0,0,0]} \oplus (4) \CMTB {[3,0,1,0,0]} \oplus (3) \CMTB {[2,2,0,0,0]} \oplus (3) \CMTB {[3,0,0,2,0]} \oplus (2) \CMTB {[3,0,0,0,2]} \oplus (5) \CMTB {[1,1,1,0,0]} \oplus (2) \CMTB {[1,1,0,2,0]} \oplus (4) \CMTB {[1,1,0,0,2]} \oplus \CMTB {[1,0,0,1,3]} \oplus \CMTB {[0,1,2,0,0]} \oplus (3) \CMTB {[1,0,1,1,1]} \oplus (2) \CMTB {[0,2,0,1,1]} \oplus (3) \CMTB {[4,0,0,1,1]} \oplus \CMTB {[5,0,1,0,0]} \oplus (2) \CMTB {[2,0,2,0,0]} \oplus \CMTB {[4,2,0,0,0]} \oplus \CMTB {[2,3,0,0,0]} \oplus \CMTB {[5,0,0,2,0]} \oplus \CMTB {[0,1,1,0,2]} \oplus (6) \CMTB {[2,1,0,1,1]} \oplus (2) \CMTB {[1,2,1,0,0]} \oplus (2) \CMTB {[2,0,1,2,0]} \oplus \CMTB {[2,0,1,0,2]} \oplus (3) \CMTB {[3,1,1,0,0]} \oplus \CMTB {[1,2,0,2,0]} \oplus \CMTB {[1,2,0,0,2]} \oplus \CMTB {[2,0,0,2,2]} \oplus (3) \CMTB {[3,1,0,2,0]} \oplus \CMTB {[3,1,0,0,2]} \oplus \CMTB {[3,0,0,3,1]} \oplus \CMTB {[3,0,1,1,1]} \oplus \CMTB {[1,1,1,1,1]} \oplus \CMTB {[4,1,0,1,1]} \oplus \CMTB {[2,2,0,1,1]}$
\item Level-12: $\CMTred {[1,0,0,0,1]} \oplus \CMTred {[0,1,0,1,0]} \oplus \CMTred {[0,0,0,0,3]} \oplus (2) \CMTred {[2,0,0,1,0]} \oplus \CMTred {[0,0,1,0,1]} \oplus (3) \CMTred {[3,0,0,0,1]} \oplus (4) \CMTred {[1,1,0,0,1]} \oplus (2) \CMTred {[4,0,0,1,0]} \oplus \CMTred {[0,2,0,1,0]} \oplus (2) \CMTred {[1,0,1,1,0]} \oplus (2) \CMTred {[1,0,0,1,2]} \oplus (4) \CMTred {[2,1,0,1,0]} \oplus \CMTred {[0,1,0,0,3]} \oplus \CMTred {[2,0,0,0,3]} \oplus (2) \CMTred {[0,1,1,0,1]} \oplus (4) \CMTred {[2,0,1,0,1]} \oplus (2) \CMTred {[1,2,0,0,1]} \oplus (3) \CMTred {[3,1,0,0,1]} \oplus (2) \CMTred {[2,0,0,2,1]} \oplus \CMTred {[3,0,0,3,0]} \oplus (3) \CMTred {[3,0,1,1,0]} \oplus (2) \CMTred {[4,1,0,1,0]} \oplus (2) \CMTred {[1,1,1,1,0]} \oplus (2) \CMTred {[2,2,0,1,0]} \oplus \CMTred {[3,0,0,1,2]} \oplus \CMTred {[1,0,2,0,1]} \oplus \CMTred {[1,1,0,1,2]} \oplus \CMTred {[3,2,0,0,1]} \oplus \CMTred {[4,0,0,2,1]} \oplus \CMTred {[2,1,1,0,1]} \oplus \CMTred {[2,1,0,2,1]}$
\item Level-13: $\CMTB {[3,0,0,0,0]} \oplus \CMTB {[1,1,0,0,0]} \oplus \CMTB {[1,0,0,1,1]} \oplus \CMTB {[0,1,1,0,0]} \oplus \CMTB {[0,1,0,0,2]} \oplus (3) \CMTB {[2,0,1,0,0]} \oplus \CMTB {[1,2,0,0,0]} \oplus (2) \CMTB {[3,1,0,0,0]} \oplus \CMTB {[2,0,0,2,0]} \oplus (2) \CMTB {[2,0,0,0,2]} \oplus \CMTB {[1,0,2,0,0]} \oplus (3) \CMTB {[3,0,0,1,1]} \oplus (2) \CMTB {[1,1,0,1,1]} \oplus \CMTB {[4,0,1,0,0]} \oplus \CMTB {[1,0,1,0,2]} \oplus \CMTB {[3,2,0,0,0]} \oplus \CMTB {[4,0,0,2,0]} \oplus (2) \CMTB {[2,1,1,0,0]} \oplus \CMTB {[2,1,0,2,0]} \oplus \CMTB {[2,1,0,0,2]} \oplus \CMTB {[2,0,1,1,1]} \oplus \CMTB {[3,1,0,1,1]}$
\item Level-14: $\CMTred {[2,0,0,0,1]} \oplus (2) \CMTred {[3,0,0,1,0]} \oplus \CMTred {[1,1,0,1,0]} \oplus \CMTred {[4,0,0,0,1]} \oplus \CMTred {[1,0,1,0,1]} \oplus (2) \CMTred {[2,1,0,0,1]} \oplus \CMTred {[2,0,1,1,0]} \oplus \CMTred {[3,1,0,1,0]} \oplus \CMTred {[2,0,0,1,2]} \oplus \CMTred {[3,0,1,0,1]}$
\item Level-15: $\CMTB {[4,0,0,0,0]} \oplus \CMTB {[2,1,0,0,0]} \oplus \CMTB {[2,0,0,1,1]} \oplus \CMTB {[3,0,1,0,0]} \oplus \CMTB {[3,0,0,0,2]}$
\item Level-16: $\CMTred {[3,0,0,0,1]}$
\end{itemize}

\newpage
{\subsection{Dynkin Label Library of ${\cal V}_{\CMTB{[4,0,0,0,0]}}$   \label{appen:L4}}
\begin{itemize} \sloppy
\item Level-0: $\CMTB {[4,0,0,0,0]}$
\item Level-1: $\CMTred {[3,0,0,0,1]} \oplus \CMTred {[4,0,0,1,0]}$
\item Level-2: $\CMTB {[2,0,1,0,0]} \oplus \CMTB {[3,1,0,0,0]} \oplus \CMTB {[3,0,0,1,1]} \oplus \CMTB {[4,0,1,0,0]}$
\item Level-3: $\CMTred {[3,0,0,1,0]} \oplus \CMTred {[1,1,0,1,0]} \oplus \CMTred {[4,0,0,0,1]} \oplus \CMTred {[2,1,0,0,1]} \oplus \CMTred {[2,0,1,1,0]} \oplus \CMTred {[3,1,0,1,0]} \oplus \CMTred {[3,0,1,0,1]} \oplus \CMTred {[4,1,0,0,1]}$
\item Level-4: $\CMTB {[4,0,0,0,0]} \oplus \CMTB {[0,2,0,0,0]} \oplus \CMTB {[1,0,0,2,0]} \oplus \CMTB {[2,1,0,0,0]} \oplus \CMTB {[2,0,0,1,1]} \oplus \CMTB {[4,1,0,0,0]} \oplus \CMTB {[3,0,1,0,0]} \oplus \CMTB {[2,2,0,0,0]} \oplus \CMTB {[3,0,0,2,0]} \oplus \CMTB {[3,0,0,0,2]} \oplus \CMTB {[1,1,1,0,0]} \oplus \CMTB {[1,1,0,2,0]} \oplus \CMTB {[4,0,0,1,1]} \oplus \CMTB {[2,0,2,0,0]} \oplus \CMTB {[4,2,0,0,0]} \oplus \CMTB {[5,0,0,0,2]} \oplus \CMTB {[2,1,0,1,1]} \oplus \CMTB {[3,1,1,0,0]} \oplus \CMTB {[3,1,0,0,2]}$
\item Level-5: $\CMTred {[0,1,0,1,0]} \oplus \CMTred {[2,0,0,1,0]} \oplus \CMTred {[3,0,0,0,1]} \oplus \CMTred {[1,1,0,0,1]} \oplus \CMTred {[1,0,0,3,0]} \oplus \CMTred {[4,0,0,1,0]} \oplus \CMTred {[0,2,0,1,0]} \oplus \CMTred {[1,0,1,1,0]} \oplus (2) \CMTred {[2,1,0,1,0]} \oplus \CMTred {[5,0,0,0,1]} \oplus \CMTred {[2,0,1,0,1]} \oplus \CMTred {[1,2,0,0,1]} \oplus (2) \CMTred {[3,1,0,0,1]} \oplus \CMTred {[2,0,0,2,1]} \oplus \CMTred {[3,0,1,1,0]} \oplus \CMTred {[4,1,0,1,0]} \oplus \CMTred {[1,1,1,1,0]} \oplus \CMTred {[2,2,0,1,0]} \oplus \CMTred {[3,0,0,1,2]} \oplus \CMTred {[4,0,0,0,3]} \oplus \CMTred {[5,1,0,0,1]} \oplus \CMTred {[4,0,1,0,1]} \oplus \CMTred {[3,2,0,0,1]} \oplus \CMTred {[2,1,1,0,1]}$
\item Level-6: $\CMTB {[0,0,1,0,0]} \oplus \CMTB {[1,1,0,0,0]} \oplus \CMTB {[1,0,0,1,1]} \oplus \CMTB {[0,1,1,0,0]} \oplus \CMTB {[0,1,0,2,0]} \oplus (2) \CMTB {[2,0,1,0,0]} \oplus \CMTB {[1,2,0,0,0]} \oplus \CMTB {[3,1,0,0,0]} \oplus \CMTB {[2,0,0,2,0]} \oplus (2) \CMTB {[3,0,0,1,1]} \oplus \CMTB {[5,1,0,0,0]} \oplus \CMTB {[0,2,1,0,0]} \oplus (2) \CMTB {[1,1,0,1,1]} \oplus \CMTB {[1,3,0,0,0]} \oplus (2) \CMTB {[4,0,1,0,0]} \oplus \CMTB {[1,0,1,2,0]} \oplus \CMTB {[3,2,0,0,0]} \oplus \CMTB {[4,0,0,0,2]} \oplus (2) \CMTB {[2,1,1,0,0]} \oplus \CMTB {[2,1,0,2,0]} \oplus \CMTB {[2,1,0,0,2]} \oplus \CMTB {[5,0,0,1,1]} \oplus \CMTB {[6,0,1,0,0]} \oplus \CMTB {[2,0,1,1,1]} \oplus \CMTB {[1,2,0,1,1]} \oplus (2) \CMTB {[3,1,0,1,1]} \oplus \CMTB {[4,1,1,0,0]} \oplus \CMTB {[2,2,1,0,0]} \oplus \CMTB {[3,0,1,0,2]} \oplus \CMTB {[4,1,0,0,2]}$
\item Level-7: $\CMTred {[0,0,0,0,1]} \oplus \CMTred {[1,0,0,1,0]} \oplus \CMTred {[0,1,0,0,1]} \oplus \CMTred {[2,0,0,0,1]} \oplus \CMTred {[0,0,1,1,0]} \oplus \CMTred {[3,0,0,1,0]} \oplus (2) \CMTred {[1,1,0,1,0]} \oplus \CMTred {[4,0,0,0,1]} \oplus \CMTred {[0,2,0,0,1]} \oplus \CMTred {[1,0,1,0,1]} \oplus \CMTred {[1,0,0,2,1]} \oplus (2) \CMTred {[2,1,0,0,1]} \oplus \CMTred {[5,0,0,1,0]} \oplus \CMTred {[0,1,1,1,0]} \oplus (2) \CMTred {[2,0,1,1,0]} \oplus (2) \CMTred {[1,2,0,1,0]} \oplus (2) \CMTred {[3,1,0,1,0]} \oplus \CMTred {[6,0,0,0,1]} \oplus \CMTred {[2,0,0,1,2]} \oplus \CMTred {[0,3,0,0,1]} \oplus \CMTred {[7,0,0,1,0]} \oplus (2) \CMTred {[3,0,1,0,1]} \oplus (2) \CMTred {[4,1,0,0,1]} \oplus \CMTred {[1,1,1,0,1]} \oplus (2) \CMTred {[2,2,0,0,1]} \oplus \CMTred {[3,0,0,2,1]} \oplus \CMTred {[1,1,0,2,1]} \oplus \CMTred {[5,1,0,1,0]} \oplus \CMTred {[1,3,0,1,0]} \oplus \CMTred {[4,0,1,1,0]} \oplus \CMTred {[3,2,0,1,0]} \oplus \CMTred {[4,0,0,1,2]} \oplus \CMTred {[2,1,1,1,0]} \oplus \CMTred {[2,1,0,1,2]} \oplus \CMTred {[5,0,1,0,1]} \oplus \CMTred {[3,1,1,0,1]}$
\item Level-8: $\CMTB {[0,0,0,0,0]} \oplus \CMTB {[0,1,0,0,0]} \oplus \CMTB {[2,0,0,0,0]} \oplus \CMTB {[0,0,0,1,1]} \oplus \CMTB {[4,0,0,0,0]} \oplus \CMTB {[0,2,0,0,0]} \oplus \CMTB {[1,0,1,0,0]} \oplus \CMTB {[1,0,0,2,0]} \oplus \CMTB {[1,0,0,0,2]} \oplus \CMTB {[2,1,0,0,0]} \oplus \CMTB {[0,0,2,0,0]} \oplus \CMTB {[6,0,0,0,0]} \oplus \CMTB {[0,1,0,1,1]} \oplus \CMTB {[0,3,0,0,0]} \oplus (2) \CMTB {[2,0,0,1,1]} \oplus \CMTB {[4,1,0,0,0]} \oplus \CMTB {[3,0,1,0,0]} \oplus (2) \CMTB {[2,2,0,0,0]} \oplus \CMTB {[3,0,0,2,0]} \oplus \CMTB {[3,0,0,0,2]} \oplus (2) \CMTB {[1,1,1,0,0]} \oplus \CMTB {[8,0,0,0,0]} \oplus \CMTB {[1,1,0,2,0]} \oplus \CMTB {[1,1,0,0,2]} \oplus \CMTB {[0,4,0,0,0]} \oplus \CMTB {[6,1,0,0,0]} \oplus \CMTB {[1,0,1,1,1]} \oplus \CMTB {[0,2,0,1,1]} \oplus (2) \CMTB {[4,0,0,1,1]} \oplus \CMTB {[5,0,1,0,0]} \oplus \CMTB {[2,0,2,0,0]} \oplus \CMTB {[4,2,0,0,0]} \oplus \CMTB {[2,3,0,0,0]} \oplus \CMTB {[5,0,0,2,0]} \oplus \CMTB {[5,0,0,0,2]} \oplus (3) \CMTB {[2,1,0,1,1]} \oplus \CMTB {[1,2,1,0,0]} \oplus (2) \CMTB {[3,1,1,0,0]} \oplus \CMTB {[1,2,0,2,0]} \oplus \CMTB {[1,2,0,0,2]} \oplus \CMTB {[2,0,0,2,2]} \oplus \CMTB {[3,1,0,2,0]} \oplus \CMTB {[3,1,0,0,2]} \oplus \CMTB {[6,0,0,1,1]} \oplus \CMTB {[4,0,2,0,0]} \oplus \CMTB {[3,0,1,1,1]} \oplus \CMTB {[4,1,0,1,1]} \oplus \CMTB {[2,2,0,1,1]}$
\item Level-9: $\CMTred {[0,0,0,1,0]} \oplus \CMTred {[1,0,0,0,1]} \oplus \CMTred {[0,1,0,1,0]} \oplus \CMTred {[2,0,0,1,0]} \oplus \CMTred {[0,0,1,0,1]} \oplus \CMTred {[3,0,0,0,1]} \oplus (2) \CMTred {[1,1,0,0,1]} \oplus \CMTred {[4,0,0,1,0]} \oplus \CMTred {[0,2,0,1,0]} \oplus \CMTred {[1,0,1,1,0]} \oplus \CMTred {[1,0,0,1,2]} \oplus (2) \CMTred {[2,1,0,1,0]} \oplus \CMTred {[5,0,0,0,1]} \oplus \CMTred {[0,1,1,0,1]} \oplus (2) \CMTred {[2,0,1,0,1]} \oplus (2) \CMTred {[1,2,0,0,1]} \oplus (2) \CMTred {[3,1,0,0,1]} \oplus \CMTred {[6,0,0,1,0]} \oplus \CMTred {[2,0,0,2,1]} \oplus \CMTred {[0,3,0,1,0]} \oplus \CMTred {[7,0,0,0,1]} \oplus (2) \CMTred {[3,0,1,1,0]} \oplus (2) \CMTred {[4,1,0,1,0]} \oplus \CMTred {[1,1,1,1,0]} \oplus (2) \CMTred {[2,2,0,1,0]} \oplus \CMTred {[3,0,0,1,2]} \oplus \CMTred {[1,1,0,1,2]} \oplus \CMTred {[5,1,0,0,1]} \oplus \CMTred {[1,3,0,0,1]} \oplus \CMTred {[4,0,1,0,1]} \oplus \CMTred {[3,2,0,0,1]} \oplus \CMTred {[4,0,0,2,1]} \oplus \CMTred {[2,1,1,0,1]} \oplus \CMTred {[2,1,0,2,1]} \oplus \CMTred {[5,0,1,1,0]} \oplus \CMTred {[3,1,1,1,0]}$
\item Level-10: $\CMTB {[0,0,1,0,0]} \oplus \CMTB {[1,1,0,0,0]} \oplus \CMTB {[1,0,0,1,1]} \oplus \CMTB {[0,1,1,0,0]} \oplus \CMTB {[0,1,0,0,2]} \oplus (2) \CMTB {[2,0,1,0,0]} \oplus \CMTB {[1,2,0,0,0]} \oplus \CMTB {[3,1,0,0,0]} \oplus \CMTB {[2,0,0,0,2]} \oplus (2) \CMTB {[3,0,0,1,1]} \oplus \CMTB {[5,1,0,0,0]} \oplus \CMTB {[0,2,1,0,0]} \oplus (2) \CMTB {[1,1,0,1,1]} \oplus \CMTB {[1,3,0,0,0]} \oplus (2) \CMTB {[4,0,1,0,0]} \oplus \CMTB {[1,0,1,0,2]} \oplus \CMTB {[3,2,0,0,0]} \oplus \CMTB {[4,0,0,2,0]} \oplus (2) \CMTB {[2,1,1,0,0]} \oplus \CMTB {[2,1,0,2,0]} \oplus \CMTB {[2,1,0,0,2]} \oplus \CMTB {[5,0,0,1,1]} \oplus \CMTB {[6,0,1,0,0]} \oplus \CMTB {[2,0,1,1,1]} \oplus \CMTB {[1,2,0,1,1]} \oplus (2) \CMTB {[3,1,0,1,1]} \oplus \CMTB {[4,1,1,0,0]} \oplus \CMTB {[2,2,1,0,0]} \oplus \CMTB {[3,0,1,2,0]} \oplus \CMTB {[4,1,0,2,0]}$
\item Level-11: $\CMTred {[0,1,0,0,1]} \oplus \CMTred {[2,0,0,0,1]} \oplus \CMTred {[3,0,0,1,0]} \oplus \CMTred {[1,1,0,1,0]} \oplus \CMTred {[1,0,0,0,3]} \oplus \CMTred {[4,0,0,0,1]} \oplus \CMTred {[0,2,0,0,1]} \oplus \CMTred {[1,0,1,0,1]} \oplus (2) \CMTred {[2,1,0,0,1]} \oplus \CMTred {[5,0,0,1,0]} \oplus \CMTred {[2,0,1,1,0]} \oplus \CMTred {[1,2,0,1,0]} \oplus (2) \CMTred {[3,1,0,1,0]} \oplus \CMTred {[2,0,0,1,2]} \oplus \CMTred {[3,0,1,0,1]} \oplus \CMTred {[4,1,0,0,1]} \oplus \CMTred {[1,1,1,0,1]} \oplus \CMTred {[2,2,0,0,1]} \oplus \CMTred {[3,0,0,2,1]} \oplus \CMTred {[4,0,0,3,0]} \oplus \CMTred {[5,1,0,1,0]} \oplus \CMTred {[4,0,1,1,0]} \oplus \CMTred {[3,2,0,1,0]} \oplus \CMTred {[2,1,1,1,0]}$
\item Level-12: $\CMTB {[4,0,0,0,0]} \oplus \CMTB {[0,2,0,0,0]} \oplus \CMTB {[1,0,0,0,2]} \oplus \CMTB {[2,1,0,0,0]} \oplus \CMTB {[2,0,0,1,1]} \oplus \CMTB {[4,1,0,0,0]} \oplus \CMTB {[3,0,1,0,0]} \oplus \CMTB {[2,2,0,0,0]} \oplus \CMTB {[3,0,0,2,0]} \oplus \CMTB {[3,0,0,0,2]} \oplus \CMTB {[1,1,1,0,0]} \oplus \CMTB {[1,1,0,0,2]} \oplus \CMTB {[4,0,0,1,1]} \oplus \CMTB {[2,0,2,0,0]} \oplus \CMTB {[4,2,0,0,0]} \oplus \CMTB {[5,0,0,2,0]} \oplus \CMTB {[2,1,0,1,1]} \oplus \CMTB {[3,1,1,0,0]} \oplus \CMTB {[3,1,0,2,0]}$
\item Level-13: $\CMTred {[3,0,0,0,1]} \oplus \CMTred {[1,1,0,0,1]} \oplus \CMTred {[4,0,0,1,0]} \oplus \CMTred {[2,1,0,1,0]} \oplus \CMTred {[2,0,1,0,1]} \oplus \CMTred {[3,1,0,0,1]} \oplus \CMTred {[3,0,1,1,0]} \oplus \CMTred {[4,1,0,1,0]}$
\item Level-14: $\CMTB {[2,0,1,0,0]} \oplus \CMTB {[3,1,0,0,0]} \oplus \CMTB {[3,0,0,1,1]} \oplus \CMTB {[4,0,1,0,0]}$
\item Level-15: $\CMTred {[3,0,0,1,0]} \oplus \CMTred {[4,0,0,0,1]}$
\item Level-16: $\CMTB {[4,0,0,0,0]}$
\end{itemize}

\newpage
$$~~$$

\end{document}